\documentclass[iop,twocolumn]{emulateapj}

\usepackage{graphicx,amssymb,amsmath,amsfonts}
\usepackage{color,subfigure}
\usepackage{natbib}

\newcommand{\fluxunits}{{\rm erg}\;{\rm s}^{-1}{\rm cm}^{-2}}
\newcommand{\Lhard}{L_{[2-10]{\rm keV}}}
\newcommand{\Lsoft}{L_{[0.5-2]{\rm keV}}}
\newcommand{\Log}{{\rm Log}~}
\newcommand{\NH}{N_{\rm H}}
\newcommand{\Lbol}{L_{\rm bol}}
\newcommand{\kbol}{k_{\rm bol}}
\newcommand{\Lbolobs}{L_{\rm bol,obs}}
\newcommand{\Lx}{L_{\rm X}}
\newcommand{\Mbh}{M_{\rm BH}}
\newcommand{\Lopt}{L_{\rm opt,obs}}
\newcommand{\Lir}{L_{\rm IR,obs}}
\newcommand{\Ldisk}{L_{\rm disk}}
\newcommand{\Ltorus}{L_{\rm torus}}
\newcommand{\f}{f_{\rm obsc}}
\newcommand{\ebvq}{E(B-V)_{\rm qso}}

\DeclareRobustCommand{\ion}[2]{%
\relax\ifmmode
\ifx\testbx\f@series
{\mathbf{#1\,\mathsc{#2}}}\else
{\mathrm{#1\,\mathsc{#2}}}\fi
\else\textup{#1\,{\mdseries\textsc{#2}}}%
\fi}
\newcommand{\rev}[1]{{ #1}}
\newcommand{\revs}[1]{{  #1}}

\begin{document}

\title{The Obscured Fraction of AGN in the XMM-COSMOS Survey: A
  Spectral Energy Distribution Perspective}
\author{E.~Lusso\altaffilmark{1}, J.~F.~Hennawi\altaffilmark{1},
  A.~Comastri\altaffilmark{2}, G.~Zamorani\altaffilmark{2},
  G.~T.~Richards\altaffilmark{3,1}, C.~Vignali\altaffilmark{2,4},
  E.~Treister\altaffilmark{5}, K.~Schawinski\altaffilmark{6},
  M.~Salvato\altaffilmark{7,8}, R.~Gilli\altaffilmark{2}}
\altaffiltext{1}{Max Planck Institut f\"{u}r Astronomie,
  K\"{o}nigstuhl 17, D-69117, Heidelberg, Germany}
\altaffiltext{2}{INAF--Osservatorio Astronomico di Bologna, via
  Ranzani 1, I-40127 Bologna, Italy} \altaffiltext{3}{Department of
  Physics, Drexel University, 3141 Chestnut Street, Philadelphia, PA
  19104, USA} \altaffiltext{4}{Dipartimento di Astronomia,
  Universit\`{a} di Bologna, via Ranzani 1, I-40127 Bologna, Italy}
\altaffiltext{5}{Universidad de Concepci\'{o}n, Departamento de
  Astronom\'{i}a,Casilla 160-C, Concepci\'{o}n, Chile}
\altaffiltext{6}{ETH Zurich Institute for Astronomy HIT,
  Wolfgang-Pauli-Strasse 27, 8093 Zurich, Switzerland}
\altaffiltext{7}{Max Planck Institut f\"{u}r Extraterrestische Physik,
  Postfach 1312, 85741 Garching, Germany} \altaffiltext{8}{Max Planck
  Institut f\"{u}r Plasma Physik and Excellence Cluster, 85748
  Garching, Germany}

\email{lusso@mpia.de}


\begin{abstract}
The fraction of AGN luminosity obscured by dust and re-emitted in the mid-IR is critical for understanding AGN evolution, unification, and parsec-scale AGN physics. For unobscured (Type-1) AGN, where we have a direct view of the accretion disk, the dust covering factor can be measured by computing the ratio of re-processed mid-IR emission to intrinsic nuclear bolometric luminosity.   We use this technique to estimate the obscured AGN fraction as a function of luminosity and redshift for 513 Type-1 AGN from the XMM--COSMOS survey. The re-processed and intrinsic luminosities are computed by fitting the 18-band COSMOS photometry with a custom SED-fitting code, which jointly models emission from: hot-dust in the AGN torus,  the accretion disk, and the host-galaxy. We find a relatively shallow decrease of the luminosity ratio as a function of $\Lbol$, which we interpret as a corresponding decrease in the obscured fraction. In the context of the receding torus model, where dust sublimation reduces the covering factor of more luminous AGN, our measurements require a torus height which increases with luminosity as $h\propto \Lbol^{\quad0.3-0.4}$.  Our obscured fraction-luminosity relation agrees with determinations from SDSS censuses of Type-1 and Type-2 quasars, and favors a torus optically thin to mid-IR radiation. We find a much weaker dependence of obscured fraction on 2--10 keV luminosity than previous determinations from X-ray surveys, and argue that X-ray surveys miss a significant population of highly obscured Compton-thick AGN.  Our analysis shows no clear evidence for evolution of obscured fraction with redshift.
\end{abstract}


\section{Introduction} 
\label{Introduction}
The spectral energy distribution (SED) of AGN covers the full
electromagnetic spectrum from radio to hard X--rays. The most
prominent features are the "near-infrared bump" at $\sim10$ $\mu$m,
and an upturn in the optical-UV, the so-called "big-blue bump"
(BBB; \citealt{1989ApJ...347...29S,1994ApJS...95....1E,2006ApJS..166..470R,2011ApJS..196....2S,2012ApJ...759....6E,2013ApJS..206....4K}).
The BBB is thought to be representative of the emission from the
accretion disc around the supermassive black hole (SMBH), while the
near-infrared bump is due to the presence of dust which re-radiates a
fraction of the optical-UV disc photons at infrared wavelengths.
\par
The presence of this screen of gas and dust surrounding the accretion
disc is the foundation of the unified model for AGN
\citep{1993ARA&A..31..473A,1995PASP..107..803U}, according to which
the observed AGN SED shape arises from different viewing angles
relative to the obscuring material, under the simple assumption
that the dust is smoothly distributed in a toroidal shape.  This
gives rise to the classification distinction between unobscured
(Type-1) AGN and obscured (Type-2) AGN. In the former case, the
observer has a direct view into the nuclear accretion disc region,
while in the latter the optical-UV emission is completely or partially
extincted depending on the inclination angle.  A key ingredient of this
model is the spatial distribution of this dust, which governs the amount of 
obscuration. This can be parametrized by a dust covering factor and its dependence on
nuclear luminosity. In the simplest model where the obscuring medium
is a dusty torus, this covering fraction is directly related to the
opening angle of the torus, or equivalently its height and distance
from the nucleus.  AGN unification postulates that 
the disk and broad
line region emit anisotropically towards the observer for a Type-1 AGN while, \rev{in the optically thin case,}
the re-processed optical--UV emission is re-radiated isotropically in
the infrared. 
\rev{For a toroidal distribution of dust, the mid-infrared luminosity is altered by inclination effects as well (\citealt{2005A&A...436...47D,1994MNRAS.268..235G}).}
\par
Thus the ratio of mid-infrared to bolometric luminosity,
which we define as $R$, provides an estimate of the covering factor of
the dust, and, therefore, it is used to infer the fraction, $\f$, of AGN which are obscured.
The parameter $\f$ is a function of the luminosity ratio $R$, and in the simplest model where the torus is optically thin to its own infrared radiation (see
\citealt{1994MNRAS.268..235G} and our detailed discussion in
Sect.~\ref{AGN obscured fraction: optically thin versus optically thick tori}), it can be written as 
$$\f \simeq R.$$ 
In this paper, we
estimate $R$ for a large sample of X-ray selected Type-1 AGN, which is
used to determine $\f$. \rev{Note that the terms {\it covering factor} and {\it obscured fraction} refer to the same physical quantity, and the amount of AGN obscuration is equivalent to $R$ in the optically thin case only (see also Sect. 2.5 in \citealt{2008ApJ...685..147N}).}
\par
In the context of AGN unification, the obscured fraction $\f$ is
simply a free parameter, and unification makes no prediction for its
dependence on luminosity, although there are compelling
physical arguments for why such a relationship might exist.  
In more luminous AGN the distance of the torus from the central source is
larger, hence the opening angle \revs{(defined as the minimum angle between the perpendicular to the disk and a line of sight which intersect the torus, see Fig.1 in \citealt{1998MNRAS.297L..39S})} is also larger, assuming a constant height of the torus. 
Dust grains in the inner part of the torus are heated by the primary optical--UV continuum radiation, 
and destroyed by evaporation, hence the torus 
extends from the dust sublimation radius outwards (\citealt{1987ApJ...320..537B}, see also \citealt{2007MNRAS.380.1172H}).
We therefore expect a decrease of $\f$
with increasing luminosity. 
This is the so-called {\it receding torus} model
\citep{1991MNRAS.252..586L}.
\par
Another explanation of AGN obscuration does not involve the presence of
the dusty torus, but rather a parsec scale wind/outflow. In this scenario
the torus is identified with the outer region of a hydromagnetic
disk-driven outflow
(\citealt{1994ApJ...434..446K,2006ApJ...648L.101E}; but see also the
recent works done by
\citealt{2011ApJ...741...29D,2012ApJ...761...70D}). Such winds can
also provide an important link between SMBH and host-galaxy (i.e., AGN
feedback). From a theoretical point of view, initial attempts to model the AGN
torus have assumed a smooth dust distribution
\citep{1992ApJ...399L..23P,1994MNRAS.268..235G,2003A&A...404....1V},
but such prescription fails to fully reproduce the mid-infrared SED.
\citet{2002ApJ...570L...9N,2008ApJ...685..160N} suggest
that a clumpy distribution better represents the observations, with
average covering factor value of $\sim0.6$.  Thus an empirical calibration 
of the amount of re-processed optical-UV emission, and its dependence on luminosity, 
would greatly inform theoretical models of parsec scale physics and the 
environment of AGN. 
\par
Broadly speaking, there is mounting evidence that supermassive black holes may be characterized by significant differences in their accretion processes, triggers, and environment as a function of luminosity and/or cosmic time 
(\citealt{hopkins07}). According to the classic {\it Soltan argument} \citep{1982MNRAS.200..115S}, the luminosity density of AGN emission over the history of the Universe \rev{should be commensurate with} the local 
mass density of supermassive black holes. However, this commensurability requires that the census
of cosmic AGN accretion accounts for the amount of obscured accretion, which can depend on both 
luminosity and cosmic time. 
The amount of obscuration is therefore a fundamental ingredient in order to understand the formation history of SMBHs (\citealt{hopkins07}).
\par
The AGN obscured fraction can be determined via multiple independent
methods. One is to use AGN {\it demographics}, that is, by conducting
a census of AGN at a wavelength which is agnostic to viewing angle
(such that both Type-1 and Type-2 AGN are selected), and then simply
computing the obscuring fraction as $\f\simeq$~N$_{\rm
  Type-2}$/(N$_{\rm Type-1}$ + N$_{\rm Type-2}$). Such determinations
typically require spectroscopic follow-up to distinguish Type-1
(broad-line) from Type-2 (narrow line) AGN, but if multi-wavelength
data are available, an SED-based classification can be adopted.  The
amount of obscuration is also a crucial ingredient for models of the
cosmic X--ray background (XRB, \citealt{2007A&A...463...79G}). Indeed,
the XRB amplitude and spectral shape depends on the relative
contributions of unobscured, moderately obscured (Compton-thin) AGN,
and highly obscured (Compton-thick) AGN. Thus XRB synthesis models
which attempt to reproduce observations of the XRB, give another
independent estimate of the obscured fraction given assumptions about
the Compton-thin populations.  Finally, the obscured fraction can be
determined from an analysis of AGN \emph{SEDs}, because the ratio of
the reprocessed infrared luminosity to bolometric AGN emission depends
on the covering factor of the obscuring medium
(e.g., \citealt{2007A&A...468..979M}, hereafter M07; \citealt{2008ApJ...679..140T,2012ApJ...757..181S}).  It is the latter approach which we pursue in this paper.
\par
The first evidence for a luminosity dependent obscured fraction was
reported in \citet{1982ApJ...256..410L}, where $\f$ is parametrized by the hard-to-soft X--ray luminosity ratio ($\Lhard/\Lsoft$); while
\citet{2003ApJ...598..886U} first measured a
significant decrease in the fraction of obscured AGN with increasing
X--ray luminosity, by employing a demographic approach (similar
results have been obtained, with the same methodology, from
independent X-ray selected AGN samples by
\citealt{2003ApJ...596L..23S,2004NuPhS.132...86H,2005ApJ...635..864L,2006ApJ...652L..79T,2008A&A...490..905H}).
The obscured fraction in these studies decreases from 0.8 to 0.1 with
$\Lhard$ increasing from $10^{42}$ erg s$^{-1}$ to $10^{46}$ erg
s$^{-1}$.  Demographic analyses, employing optically selected samples
from SDSS (\citealt{2005MNRAS.360..565S}), and radio-selected AGN
samples (\citealt{2004MNRAS.349..503G}), have found
a similar trend between $\f$ and [\ion{O}{iii}] luminosity, with an
obscured fraction decreasing from 0.85 to 0.5 with increasing
[\ion{O}{iii}] luminosity ($10^{44-46}$ erg s$^{-1}$, see also \citealt{2000MNRAS.316..449W} for similar results by employing the [\ion{O}{ii}] emission-line luminosity).
The demographic analysis presented by \citet{2012arXiv1209.6055A} for a sample of WISE selected AGN in the Bo$\rm\ddot{o}$tes field confirms that a relation between the Type-1 AGN fraction and the AGN bolometric luminosity also exists. The Type-1 AGN fraction estimated by \citet{2012arXiv1209.6055A} ranges from $\sim0.3$ to $\sim0.64$ with $\Lbol=10^{44.3-45.6}$ erg s$^{-1}$, which can be translated into an obscured fraction decreasing from 0.7 to 0.36 with increasing $\Lbol$.
\par
Inherent in 
all of these demographic studies are ambiguities in how Type-1/2 AGN are identified and 
defined, and therefore their selection of AGN could be biased. 
For example, X--ray surveys miss Compton-thick AGN as pointed out by
\citet{2007A&A...463...79G}, where, although the fraction of obscured
AGN is assumed to decrease with X--ray luminosity, a non-negligible
population of obscured AGN is still required to properly model the XRB
spectrum. Optical selection of Type-2 AGN may miss obscured objects
which do not show strong narrow emission line regions
(\citealt{2008AJ....136.2373R}), while radio-loud objects might have
different obscured fraction than radio-quiet ones in radio selected
AGN samples (\citealt{2000MNRAS.316..449W}). 
\par
Previous work using SED-based analyses to constrain the obscured
fraction has been
performed by several studies using AGN samples selected from SDSS
(e.g.,
\citealt{2007ApJ...661...30G,2008MNRAS.386.1252H,2012arXiv1205.4543R,2012arXiv1212.4245M}),
and {\it Spitzer} (e.g., M07;
\citealt{2008ApJ...679..140T,2009MNRAS.399.1206H}). 
SED-based studies have found covering factors of the
order of $0.3-0.4$ , considering optically
selected AGN samples from SDSS (\citealt{2012arXiv1205.4543R})
and the well-studied CLASXS \textit{Chandra} survey
\citep{2009MNRAS.397.1326R}.  \textit{Spitzer}-selected Type-1
AGN samples show average covering factors from hot-dust
clouds\footnote{In the following we will not make any
distinction between emission from the \textit{torus} and
the emission from the \textit{hot-dust}.} that range from
$0.15$ \citep{2012MNRAS.420..526M} to $\sim0.4$
(\citealt{2008MNRAS.386..697R}; see also
\citealt{2009MNRAS.399.1206H} for a sample of Type-2 AGN).
All SED-based studies need photometric coverage over a broad
range of wavelength, from infrared to \rev{optical--UV.}
Optical and infrared selected AGN samples did not have the
 X--ray coverage for a large number of objects. 
 Therefore, in order to compare the results coming from X--ray analyses with
infrared/optical ones, a bolometric correction ($\kbol$) needs
to be assumed.  Corrections are also often assumed
to estimate the total infrared and/or optical luminosity when
the necessary multi wavelength coverage is not present (e.g.,
M07; \citealt{2008ApJ...679..140T}).  Very few studies, so
far, have presented the obscured fractions corrected for the
host-galaxy emission, subtracting this component from the
bolometric budget (see, for example, the recent work by
\citealt{2012ApJ...757..181S}).  However, even in these works, the
specific assumptions about dust geometry and emission in the
parametrization of covering factor are usually not discussed,
and no reddening correction of the disk emission in the
optical is applied.
\par
The evolution of the AGN obscured fraction with redshift is even more
uncertain.  \citet{2003ApJ...598..886U} did not find clear evidence
for a redshift dependence (see also \citealt{2007A&A...463...79G}), while
recently \citet{2008A&A...490..905H} have argued for a significant
increase of the obscured fraction with redshift (\citealt{2006ApJ...653.1070B,2006ApJ...652L..79T}; but see also \citealt{2010AIPC.1248..359G}).
\citet{2012A&A...546A..84I} also show that the fraction of absorbed
AGN at high luminosity may be higher at high redshift than in the
local Universe considering a rest-frame 9--20 keV selection of heavily
obscured AGN at $z>1.7$ from the deep XMM-CDFS survey
(see also \citealt{2013MNRAS.428..354V} for similar results at $z>3$).

The goal of the present study is to measure the covering factor of the
AGN obscuring medium (i.e. the obscured fraction), and its dependence
on luminosity and redshift. Our approach is to consider a sample of
Type-1 AGN, such that both optical-UV emission from the accretion disk
and reprocessed infrared emission from the torus can be measured. The
ratio of infrared to optical-UV luminosity can then be used to
determine the covering factor of dust. We utilize SED fitting
(\citealt{2011A&A...534A.110L,2012MNRAS.425..623L}, hereafter L11 and L12, respectively), to conduct a spectral
decomposition of the various emission components in the AGN, and thus
obtain robust estimate of the nuclear and torus emission.  We thus
require
a well sampled SED over a broad range of wavelength. In particular,
far-infrared data are fundamental to probe the star formation
activity, while mid-infrared observations are necessary to cover the
wavelengths where most of the re-processed AGN optical-UV luminosity is expected to
be emitted.  For these reasons, we carried out our analysis over a
sample of Type-1 AGN drawn from the XMM-COSMOS survey, which is a
unique area given its deep and comprehensive multi-wavelength coverage:
infrared coverage from {\it Spitzer} and \textit{Herschel}, optical
bands with {\it Hubble}, {\it Subaru}, SDSS and other ground-based
telescopes, near- and far-ultraviolet bands with the \textit{Galaxy
  Evolution Explorer} (GALEX) and X-rays with XMM--{\it Newton}.  Our
approach exploits the best available multi wavelength coverage, is less biased against Type-2 AGN classification inherent to demographic studies,
and makes no assumptions about bolometric corrections since we
directly fit the full AGN SED.  Furthermore, our SED-fitting approach
explicitly corrects for the effect of intrinsic AGN reddening, and
subtracts off the contaminating emission from the host-galaxy.

A comparison of our measurements of the obscured fraction with that
determined independently from AGN demographics or synthesis of the XRB
is interesting for several reasons. Naturally, these obscured
fractions should all agree, but each approach depends on an
independent set of assumptions. For example, as we will see in
Sect.~\ref{AGN obscured fraction: optically thin versus optically thick tori}, the results
of our SED-fitting approach depend somewhat on the properties of the
dust in the torus (i.e. optically thick or thin, anisotropic emission,
etc.), which can thus be constrained by comparing to an independent
determination from another technique. Demographic measurements of the obscured fraction
require an unbiased survey of the total AGN population; hence
a disagreement with our SED based obscured fraction estimates could
reveal biases in these surveys. For example X--ray surveys could be
missing Compton thick Type-2 AGN, mid-IR selection could be severely
contaminated by star-forming galaxies, and optical Type-2 AGN surveys
might be missing sources lacking strong narrow line emission.
\par
The structure of this paper is as follows. In Section~\ref{Sample
  description} we discuss the sample selection and the employed
photometric data-sets. In Sect.~\ref{SED-fitting method} we describe
the SED-fitting code and, in particular, the new developments with
respect to L11 and L12. The luminosities computed at different
wavelengths will be presented in Section~\ref{Analysis}. Our $R$
estimates and their variation with $\Lbol$ are presented in
Section~\ref{Mid-infrared to bolometric luminosity ratio versus bolometric luminosity}, while the results of our analysis are
discussed in Section~\ref{Discussion}. Conclusions are outlined in
Sect.~\ref{Summary and Conclusions}.  
Details regarding the SED-fitting code and discussion about outliers are presented in the Appendix.
We adopt a concordance
$\Lambda-$cosmology with $H_{0}=70\, \rm{km \,s^{-1}\, Mpc^{-1}}$,
$\Omega_{M}=0.27$, $\Omega_{\Lambda}=1-\Omega_{M}$
(\citealt{komatsu09}).

\section{Sample description}
\label{Sample description}
The Type-1 AGN sample discussed in this paper is extracted from the XMM-COSMOS catalog which comprises $1822$ point--like X--ray sources detected by XMM-\textit{Newton} over an area of $\sim 2~\rm deg^2$ \citep{hasinger07,cappelluti09}.
All the details about the catalog are reported in \cite{2010ApJ...716..348B}.
We consider in this analysis 1577 X--ray selected sources for which a reliable optical counterpart can be associated (see discussion in \citealt{2010ApJ...716..348B}, Table 1\footnote{The multi-wavelength XMM-COSMOS catalog can be retrieved from: http://www.mpe.mpg.de/XMMCosmos/xmm53\_release/, version $1^{\rm st}$ November 2011. This is an updated version of the catalog already published by Brusa et al. (2010), which includes the photometric redshift catalog by Salvato et al. (2011), and new spectroscopic redshift measurements.}).
We have selected 1375 X--ray sources detected in the [0.5-2]~keV band at a flux larger than $5\times10^{-16}\fluxunits$ (see \citealt{2010ApJ...716..348B}). From this sample, 403
objects are spectroscopically classified as broad-line AGN on the basis of broad emission lines ($FWHM > 2000 \,{\rm km \; s^{-1}}$) in their optical spectra (see \citealt{lilly07,trump09}). The origin of spectroscopic redshifts for the $403$ sources is as follows: $64$ objects from the SDSS archive (\citealt{adelmanmccarthy05,kauffmann03}), $75$ from MMT observations (\citealt{prescott06}), $71$ from the IMACS observation campaign (\citealt{trump07}), $143$ from the zCOSMOS bright $20$k sample (see \citealt{lilly07}), $44$ from the zCOSMOS faint catalog, and $5$ from individual Keck runs  \citep{2010ApJ...716..348B}. We will refer to this sample as the ``spectro-z" Type-1 AGN sample.
For a detailed description of the sample properties see \citet{2012ApJ...759....6E} and \citet{2012arXiv1210.3033H}.
\par
\revs{Unfortunately, the spectroscopic information is available only for a fraction of the objects and, the AGN with spectroscopy data are, on average, brighter in the optical bands than those without spectroscopy. This would potentially introduce a bias in our results if we used only AGN with spectroscopy. To avoid this bias, we added to the spectro-z sample a sample of 136 Type-1 AGN defined as such via SED fitting (``photo-z" sample hereafter).
The information on the nature of these sources is retrieved from \citet[S09 hereafter]{salvato09} and \citet[S11 hereafter]{2011ApJ...742...61S}. In these papers the authors compute reliable redshifts for the entire XMM--COSMOS survey. While the reader should refer to the original papers for more detailed description, in the following we provide the relevant information.}

\rev{Each source in XMM-COSMOS has been fitted with different libraries and priors, depending on the morphology \revs{(from deep COSMOS HST/ACS images, see Sect.~3 in S11)}, variability \revs{(see Sect.~3 in S11 for details)}, and X--ray flux. If the object was morphologically extended in the optical bands than it was assumed to be galaxy dominated, and a library of normal galaxies with emission lines was considered (i.e., \citealt{ilbert09}). Otherwise, a dedicated library including few galaxies, local AGN, QSO, and hybrids was created, with different luminosity priors (see S09 for details on the priors and Fig.~8 in S11). 
In particular, hybrids were created by assuming a varying ratio between AGN and galaxy templates (see Table~2 in S09). 
\revs{The library and the priors adopted, allowed the authors to obtain reliable photometric redshift for the entire sample. More specifically, using a sample of 590 sources in COSMOS, S11 obtained an accuracy of $\sigma_{\Delta z/(1+z)}\sim0.015$ and a fraction of outliers of 6\% for the entire Chandra--COSMOS sample.}
Moreover, the classification done via SED fitting with the one obtained via hardness ratio was consistent. The sources classified as AGN dominated are clearly locate were they were expected to be (Fig.~10 in S09). 
In this work we selected all sources with a best-fit photometric classification consistent with an AGN-dominated SED (i.e., $19\leq{\rm SED-Type}\leq30$ as presented by S09).
We double-check the photo-z classification by comparing the spectroscopic class with the photometric one. 
The large majority of the broad emission line AGN in the spectro-z sample are classified as Type-1 AGN by the SED fitting as well (342/403; 85\%), while the number of spectroscopic Type-1 AGN which have SED--Type different from the one mentioned above is relatively small (61/403).

\revs{The expected contamination and incompleteness in the classification method for the analyzed Type-1 AGN sample is already presented by L12. They found that the Type-1 sample is expected to be contaminated (i.e. Type-2 AGN misclassified as Type 1 AGN from the SED analysis) at the level of $\sim$1.6\% and incomplete (i.e. Type-1 AGN misclassified as Type-2 AGN, and therefore not included in our Type-1 sample) at the level of $\sim$9.2\% (see their Sect. 2.1 for details).}

The SED-fitting classification is not necessarily incorrect, but rather been more sensitive of the global properties of the sources. It is also important to note that the fitting procedure covers a wide wavelength range (from U-band to 8~$\mu$m), while the spectrum covers few thousand Angstrom. 
Therefore, the AGN classification provided by Salvato et al.~for the photo-z sample is reasonably robust. All the results presented in the paper still hold without considering the photo-z, but their inclusion increases the sample statistics and allows us to extend our sample at fainter magnitudes.
}

In the following, we assume that 136 X-ray sources, classified by the SED fitting with an AGN-dominated SED are Type-1 AGN.
We will refer to this sample as the ``photo-z" Type-1 AGN sample.
The final Type-1 AGN sample used in our analysis comprises 539 X-ray selected AGN and it spans a wide range of redshifts ($0.04<z<4.25$, with a median redshift of 1.73).
\rev{As pointed out by \citet{2008ApJ...679..140T}, considering sources over a broad range of redshifts may introduce a bias in the $\f-L$ relationship. We have discussed this issue in Sects. \ref{Dependence of obscured AGN fraction with bolometric luminosity} and \ref{Evolution with redshift}.}

\subsection{Multi-wavelength coverage}
\label{Multi-wavelength coverage}
The catalog includes multi-wavelength data from \rev{far-infrared} to hard X-rays: \textit{Herschel} data at 160~$\mu$m and 100~$\mu$m (\citealt{2011A&A...532A..90L}), 70 $\mu$m and 24 $\mu$m MIPS GO3 data (\citealt{2009ApJ...703..222L}), IRAC flux densities (\citealt{sanders07,2010ApJ...709..644I}), near-infrared J UKIRT (\citealt{2008yCat.2284....0C}), H-band \citep{2010ApJ...708..202M}, CFHT/K-band data (\citealt{mccraken08}), HST/ACS F814W imaging of the COSMOS field (\citealt{koekemoer07}), optical multiband photometry (SDSS, Subaru, \citealt{capak07}), near- and far-ultraviolet bands with GALEX (\citealt{2007ApJS..172..468Z}).
The observations in the \revs{optical-UV and near-infrared} bands are not simultaneous, as they span a time interval of about 5 years: 2001 (SDSS), 2004 (Subaru and CFHT) and 2006 (IRAC). 
\rev{In order to reduce possible variability effects, we have selected the bands closest in time to the IRAC observations (i.e., we excluded SDSS data, that in any case are less deep than other data available in similar bands).}
All the data for the SED computation were shifted to the rest frame, so that no K-corrections were needed.
Galactic reddening has been taken into account: we used the selective attenuation of the stellar continuum $k(\lambda)$ taken from Table 11 of \cite{capak07}. Galactic extinction is estimated from \cite{schlegel98} for each object. 
We decided to \rev{ignore} the near-UV GALEX (NUV) band for objects with redshift \rev{higher} than 1, and far-UV GALEX (FUV) band for sources with redshift \rev{higher} than 0.3 in order to avoid IGM absorption at wavelengths below 1216~\AA{} in the rest-frame.
\par
In the far-infrared, the inclusion of \textit{Herschel} data at 100~$\mu$m and 160~$\mu$m (\citealt{2011A&A...532A..90L}) better constrains the AGN emission in the mid-infrared. The number of detections at 100~$\mu$m is 77 (14\%, 67 and 10 detections in the spectro-z and photo-z, respectively), while at 160~$\mu$m it is 68 (13\%, 58 and 10 detections in the spectro-z and photo-z, respectively). 

Count rates in the 0.5-2 keV and 2-10 keV are converted into monochromatic X--ray fluxes in the observed frame at 1 and 4 keV, respectively, considering a Galactic column density $\NH = 2.5 \times 10^{20}\,\rm cm^{-2}$ (see \citealt{1990ARA&A..28..215D,2005A&A...440..775K}).
We have computed the integrated unabsorbed luminosity in the [0.5-2]keV and [2-10]keV bands, for a sub-sample of 133 Type-1 AGN (25\%, 102 spectro-z and 31 photo-z), for which we have an estimate of the column density $\NH$ from spectral analysis (see \citealt{2007ApJS..172..368M,2010A&A...514A..85M}), while for 75 AGN (48 spectro-z and 27 photo-z) absorption is estimated from hardness ratios \citep{2010ApJ...716..348B}. For the rest of the sample (61\%, 331/539) the hardness ratios are consistent with no intrinsic absorption.
The integrated intrinsic unabsorbed luminosity is computed assuming a power-law spectrum with slope $\Gamma=2$ and $\Gamma=1.7$ for the [0.5-2]keV and [2-10]keV bands, respectively \citep{cappelluti09}. 
\rev{The choice of two different $\Gamma$ values is consistent with the procedure adopted by \citet{cappelluti09} in order to account for the "soft excess" observed in AGN.}
The average shift induced by the correction for absorption in the Type-1 sample is, as expected, small in the soft band $\langle\Delta \Log\Lsoft\rangle=0.10\pm0.01$, and negligible in the hard band.

\section{SED-fitting method}
\label{SED-fitting method}
\begin{figure*}
\epsscale{1.16}
\plottwo{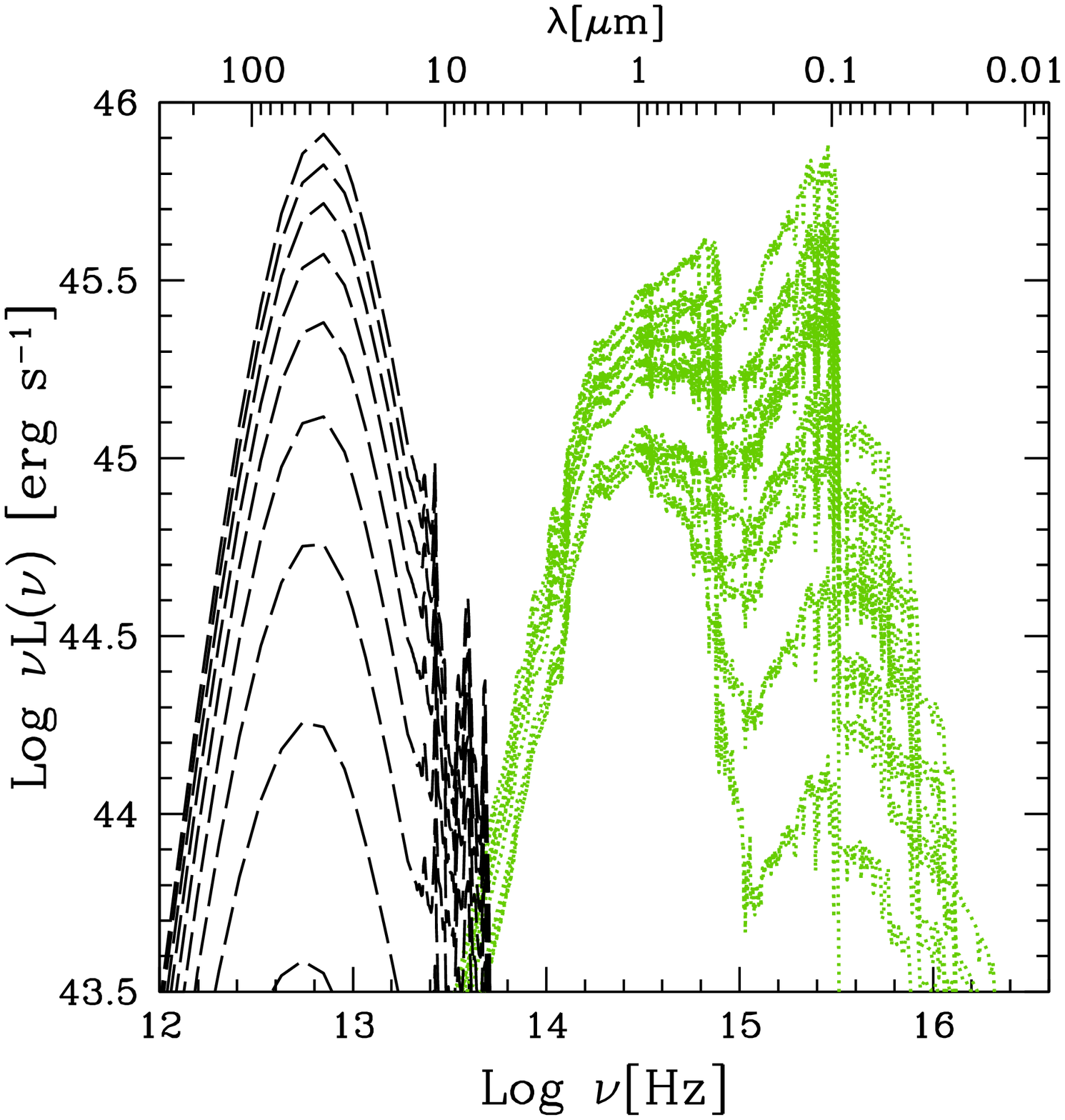}{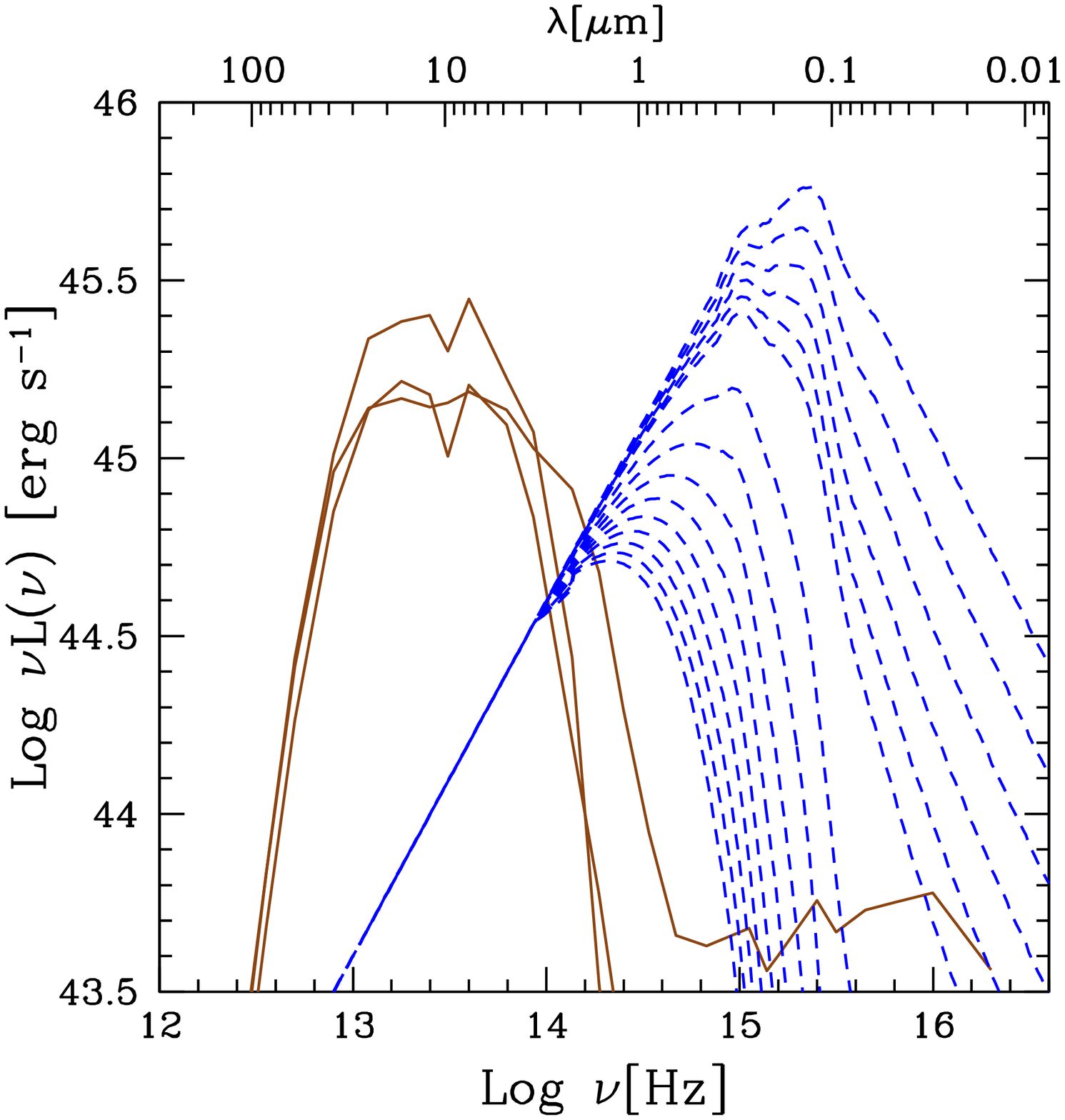}
\caption{Examples of templates employed in the SED-fitting code. \textit{Left panel}: The black long-dashed and the green dotted lines correspond to the starburst and host-galaxy templates, respectively. \textit{Right panel}: The brown solid and blue dashed lines correspond to the hot-dust from reprocessed AGN emission and BBB templates with increasing reddening, respectively.}
\label{templates}
\end{figure*}
The SED-fitting code presented in the present paper is a modified version of the one already employed in L11 (see their Section~5) and L12 for a sample of X-ray selected AGN in COSMOS over a wide range of obscuring column densities ($20.4\leq\Log \NH \,[\rm cm^{-2}]\leq24$).
The major improvement is the addition of a fourth component in the fit to those already considered, i.e. the cold-dust emission from star-forming regions, the hot-dust emission (AGN torus), and optical-UV emission from the evolving stellar population.  
The additional fourth component represents the AGN emission in the optical-UV from the BBB. 
\par
We used two different starburst template libraries for the SED-fitting: \citet{2001ApJ...556..562C} and \citet{2002ApJ...576..159D}. These template libraries represent a wide range of SED shapes and luminosities and are widely used in the literature.
The total far-infrared template sample used in our analysis is composed of 169 templates (105 from \citealt{2001ApJ...556..562C} and 64 from \citealt{2002ApJ...576..159D}), and they have been used to fit the cold dust alone, i.e. far-IR emission.
A subsample of starburst templates are plotted in Figure~\ref{templates} (black long-dashed lines).
\par
The nuclear hot-dust SED templates are taken from \citet{2004MNRAS.355..973S}. They were constructed from a large sample of Seyfert galaxies selected from the literature for which clear signatures of non-stellar nuclear emission were detected in the near-IR and mid-IR, and also using the radiative transfer code GRASIL \citep{1998ApJ...509..103S}. 
The infrared SEDs are divided into 4 intervals of absorption: $\NH<10^{22}$ cm$^{-2}$ for Seyfert 1, $10^{22}<\NH<10^{23}$ cm$^{-2}$, $10^{23}<\NH<10^{24}$ cm$^{-2}$, and $\NH>10^{24}$ cm$^{-2}$ for Seyfert 2. The latter case is neglected in our analysis. The three templates employed in the code are plotted in Fig.~\ref{templates} with the brown solid line.
The $\NH$ estimates are used to select the torus template in the SED-fitting code for each AGN in the sample. 
The mean $\NH$ of our sample is $\sim10^{21}$ cm$^{-2}$. Sixty-one percent (331/539) have $\NH$ consistent with no absorption, while the rest have a median of $7.5\times10^{21}$ cm$^{-2}$.
Four-hundred and forty-six objects (83\%) have $\NH$ lower than $10^{22}$ cm$^{-2}$; we have therefore considered the Seyfert 1 template for them.
Only 17\% (93/539) have $\NH$ greater than $10^{22}$ cm$^{-2}$. For these sources dedicated templates (constructed considering Type-1 AGN with $\NH>10^{22}$ cm$^{-2}$) are needed, but we have nevertheless employed the torus templates of Sy2 for this sub-sample of sources. The Seyfert 1 torus template is not extremely different than the Seyfert 2 ones, hence the results are not affected in any significant way.
\par
We used a set of 30 galaxy templates built from the \citet[BC03 hereafter]{2003MNRAS.344.1000B} code for spectral synthesis models, using solar metallicity and Chabrier IMF (\citealt{2003ApJ...586L.133C}).
For the purposes of this analysis a set of galaxy templates representative of the entire galaxy population from passive to star forming is selected. To this aim, 10 exponentially decaying star formation histories (SFHs) with characteristic times ranging from $\tau = 0.1$ to $10$\,Gyr and a model with constant star formation are included.
For each SFH, a subsample of ages available in BC03 models is selected, to avoid both degeneracy among parameters and speed up the computation. In particular, early-type galaxies, characterized by a small amount of ongoing star formation, are represented by models with values of $\tau$ smaller than $1$ Gyr and ages larger than $2$\,Gyr, whereas more actively star forming galaxies are represented by models with longer values of $\tau$ and a wider range of ages from $0.1$ to $10$\,Gyr.
An additional constraint on the age is that, for each source, the age has to be smaller than the age of the Universe at the redshift of the source.
Each template is reddened according to the \citet{2000ApJ...533..682C} reddening law. 
The $E(B-V)_{\rm gal}$ values range between 0 and 1 with a step of $0.05$.
A subsample of templates from star forming to passive (without reddening) is presented in Figure~\ref{templates} (green dotted lines).
\par
The BBB template representative of the accretion disk emission is taken from \citet{2006ApJS..166..470R}. 
The near-infrared bump is neglected since we have already covered the mid-infrared region of the SED with the hot-dust templates. 
This template is reddened according to the \citet{prevot84} reddening law for the Small Magellanic Clouds (SMC, which seems to be appropriate for Type-1 AGN, \citealt{2004AJ....128.1112H,salvato09}). 
The $\ebvq$ values range between 0 and 1 with a variable step ($\Delta \ebvq=0.01$ for $\ebvq$ between 0 and 0.1, and $\Delta \ebvq=0.05$ for $\ebvq$ between 0.1 and 1) for a total of 29 templates. A subsample of templates with different reddening levels is presented in Figure~\ref{templates} (blue dashed lines).
\par
\begin{figure}
 \centering\includegraphics[width=9cm,clip]{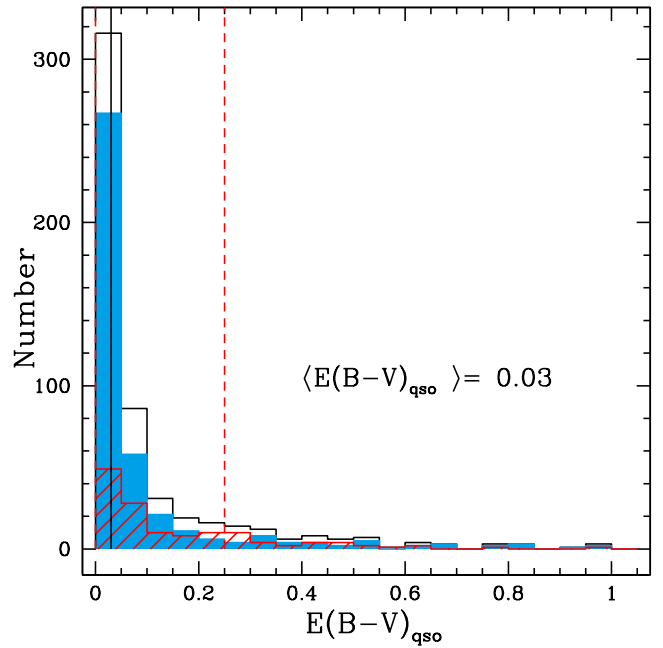}
 \caption{Disk reddening distribution for main sample ({\it open histogram}).  \rev{The spectro-z ({\it cyan filled histogram}) and photo-z ({\it red hatched histogram}) sample are also plotted.} The solid line represents the median at 0.03, 
 while the dashed lines \revs{correspond to the 16$^{\rm th}$ and the 84$^{\rm th}$ percentile at 0 and 0.25, respectively.}}
 \label{histebvq}
\end{figure}

\rev{Fig.~\ref{histebvq} shows the distribution of the best-fit $\ebvq$
for our AGN sample from SED fitting.  More than half of the sample
is well fitted with low $\ebvq$ values ($75\%$ with $\ebvq\leq 0.1$,
402 objects), with a median $\ebvq$ of 0.03.  However, 137 Type-1 AGN
(24\% \revs{of the total sample, 59 objects from the photo-z sample and 78 from the spectro-z sample}) show evidence for a significant amount of obscuration\footnote{A definition of the dust reddened Type-1 AGN
sample in XMM-COSMOS was already presented in
\citet{2010A&A...512A..34L}. Their Fig.~13 clearly shows that
$\sim10-20$\% of the AGN population in XMM-COSMOS has $\ebvq>0.1$. This dust-reddened AGN population also presents colors
that are redder than what has been found by \citet[see their
Fig.~6]{richards03}.} ($\ebvq>0.1$). 
\revs{The fraction of photo-z Type-1 AGN with $\ebvq>0.1$ is higher (43\%, 59/136) than in the spectro-z sample (19\%, 78/403). Given that the AGN in the photo-z sample are, on average, fainter in the optical bands than those in the spectro-z sample, it is not surprising that the Type-1 AGN in the photo-z sample have relatively higher reddening values than the spectroscopic Type-1s.}
In any case, the total fraction of reddened Type-1 AGN
in our sample is consistent with previous work in the literature
(e.g., \citealt{richards03,2006MNRAS.367..717M,2007ApJ...667..673G}).}

\begin{figure*}
      \begin{center}
       \vspace{-2.2cm}
        \subfigure{\label{fig:first}
            \includegraphics[width=0.4\textwidth]{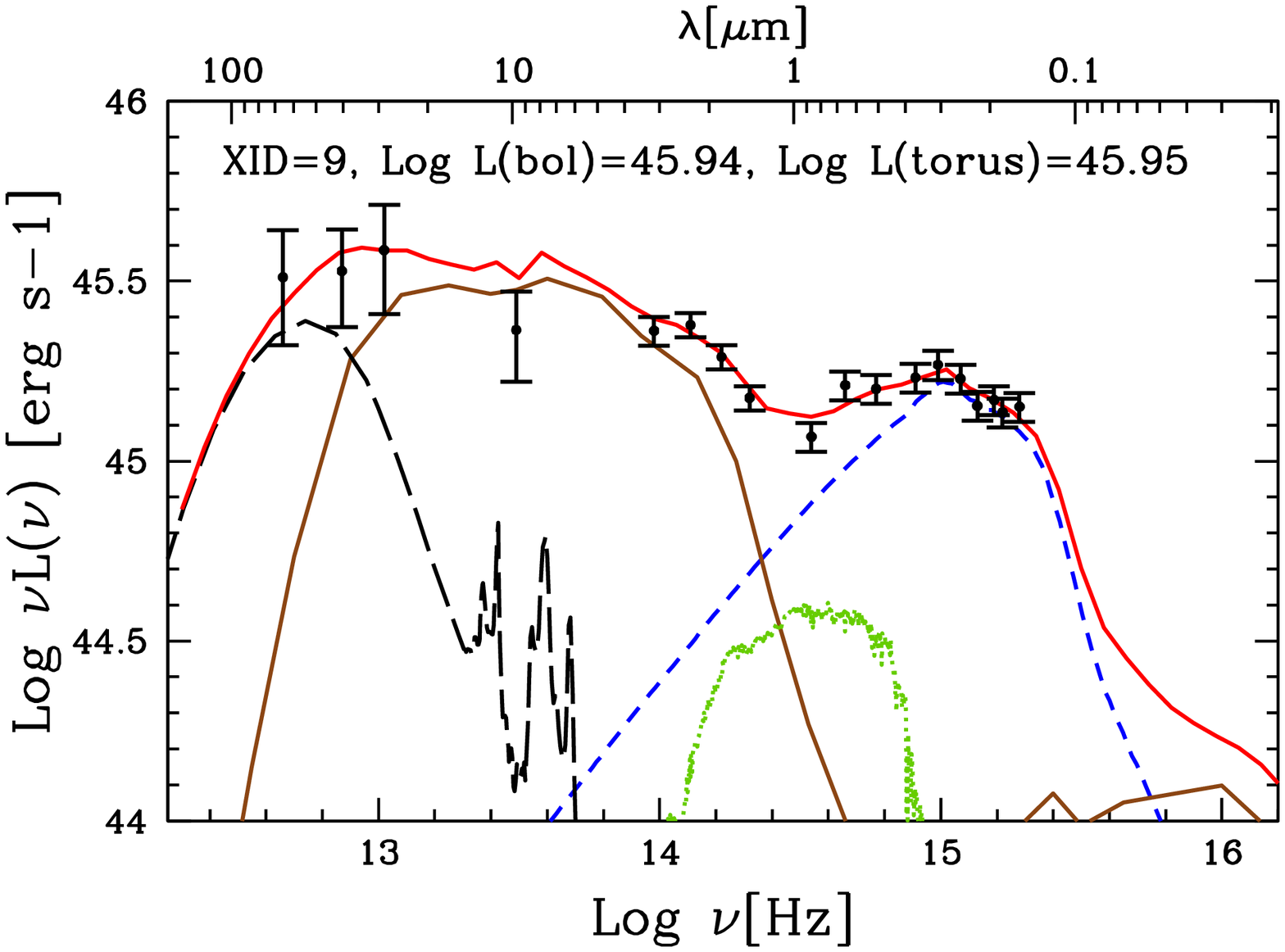}
        }
        \subfigure{\label{fig:second}
           \includegraphics[width=0.4\textwidth]{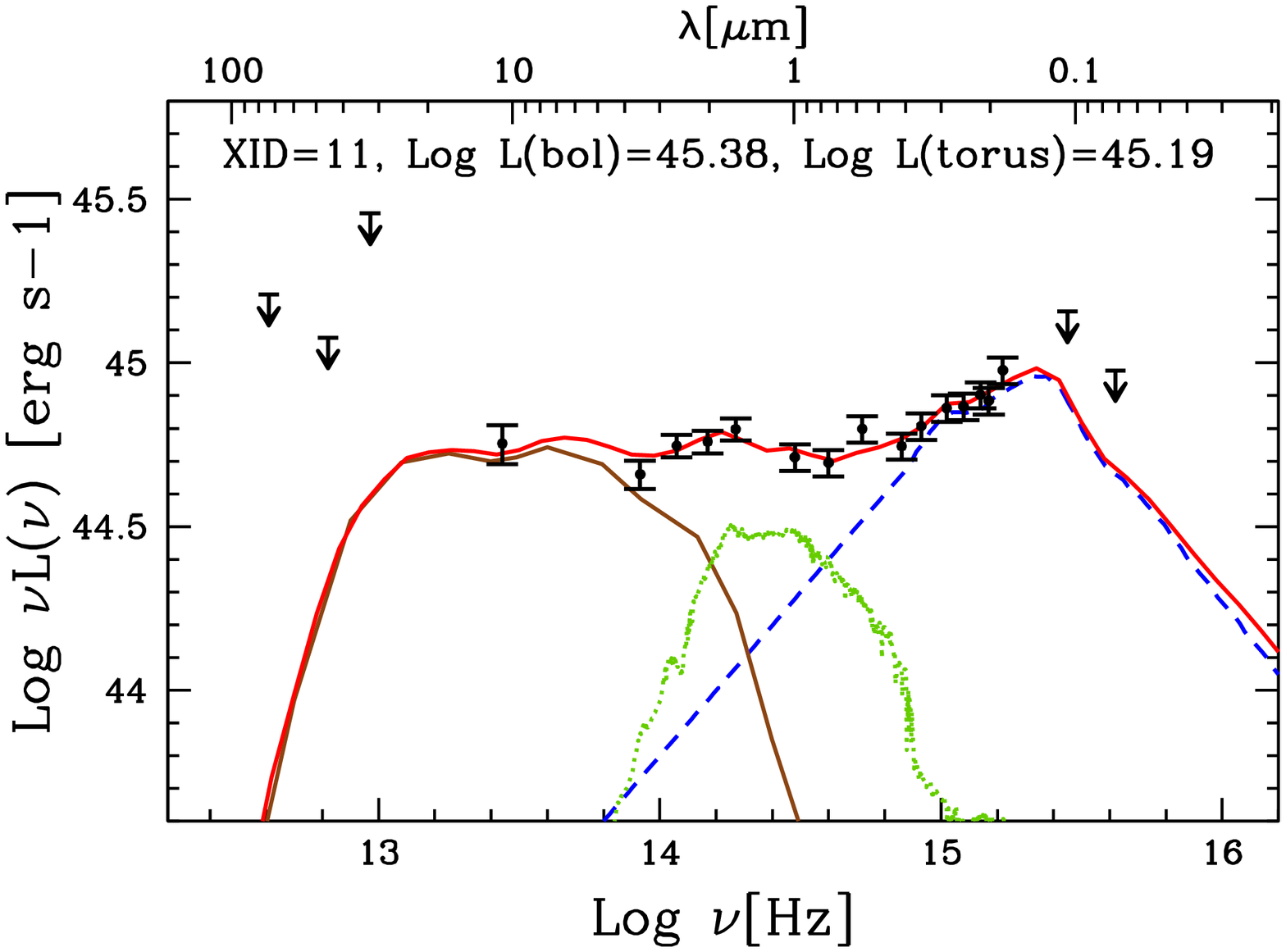}
        }\\ 
        \vspace{-2.5cm} 
        \subfigure{\label{fig:third}
            \includegraphics[width=0.4\textwidth]{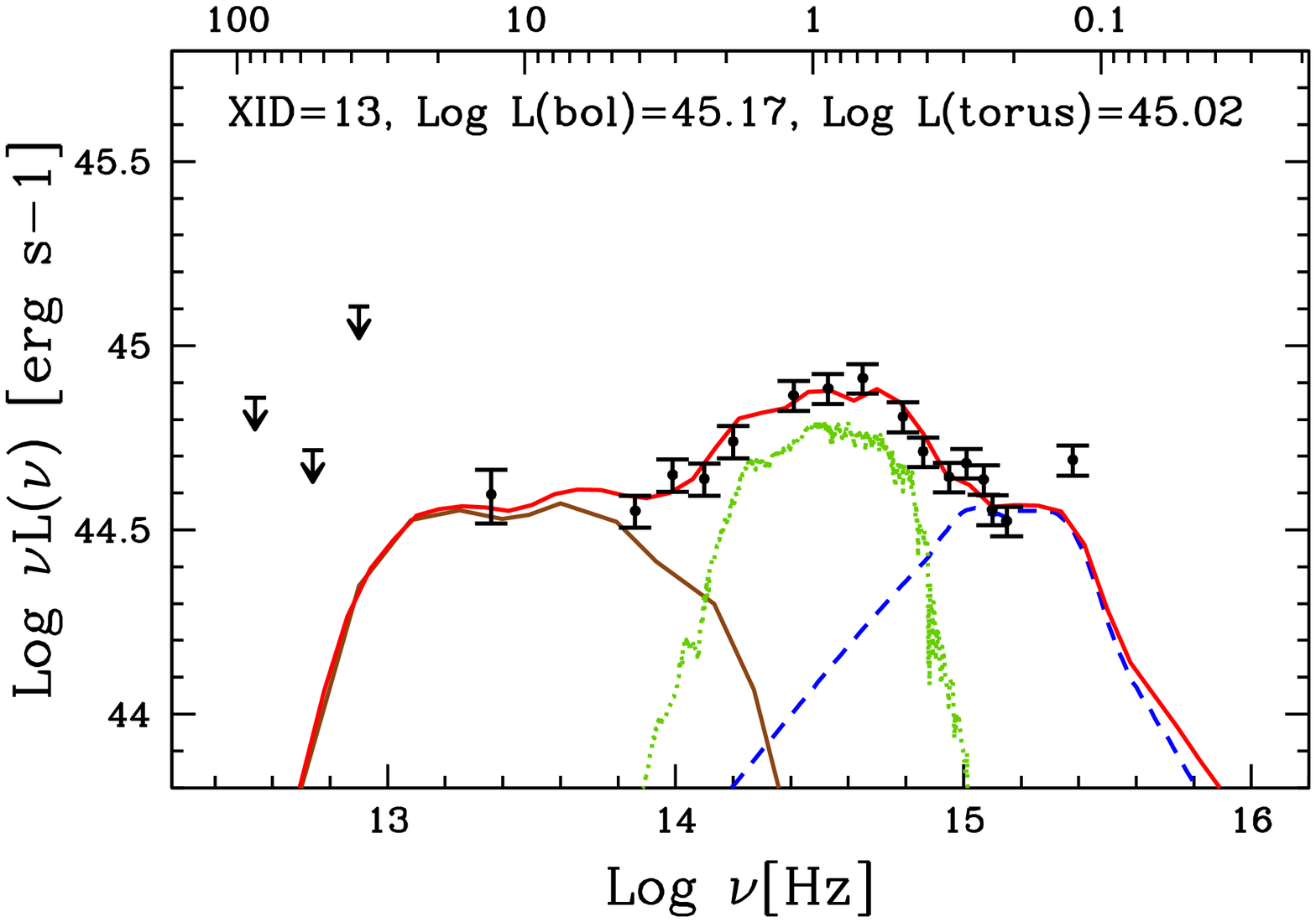}
        }
        \subfigure{\label{fig:fourth}
            \includegraphics[width=0.4\textwidth]{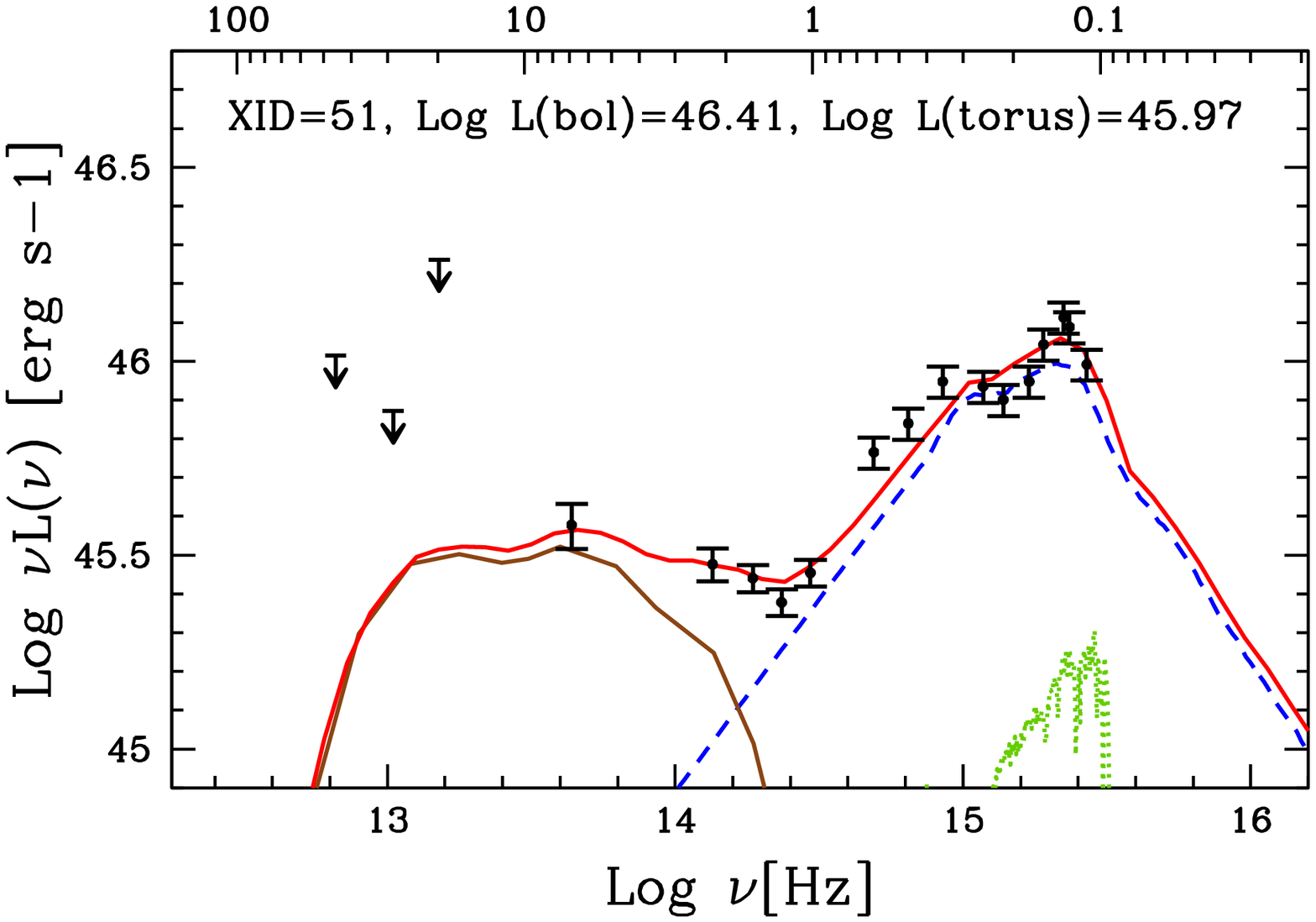}
        }\\ 
        \vspace{-2.5cm} 
        \subfigure{\label{fig:third}
            \includegraphics[width=0.4\textwidth]{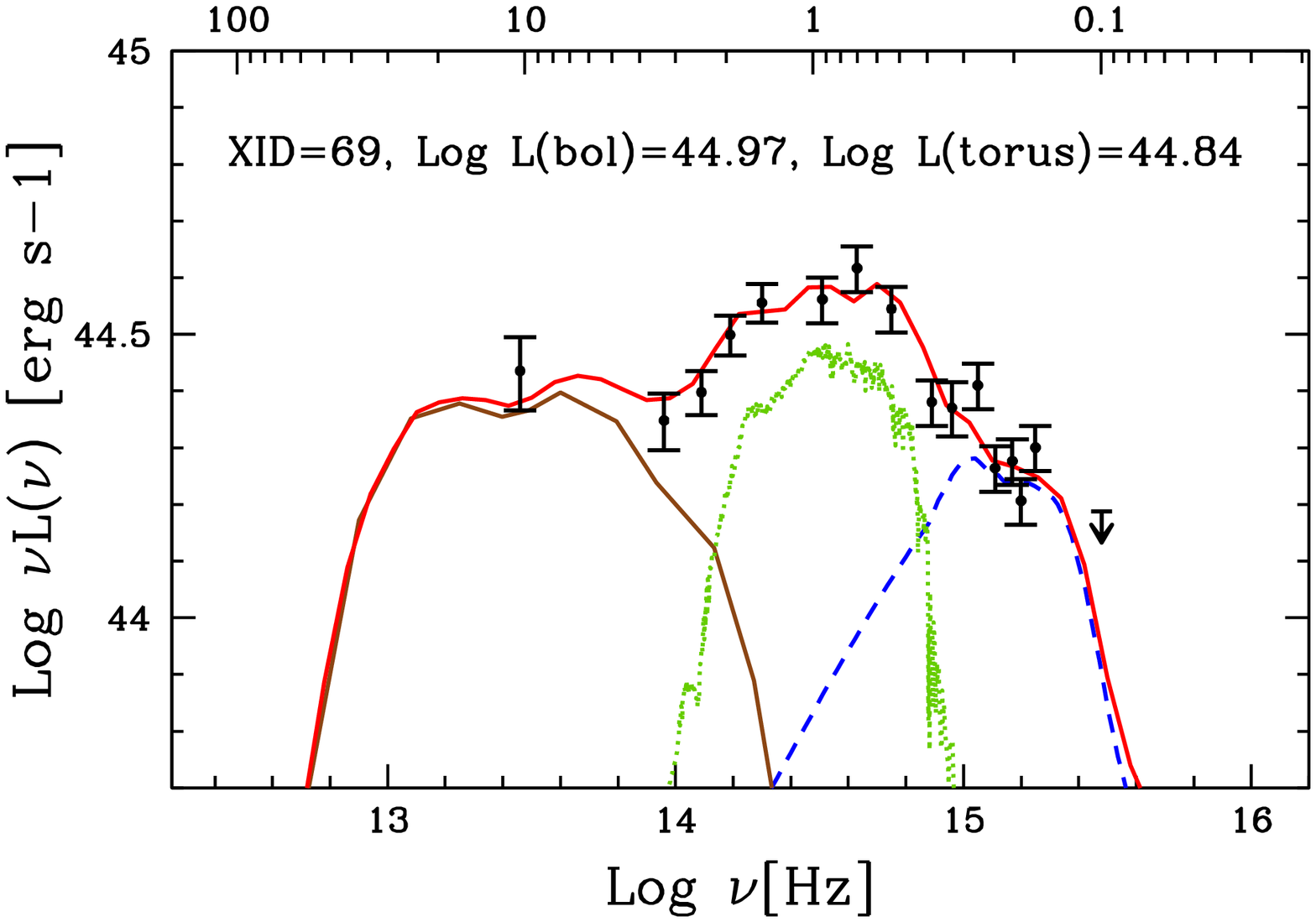}
        }
        \subfigure{\label{fig:fourth}
            \includegraphics[width=0.4\textwidth]{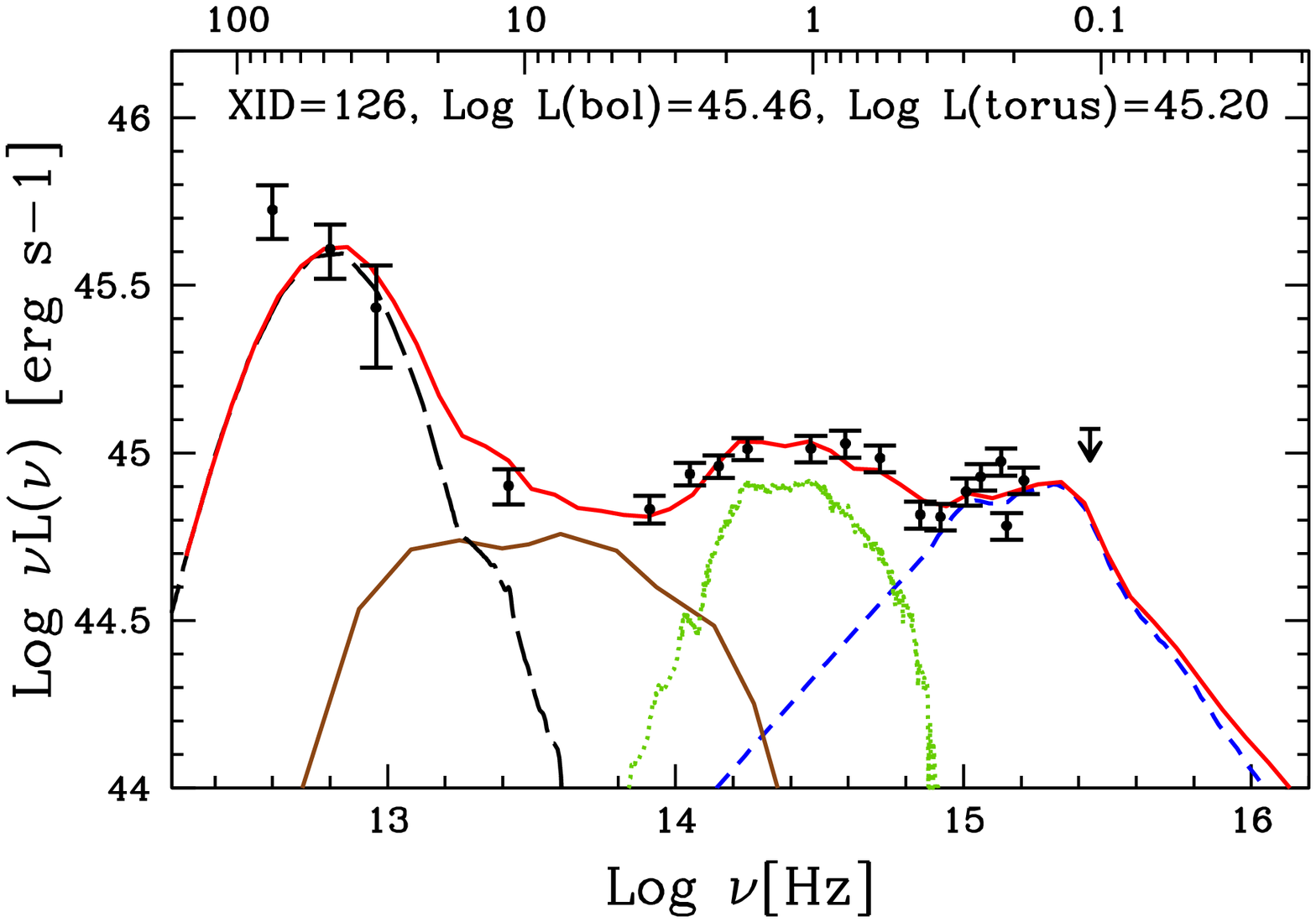}
        }\\
        \vspace{-2.5cm} 
        \subfigure{\label{fig:third}
            \includegraphics[width=0.4\textwidth]{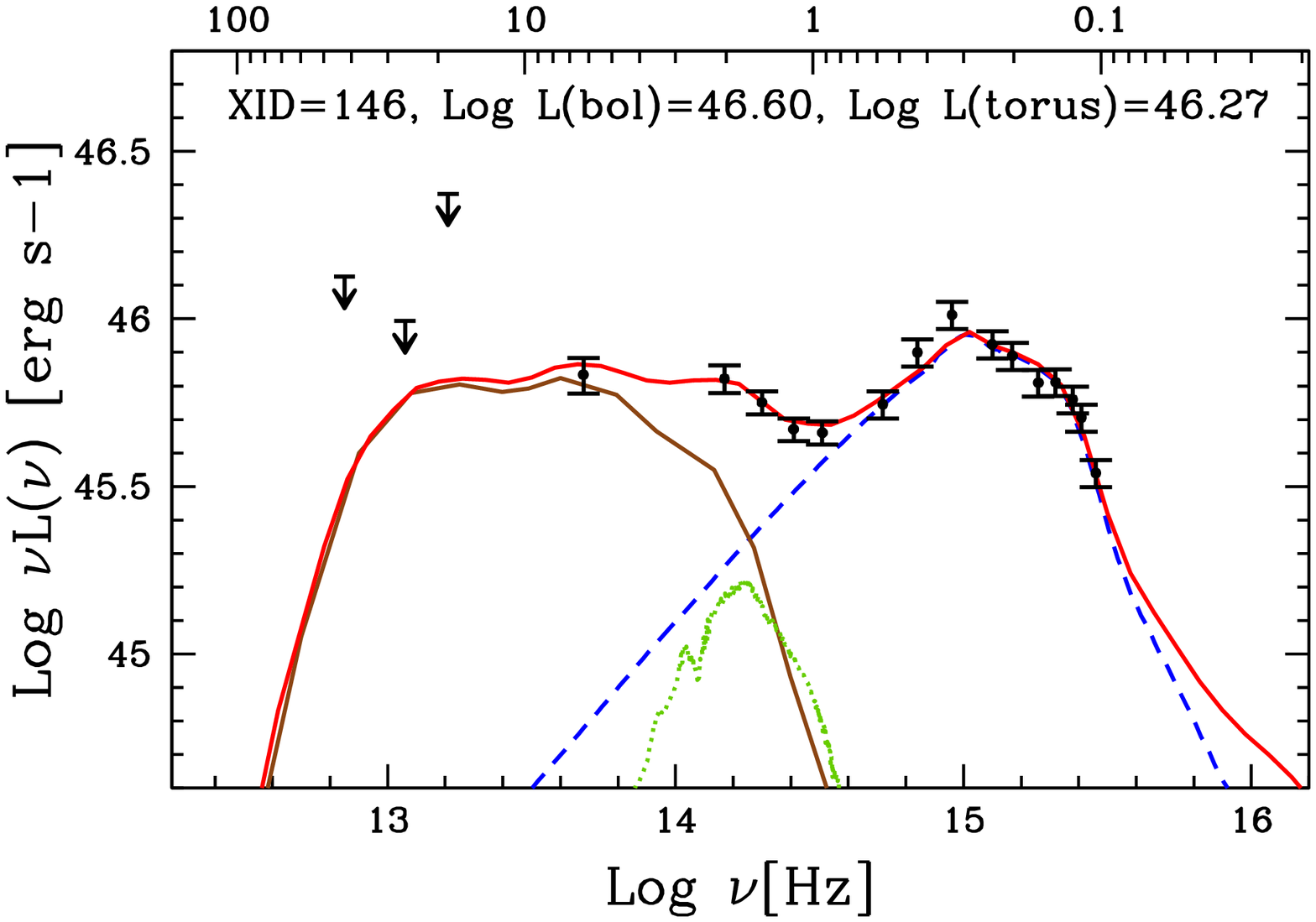}
        }
        \subfigure{\label{fig:fourth}
            \includegraphics[width=0.4\textwidth]{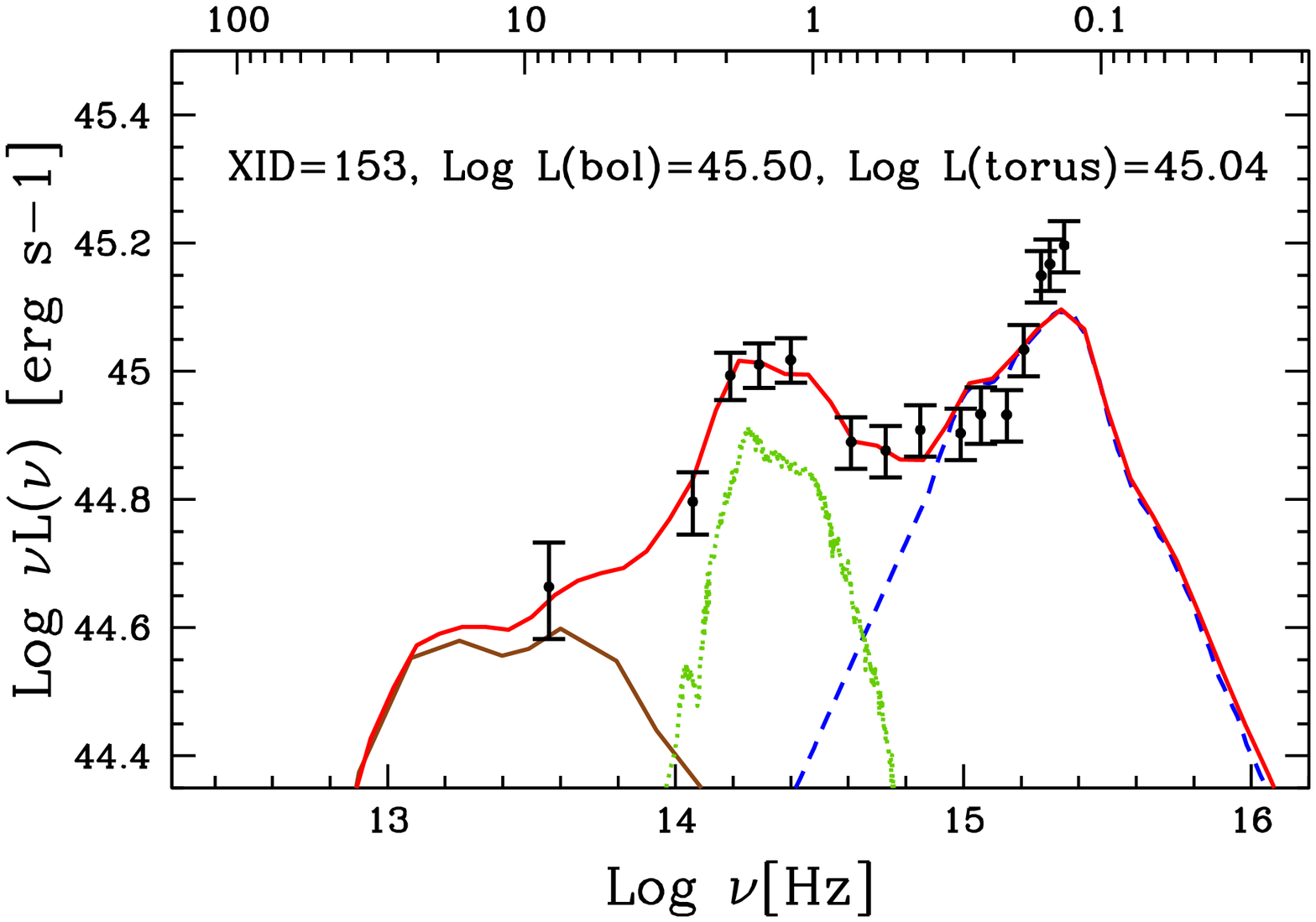}
        }
    \end{center}
 \caption{Examples of SED decompositions \rev{in the rest-frame plane}. Black circles are the observed photometry in the rest-frame (from the far-infrared to the optical-UV). The black long-dashed, brown solid, green dotted, and blue dashed lines correspond to the starburst, hot-dust from reprocessed AGN emission, host-galaxy, and BBB templates found as the best fit solution, respectively. The red line represents the best-fit SED. XID, bolometric and torus luminosities in erg s$^{-1}$ are also reported.}
 \label{panel}
\end{figure*}
In Fig.~\ref{panel} the broad-band SEDs of eight XMM-Newton Type-1 AGN are plotted as examples. 
The four components adopted in the SED-fitting code (\textit{starburst}, \textit{AGN torus}, \textit{host-galaxy} and \textit{BBB} templates) are also plotted. 
The red line represents the best-fit, while the black points represent the photometric data used in the code, from low to high frequency: Herschel/MIPS-Spitzer (160~$\mu m$, 100~$\mu m$, 70~$\mu m$, and 24~$\mu m$ if available), 4 IRAC bands, near-IR bands ($J$, $H$, and $K$), optical Subaru, CFHT and GALEX bands.
XID 9 and 126 are representative of a full SED with all detections from the far-infrared to the optical. Unfortunately, there is a limited number of detections at 160 and 70$\mu$m (see Sect.~\ref{Multi-wavelength coverage}), so that the more representative situation is shown in the other panels. 
XID 11 is representative of an AGN SED with contribution from the host-galaxy in the near-IR, while for the XID 146 and 51 the host-galaxy contribution is almost negligible. XID 153, 13, and 69 represent cases where the host-galaxy emission is significant. 
Subtracting the host-galaxy contributions, especially for those sources with high stellar contamination, is therefore essential in order to measure, for example, the $R$ parameter properly.
\par
The observed data points from infrared to optical are fitted employing a standard $\chi^2$ minimization procedure
\begin{equation}
\begin{array}{l}
\label{chisquare}
 \displaystyle\chi^2=\sum_{i=1}^{\rm{n_{filters}}}\frac{1}{\sigma_i^2} \left(F_{{\rm obs},i}-A\times F_{{\rm SB},i}-B\times F_{{\rm torus},i}+\right.\\
 \left.-C\times F_{{\rm gal},i}-D\times F_{{\rm BBB},i}\right)^2
 \end{array}
\end{equation}
where $F_{{\rm obs},i}$ and $\sigma_i$ are the monochromatic observed flux and its error in the band $i$; $F_{{\rm SB},i}$, $F_{{\rm torus},i}$, $F_{{\rm gal},i}$, and $F_{{\rm BBB},i}$ are the monochromatic template fluxes for the starburst, the torus, the host-galaxy, and the BBB component, respectively; $A$, $B$, $C$, and $D$ are the normalization constants for the starburst, the torus, the host-galaxy, and the BBB component.
The starburst component is used only when the source has a detection between $160\mu m$ and $24\mu m$. Otherwise, a three components SED-fit is used. Twenty is the maximum number of bands adopted in the SED-fitting (only detections are considered), namely: 160~$\mu m$, 100~$\mu m$, 70~$\mu m$, 24~$\mu m$, 8.0~$\mu m$, 5.8~$\mu m$, 4.5~$\mu m$, 3.6~$\mu m$, $K_S$, $J$, $H$, $z^+$, $i^*$, $r^+$, $g^+$, $V_J$, $B_J$, $u^*$, NUV and FUV.
\par
For each source the code computes several physical parameters such as SFR (from both optical and far-infrared), stellar mass and colors of the host-galaxy, AGN luminosity computed in different regions of the SED, and far-infrared luminosity of the cold-dust.
All outputs are estimated from the best-fit solution.
The upper and the lower confidence levels of the luminosities are evaluated from the distribution of the normalizations of all fit solutions corresponding to 68\% confidence level by taking the $\Delta \chi^2$ for a single parameter of interest ($\Delta \chi^2=1$, see \citealt{1976ApJ...210..642A}).  
Large uncertainties on output parameters reflect the degeneracy among the templates involved in the fit, especially between star-forming galaxies and reddened BBB templates (see Fig.~\ref{templates}).


\section{Analysis}
\label{Analysis}

In this Section we discuss the computation of the mid-infrared to bolometric luminosity ratio, $R$, and the assumptions underlying its definition. 
We then present luminosities from both observed rest-frame SEDs and model-fitting.
Finally, we compare the luminosities computed with these two methods to highlight the impact of host-galaxy and reddening correction on these measurements.

\subsection{Mid-infrared to bolometric luminosity ratio}
\label{Mid-infrared to bolometric luminosity ratio}

The mid-infrared to bolometric luminosity ratio, which we employ to
parametrize the obscured fraction, is defined as
\begin{equation}
\label{f}
R\equiv\frac{\Ltorus}{\Lbol}=\frac{\int_{\nu_{1000\mu m}}^{\nu_{1\mu m}} L_\nu {\rm d}\nu}{\int_{\nu_{1\mu m}}^{\nu_{\rm max}} L_\nu {\rm d}\nu},
\end{equation} 
where $\Ltorus$ is the infrared emission reprocessed by the dust at
the wavelength range $1-1000\mu$m, while $\Lbol$ is our definition of
the bolometric luminosity, which represents the optical--UV and
X--ray emission emitted by the nucleus and reprocessed by the dust
grains in the torus. There is some ambiguity in the literature about
whether the X--ray emission, which partially arises from the accretion
disk itself, but also from accretion disk photons Compton up-scattered
by a hot X--ray corona, should be included in the bolometric
luminosity. However, because we are here interested in all emission
being reprocessed by dust grains, we include the X--ray contribution
($\sim 10\%$) to the bolometric luminosity\footnote{We are implicitly
  assuming that the observed soft and hard X--ray emission is
  originated on scales smaller than that of the torus.}. Thus we need
to quantify the maximum frequency, $\nu_{\rm max}$, which we we define
to be the frequency at which the dust optical depth in the torus,
$\tau_{\rm d}=N_{\rm d} \sigma_{\rm d}(\nu)$, is unity, where $N_{\rm
  d}$ is the column density of dust in the torus, and $\sigma_{\rm
  d}(\nu)$ is the dust cross-section.  Photons with a frequency higher
than $\nu_{\rm max}$ should not be counted in the bolometric emission
budget, as they will just pass through the torus without being
reprocessed.
\par
\citet{2003ApJ...598.1026D} estimated the X-ray extinction and
scattering cross section per \ion{H}{} nucleon due to interstellar
dust, assumed to be a mixture of carbonaceous and amorphous silicate
grains, and absorption due to gas with interstellar gas-phase
abundances in the energy range $0.1-10$ keV (see their
Fig.~6). Absorption by \ion{H}{} and \ion{He}{} dominates at energies
lower than $\sim0.25$ keV, while above this energy value observations
of extinction and scattering by dust is significant for bright sources
with sufficient dust column densities along the line of sight. At
$E\geq 0.8$ keV extinction is mainly due to dust grains.  The energy
at which the dust optical depth is equal to one considering an average
$\NH$ value for obscured AGN of $\sim10^{22}$ cm$^{-2}$ (see L12) is
around 1 keV, which corresponds to a frequency of $2.4\times10^{17}$
Hz.  We have to integrate Eq. (\ref{f}) out to energies that are intercepted by the
torus (not along our line-of-sight), and thus the average $\NH$ for
Type-2 AGN should be more appropriate.  
The precise energy of the absorption in the X--rays will be
dependent on the chemical nature of the grain material
(\citealt{1998ApJ...505..236F,2003ARA&A..41..241D,2003ApJ...598.1026D});
given that soft X--ray emission usually contributes about 10\% of
the total bolometric output, the uncertainty in our luminosity estimates due
to the unknown X-ray opacity will be less than this. In order to quantify the
degree of variation on the $\Lbol$ estimates we have also considered
an energy cutoff of 0.4 keV and 2 keV (frequency of $10^{17}$ and
$4.8\times10^{17}$ Hz), which correspond to an $\NH$ of about
$10^{21}$ (the average value of the present Type-1 AGN sample) and
$1.6\times10^{22}$ cm$^{-2}$, respectively\footnote{Column densities \rev{lower} than $10^{22}$ cm$^{-2}$ are almost transparent to hard ($> 2$ keV) X--rays \citep{1983ApJ...270..119M}.}.  
The shift induced by considering the difference between $\Lbol$
estimated up to a maximum energy cutoff of 0.4 and 2 keV is
$\langle\Delta \Log \Lbol\rangle=-0.066\pm0.004$. Thus our assumption
of a 1 keV cutoff will result in uncertainties smaller than this, and
this effect is much smaller than other uncertainties in our
calculation, i.e. degeneracies between templates, uncertainties on the
data, etc.  We consider as our fiducial $\nu_{\rm max}$ value the
frequency corresponding to the energy at 1 keV.
\par
We can compute total luminosities in a given
range in two ways, one by integrating the actual photometry, and the
other by integrating the resulting best-fit SEDs output of the
SED-fitting code.  For the first approach, no attempt is made to
subtract off the host-galaxy contribution or to de-redden the
photometry (i.e., the BBB).  For the latter approach, we present two
cases. In the first case we have estimated $R$ considering $\Lbol$
without correcting for the intrinsic AGN reddening. This approach is
what has been used by other works on SEDs that tried to estimate
covering fractions (see \citealt{2007A&A...468..979M,2008ApJ...679..140T}). In the second case $\Lbol$ is corrected for both
host-galaxy and reddening contributions.  In the next Sections we will
proceed as follows. We will present these two ways of computing the
total luminosities in a given range, and we will dedicate a separate
discussion about the effect of the AGN reddening and host-galaxy correction on
$\Lbol$.

\subsection{Luminosities from observed rest-frame SED}
\label{Optical-UV and infrared luminosities from observed rest-frame SEDs}
\begin{figure}
 \includegraphics[width=9cm,clip]{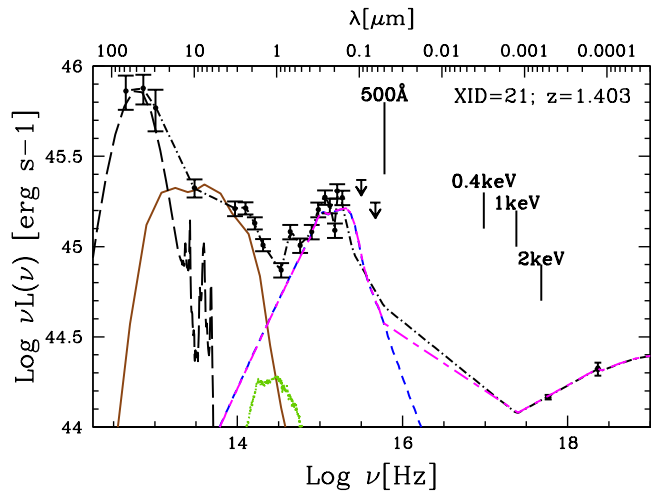}
 \caption{Example of full AGN SED from far-IR to X-rays at redshift 1.4 \revs{(XID=21)}. The rest-frame data, used to construct the observed and fitted SED, are represented with black points. The black long-dashed, brown solid, green dotted, and blue dashed lines correspond to the best-fit starburst, hot-dust from reprocessed AGN emission, host-galaxy, and BBB templates, respectively. The black dot-dashed line represents the observed rest-frame SED as described in Section~\ref{Optical-UV and infrared luminosities from observed rest-frame SEDs}, while the magenta short-long dashed line is the best-fit BBB template plus X-rays as described in Section~\ref{Disk and torus luminosities from SED-fitting}.}
 \label{sedexample}
\end{figure}

We have computed the individual observed rest-frame SEDs for all sources in the sample, following the same approach as in L10.
For the estimate of the rest-frame AGN SED we need to extrapolate the UV data to X-ray ``gap" and at high X-ray energies. The SED is extrapolated up to 1200$\text{\AA}$ with the slope computed considering the last two rest-frame optical data points at the highest frequency in each SED (only when the last \rev{UV} rest frame data point is at $\lambda>1200\text{\AA}$). Then, a power law spectrum to 500$\text{\AA}$ is assumed, as measured by HST observations for radio-quiet AGN ($f_\nu\propto \nu^{-1.8}$, see \citealt{zheng97}). 
The UV luminosity at 500 $\text{\AA}$ is then linearly connected (in the log space) to the luminosity corresponding to the frequency of 1 keV.
\rev{We note that the fraction of bolometric luminosity in the 500~\AA{} to 1 keV range depends on the model adopted, as shown by \citet{2013ApJS..206....4K}. However, these authors found that bolometric corrections estimated at 2500~\AA{}, considering different UV-X--ray extrapolations, agree within a factor of about 1.5.} 
Finally, the X-ray spectrum is extrapolated at higher energies introducing an exponential cut-off at 200 keV (e.g., \citealt{2002A&A...389..802P}). An example of observed rest-frame SED is plotted in Figure~\ref{sedexample} with the black dot-dashed line.
\rev{We checked if the extrapolation works properly by inspecting all objects visually.}
\par
Observed infrared and bolometric luminosities (hereafter $\Lir$ and $\Lbolobs$) are quantified by integrating the observed rest-frame SED in the log space from 1$~\mu$m to 24$~\mu$m \citep{2012arXiv1210.3033H}, and from 1$~\mu$m to 1 keV, respectively.

\subsection{Luminosities from SED-fitting}
\label{Disk and torus luminosities from SED-fitting}
The four-component SED-fitting code presented here allows us to have a reliable estimate of both disk and torus luminosities (hereafter $\Ldisk$ and $\Ltorus$). In Fig.~\ref{sedexample} a full AGN SED from far-infrared to X-rays, with the respective best-fits, is presented as an example. 
X-rays are not taken into account in the fit procedure. Therefore, in order to estimate $\Ldisk$, we need to include them in a separate step. 
We consider the disk template up to 500 $\text{\AA}$ and the de-absorbed X--ray spectrum (as described in the previous Section) at energies higher than 1 keV.
We then linearly connect these two curves (an example is presented with the magenta short-long dashed line in Fig.~\ref{sedexample}). The resulting disk+X--ray SED is integrated from 1~$\mu$m to 1 keV and it is our definition of bolometric luminosity. 
Only six objects (2 spectro-z and 4 photo-z) over 539 Type-1 AGN do not require any disk component in the best-fit. As a result, $\Ldisk$ is not available for these sources\footnote{The SED for XID=300 (COSMOS$\_$J100050.16+022618.5, $z=3.715$) is fitted with a starburst galaxy, but presents clear QSO features in the spectrum (\ion{Ly}{$\alpha$}, \ion{C}{iv}). The observed SED has very strong BBB with absorption at wavelength shortwards of 912~\AA{}, but the SED-fitting code is not able to reproduce this feature with a reddened disk.
The SED for XID=2394 (COSMOS$\_$J100127.53+020837.8, $z=3.333$) is very well fitted with a star-forming galaxy, and the spectrum presents faint AGN feature (\ion{C}{iv}).}. 
\par
The $\Ltorus$ values are computed by integrating the best-fit torus templates (brown solid line in Fig.~\ref{sedexample}) from 1-1000~$\mu$m. 
Information on $\Ltorus$ is available from the best-fit models for 516 out of 539 (96\%) Type-1 AGN, but three of these sources have been removed because the optical-UV photometry is fitted with a galaxy template only (and therefore $R$ cannot be estimated \rev{without constrains} on $\Ldisk$). This leads to a sub-sample of 513 Type-1 AGN (388 spectro-z and 125 photo-z) with both $\Ltorus$ and $\Ldisk$ estimates. 
\par
For the remaining 19 Type-1 AGN (4\%) a torus template is not considered in the best-fit. All of these sources are at high redshift ($1.60\leq z\leq 4.26$). Three objects are detected at 24~$\mu$m, but the code is not considering the torus model in the best-fit due of the combination of two facts: first, large error on the 24~$\mu$m detection, and second, IRAC and optical-UV photometry are nicely fitted with galaxy plus disk templates only.
No photometric coverage in the mid-infrared is present for the other 16, which do not have a 24~$\mu$m detection.
\par
As we have already pointed out, there are several factors that need to be taken into account when obtaining $\Lbolobs$ by integrating the interpolated photometry. Host-galaxy emission and reddening are both present and they can lead to over/underestimate AGN emission, respectively. 
The analysis presented here takes into account these contributions, thanks to our model fitting procedure, and we will discuss their effects on $\Lbolobs$ in the following.

\subsection{Disk luminosities: effect of the host-galaxy \& reddening}
\label{Disk luminosities: effect of the host-galaxy and reddening}
\begin{figure*}
\epsscale{1.15}
\plottwo{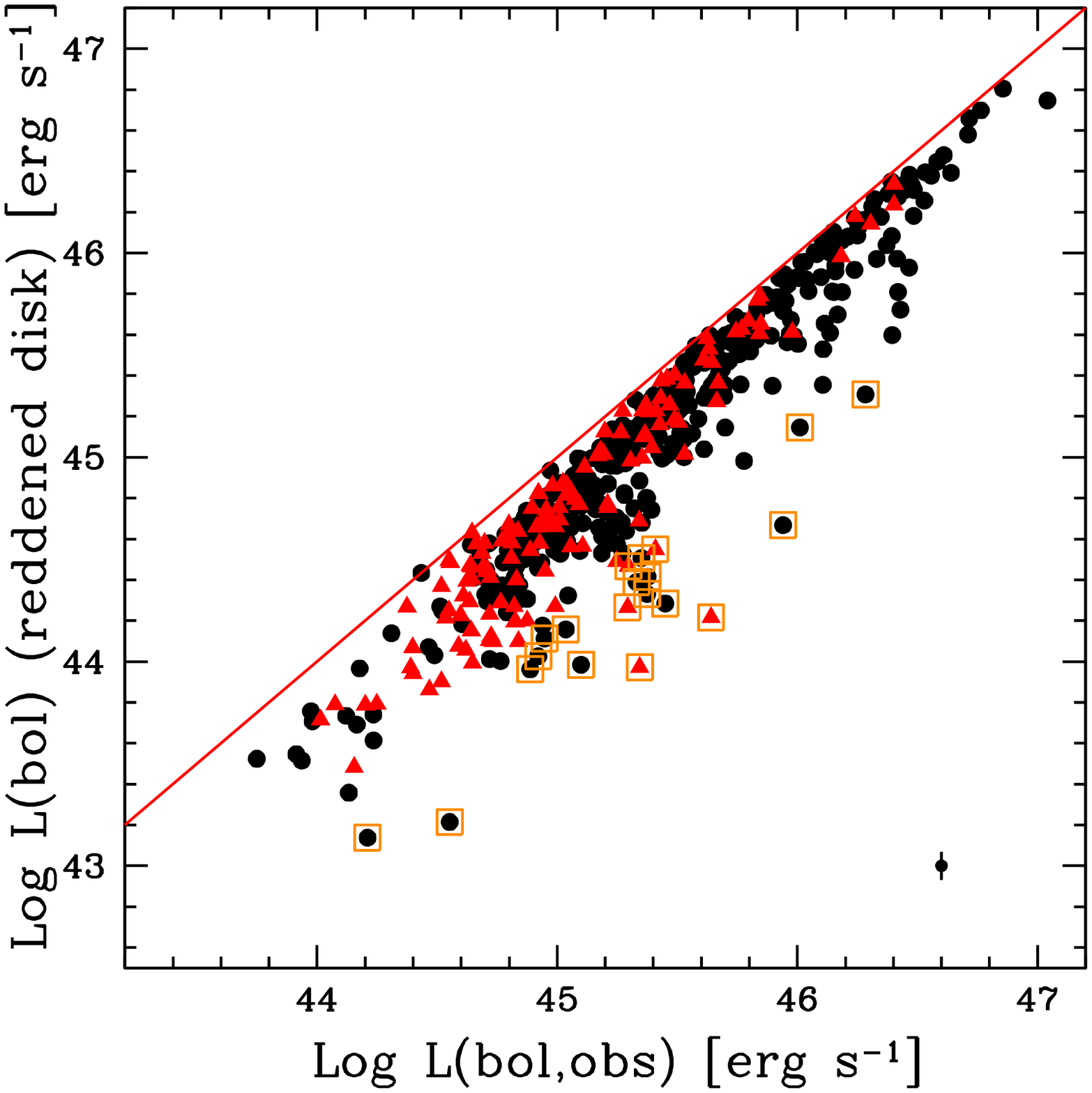}{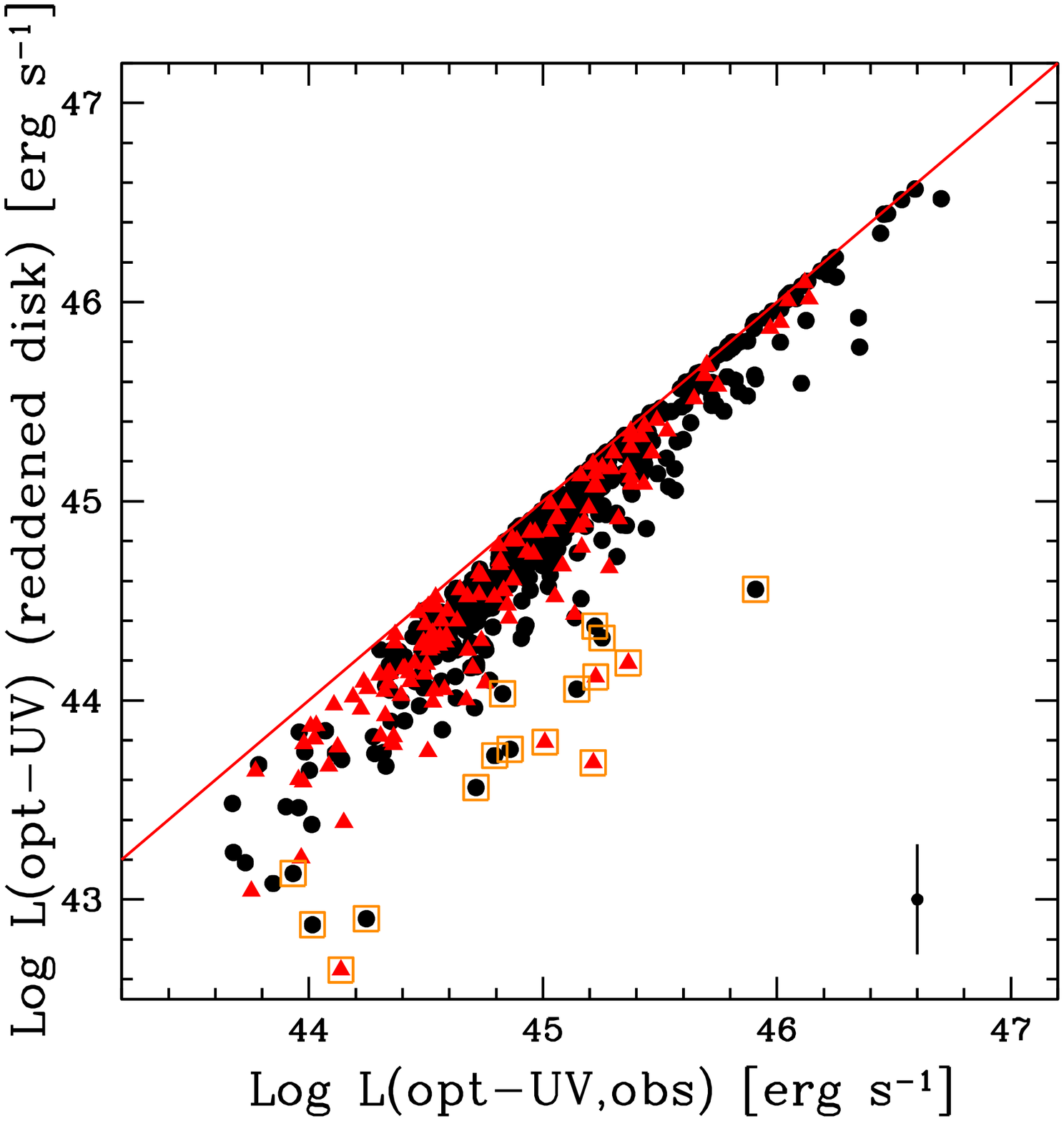}
\caption{\textit{Left panel}: comparison between the values of $\Lbol$ computed by the SED-fitting code (host-galaxy subtracted) and those obtained integrating the observed rest-frame SED from 1~$\mu$m to 1~keV (see Sect.~\ref{Optical-UV and infrared luminosities from observed rest-frame SEDs}). Reddening has not been taken into account in the $\Lbol$ values. Three-sigma outliers from the median are marked with orange open squares \revs{(20 objects)}. \rev{Black points and red triangles represent the spectro-z and photo-z sample, respectively.} The red solid line represents the one-to-one correlation. The average error on $\Lbol$ is plotted in the bottom right for clarity. \textit{Right panel}: \rev{comparison between the observed optical--UV luminosity ($L_{\rm opt-UV, obs}$) by integrating the interpolated rest-frame photometry between 1~$\mu$m and the \revs{bluest rest-frame} data point in the \revs{optical-UV}, and the fitted optical--UV luminosity ($L_{\rm opt-UV}$) by integrating the best-fit BBB template in the same wavelength range \revs{(16 outliers, open orange squares)}.} }
\label{ldiskcomphostcorrected}
\end{figure*}
Our first step in determining the intrinsic bolometric AGN emission is to subtract the host-galaxy contribution from total observed $\Lbol$ without taking into account the AGN reddening.
A comparison between $\Lbolobs$ (from interpolated photometry) and $\Lbol$ (from model fitting, host-galaxy subtracted) is presented in the left panel of Fig.~\ref{ldiskcomphostcorrected}, where the one-to-one correlation is plotted with the red solid line as reference. The median shift between $\Lbolobs$, which includes both host-galaxy and reddening contamination, and $\Lbol$ is 0.25 dex.
Sources that deviate more from the average $\Log \Lbolobs/\Lbol$ are those with higher host-galaxy contamination and reddening.
Twenty AGN (12 spectro-z and 8 photo-z) lie more than 3~$\sigma$ away from the median and are marked with orange open squares in Fig.~\ref{ldiskcomphostcorrected} \revs{(left panel)}. 
Representative examples of SEDs of these outliers are discussed in Appendix~\ref{appendixA}.  
Part of the scatter and the fact that none of the sources lie on the one-to-one correlation might be due to the different methods of extrapolation in the UV-soft X--ray gap ($15.5\lesssim\Log \nu\lesssim17.5$ Hz) adopted (see Fig.~\ref{sedexample}). 
In order to check this issue we have estimated, for each object, the observed optical--UV luminosity ($L_{\rm opt-UV, obs}$) by integrating the interpolated rest-frame photometry between 1~$\mu$m and the \revs{bluest rest-frame} data point in the \revs{optical-UV}, and the fitted optical--UV luminosity ($L_{\rm opt-UV}$) by integrating the best-fit BBB template in the same wavelength range. The result is plotted in the right panel of Fig.~\ref{ldiskcomphostcorrected}. \rev{Sources are now closer to the one-to-one relation} \revs{(with 16 outliers)}, but the scatter is still significant demonstrating that the host-galaxy contribution is important in the optical--UV.
\par
\revs{We have also computed the host-galaxy luminosity ($L_{\rm host}$) from the best-fit galaxy template for each Type-1 AGN over the same wavelength range of $L_{\rm opt-UV}$. For a significant fraction of the objects $L_{\rm host}$ is comparable to $L_{\rm opt-UV}$. For example, we found that the ratio between $L_{\rm host}$ and $L_{\rm opt-UV}$ is more than 0.7 for 34\% of the Type-1 AGN in our sample.}
\par
Overall, our model fitting procedure highlights a significant host-galaxy contamination in the bolometric emission of Type-1 AGN in XMM-COSMOS, in agreement with previous results (see also Fig.~4 in \citealt{2012MNRAS.427.3103B}).
In a recent paper
(\citealt{2012ApJ...759....6E}) the average SED of the same COSMOS
sample employed here is presented. It is clear from Figure 14 in
\citet{2012ApJ...759....6E} that the mean observed SED is quite flat,
and lacks the $1\mu$m inflection point between the UV and near-IR
bumps seen in previous analyses. This suggests that this sample has a
higher contamination from the host-galaxy light than brighter
optically selected samples (see
\citealt{2012ApJ...759....6E,2012arXiv1210.3033H}).
\par
The next step in determining the intrinsic nuclear $\Lbol$ is to correct for reddening the best-fit BBB template by employing the corresponding $\ebvq$ value for each object in our sample.
\begin{figure*}
\epsscale{1.15}
\plottwo{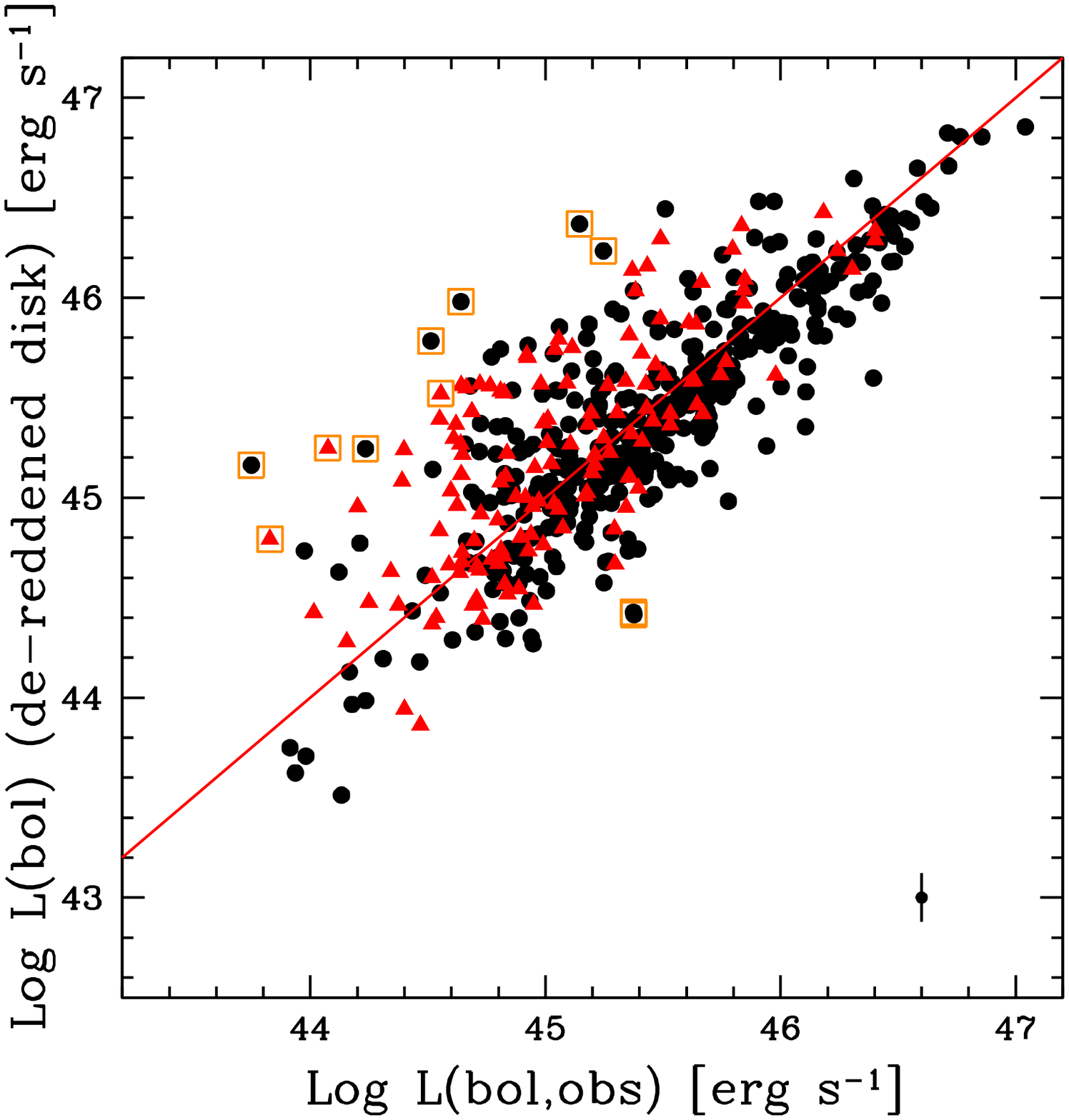}{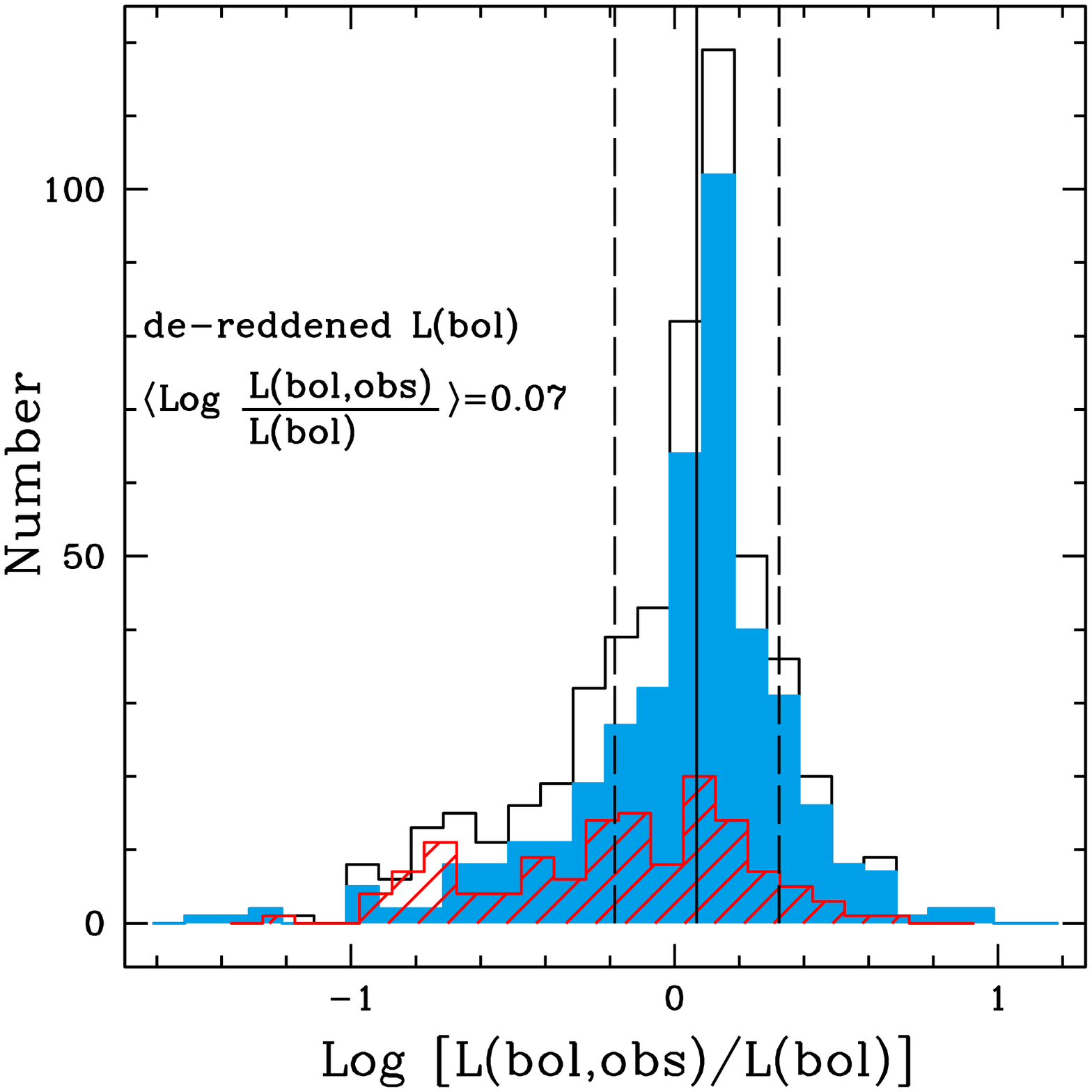}
\caption{\textit{Left panel}: comparison between the values of $\Lbol$ computed by the SED-fitting code (see Sect.~\ref{Disk and torus luminosities from SED-fitting}) and those obtained integrating the observed rest-frame SED from 1$\mu$m to 1~keV (see Sect.~\ref{Optical-UV and infrared luminosities from observed rest-frame SEDs}). Reddening has been taken into account in the $\Lbol$ values. \rev{Black points and red triangles represent the spectro-z and photo-z sample, respectively.} The red solid line represents the one-to-one correlation, while 3~$\sigma$ outliers are marked with orange open squares. \textit{Right panel}: Histogram of the ratio between $\Lbolobs$ and $\Lbol$. Key as in Fig.~\ref{histebvq}. The solid line represents the median at $\Log \Lbolobs/\Lbol=0.07$, while the dashed lines \rev{correspond to the median absolute dispersion (\revs{$1.4826\times$MAD}, see \S~\ref{Mid-infrared to bolometric luminosity ratio versus bolometric luminosity}) at 0.25 dex}.}
\label{ldderedcomp}
\end{figure*}
The reddening corrected best-fit bolometric luminosities are plotted in the left panel of Fig.~\ref{ldderedcomp} as a function of $\Lbolobs$.
As expected, AGN with high $\ebvq$ tend to have higher $\Lbol$ than $\Lbolobs$, with the ratio of the corrected $\Lbol$ and $\Lbolobs$ distributed almost symmetrically around zero (see right panel of Fig.~\ref{ldderedcomp}), but with a tail toward small values resulting from highly reddened systems. 
Outliers deviating more than 3~$\sigma$ below the median (2 objects, which are superimposed in the plot) represent high host-galaxy contamination, while outliers deviating more than 3~$\sigma$ above the median (9 objects) have large values of $\ebvq$ ($\ebvq\sim1$) and small contribution from the host-galaxy. \rev{Representative SEDs} of these outliers are presented in Appendix~\ref{appendixB}.
\par
\par
\begin{figure*}
\epsscale{1.15}
\plottwo{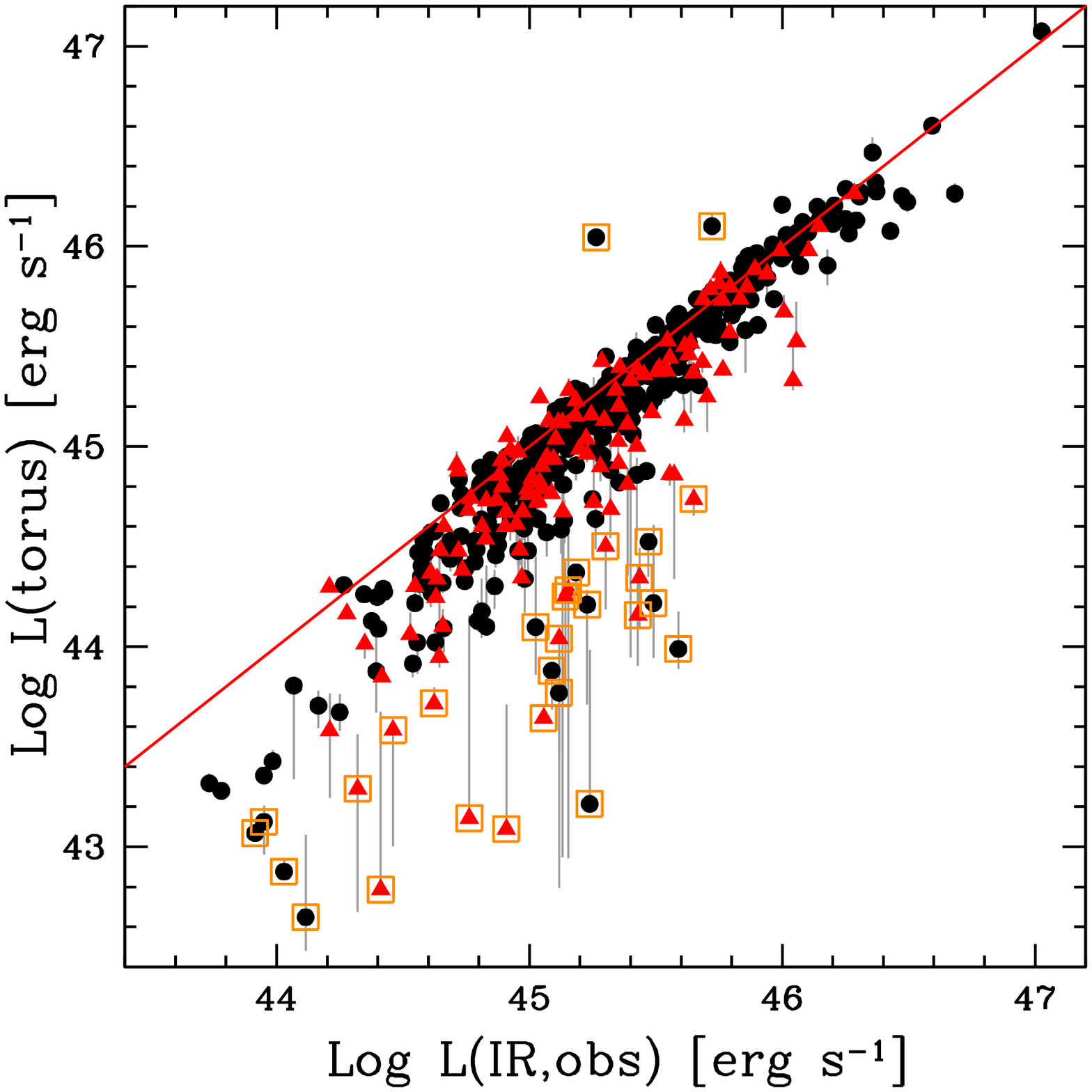}{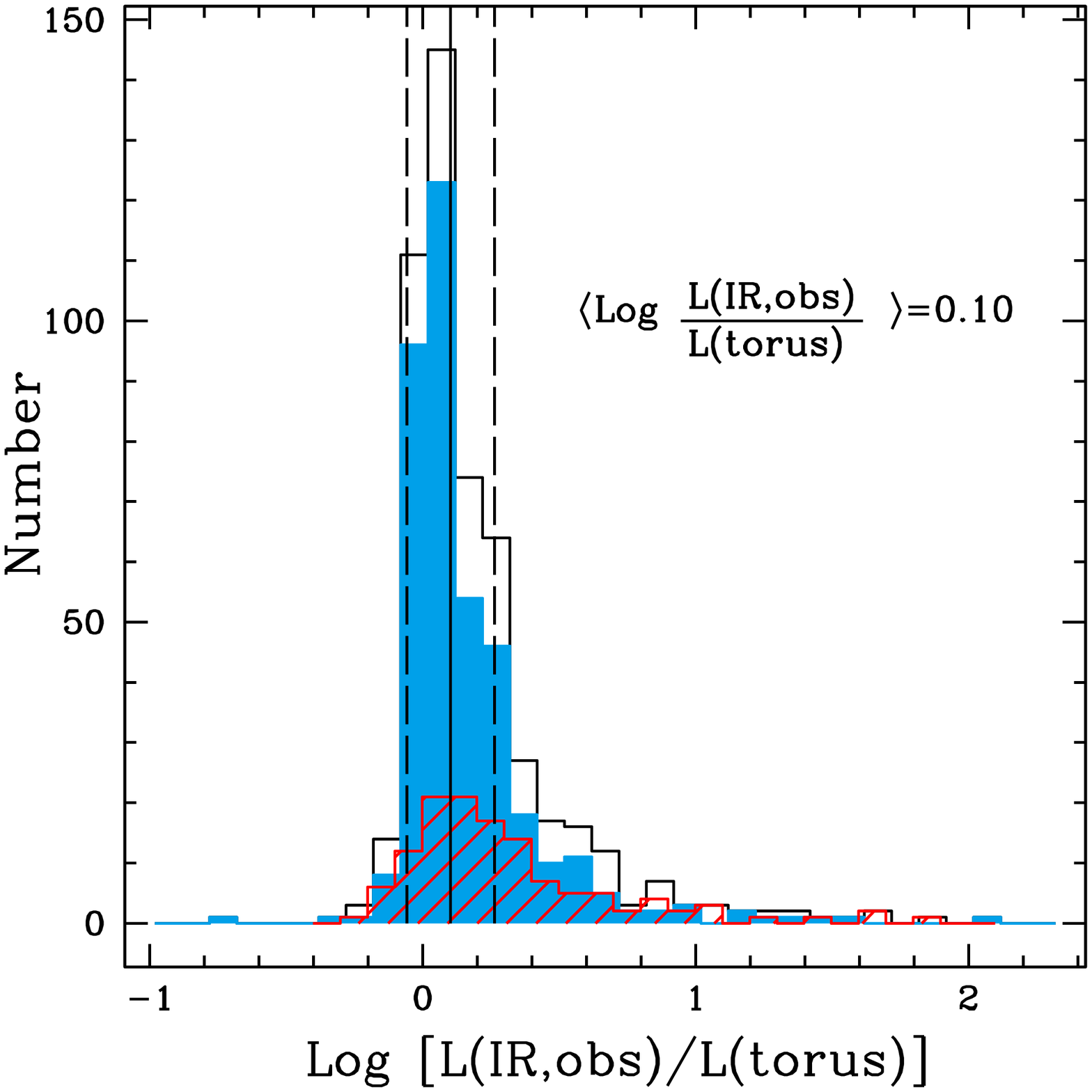}
 \caption{\textit{Left panel}: comparison between the values of torus luminosity computed by the SED-fitting code and those obtained integrating the observed rest-frame SED from 1~$\mu$m to 24~$\mu$m. Key as in Fig.~\ref{ldiskcomphostcorrected}. Grey bars represent $1-\sigma$ error as discussed in Sect.~\ref{SED-fitting method}. \textit{Right panel}: Histogram of the ratio between $\Lir$ and $\Ltorus$. \rev{Key as in Fig.~\ref{histebvq}}. The solid line represents the median at $\Log \Lir/\Ltorus=0.10$, while the dashed lines \rev{correspond to the median absolute dispersion (\revs{$1.4826\times$MAD}, see \S~\ref{Mid-infrared to bolometric luminosity ratio versus bolometric luminosity}) at 0.16 dex}.}
 \label{lircomp}
\end{figure*}

As a further check we have compared the infrared luminosities estimated from the observed rest-frame SED and those output of our SED-fitting code.
The left panel of Fig.~\ref{lircomp} shows this comparison. The majority of the sources lie along the one-to-one correlation with a median shift between the observed infrared luminosity and the torus luminosity of $\langle \Log \Lir/\Ltorus\rangle=0.10$ (see the right panel of Fig.~\ref{lircomp}), meaning that the infrared emission observed is mainly originated from hot-dust. 
The tail at high $\Log \Lir/\Ltorus$ values is likely due to some contamination from star formation, which has been subtracted using the fitting technique. 
Outliers are again defined as those at more than 3~$\sigma$ away from the median. Six percent (32/513) of the sample and only 2 sources lie below and above 3~$\sigma$ from the median, respectively. 
Their SEDs are discussed in Appendix~\ref{appendixC}. 

\par
\par
In summary, AGN bolometric luminosities need to be corrected for the effects of both host-galaxy contamination and intrinsic AGN reddening. 
Studies that do not take these factors into account may bias they results on AGN obscuring fractions.
To further emphasize this last point, in the following Section we present a comparison of the $R-\Lbol$ relations under different assumption: {\it 1)} the relation is presented without correcting for host-galaxy and reddening, {\it 2)} considering $\Lbol$ and $\Ltorus$ from the model fitting where the $\Lbol$ values are host-galaxy subtracted and, finally, {\it 3)} where both effects of the host-galaxy and reddening are considered.

\section{Mid-infrared to bolometric luminosity ratio versus $\Lbol$}
\label{Mid-infrared to bolometric luminosity ratio versus bolometric luminosity}
\begin{figure*}
\epsscale{0.55}
\plotone{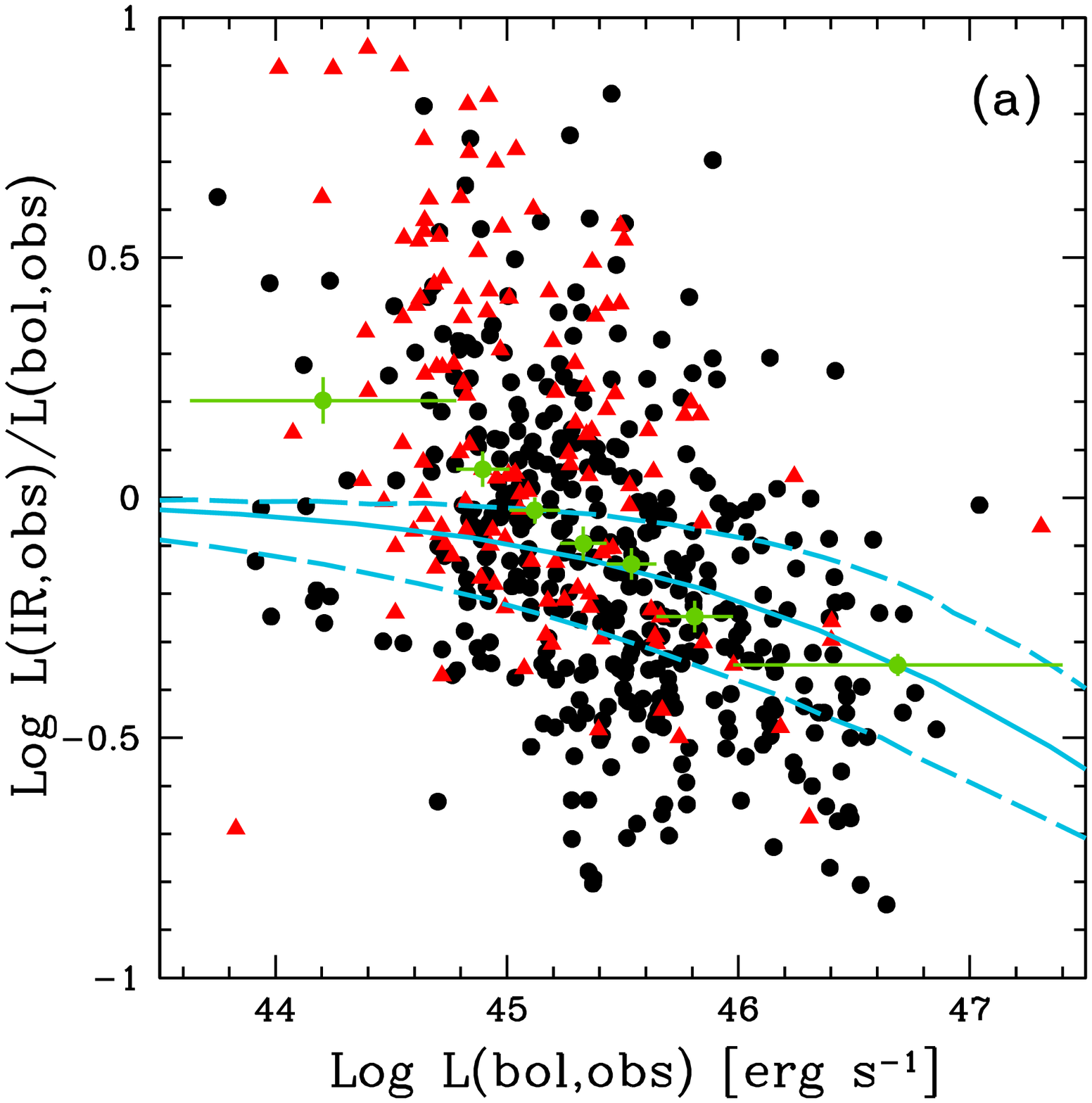}\\
\epsscale{1.1}
\plottwo{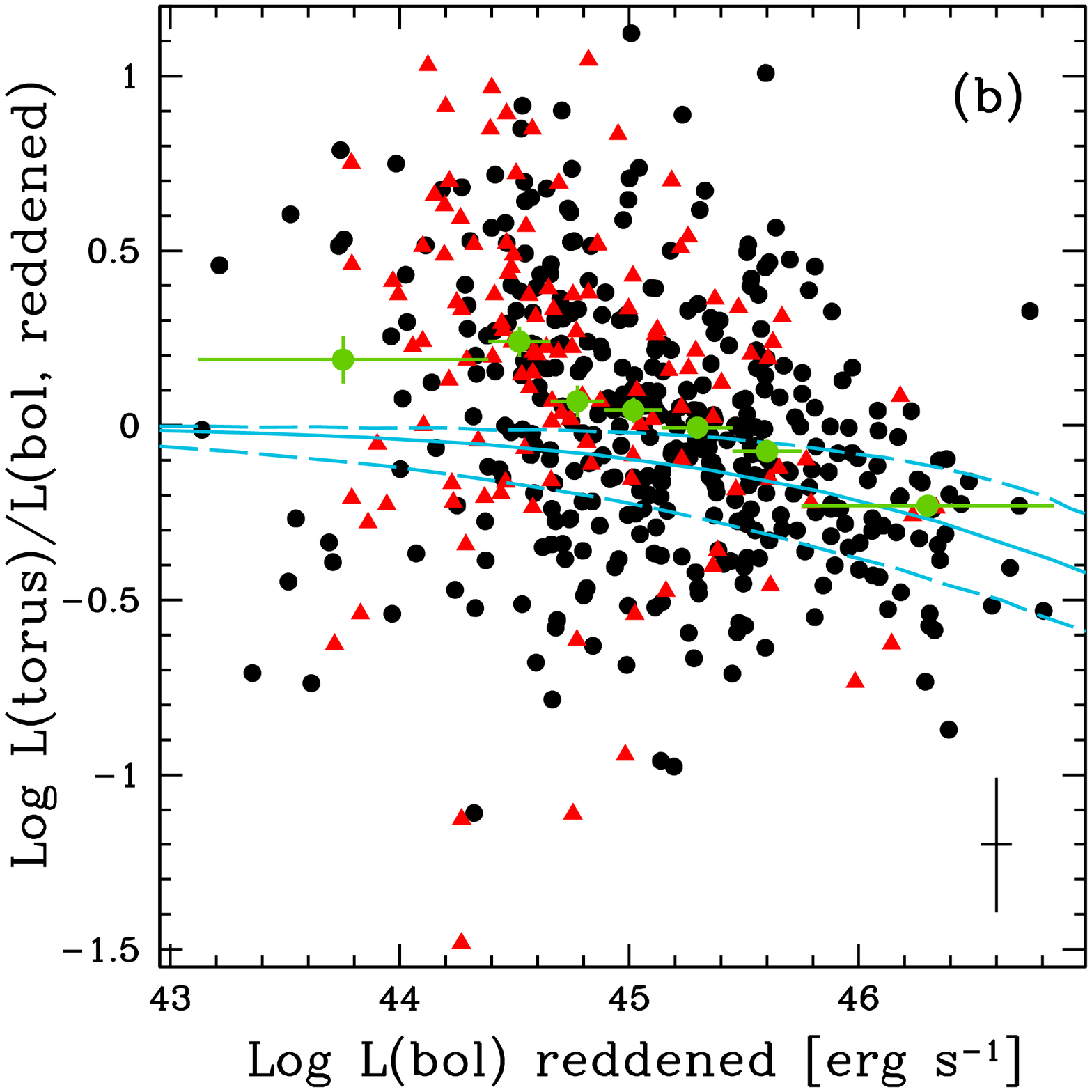}{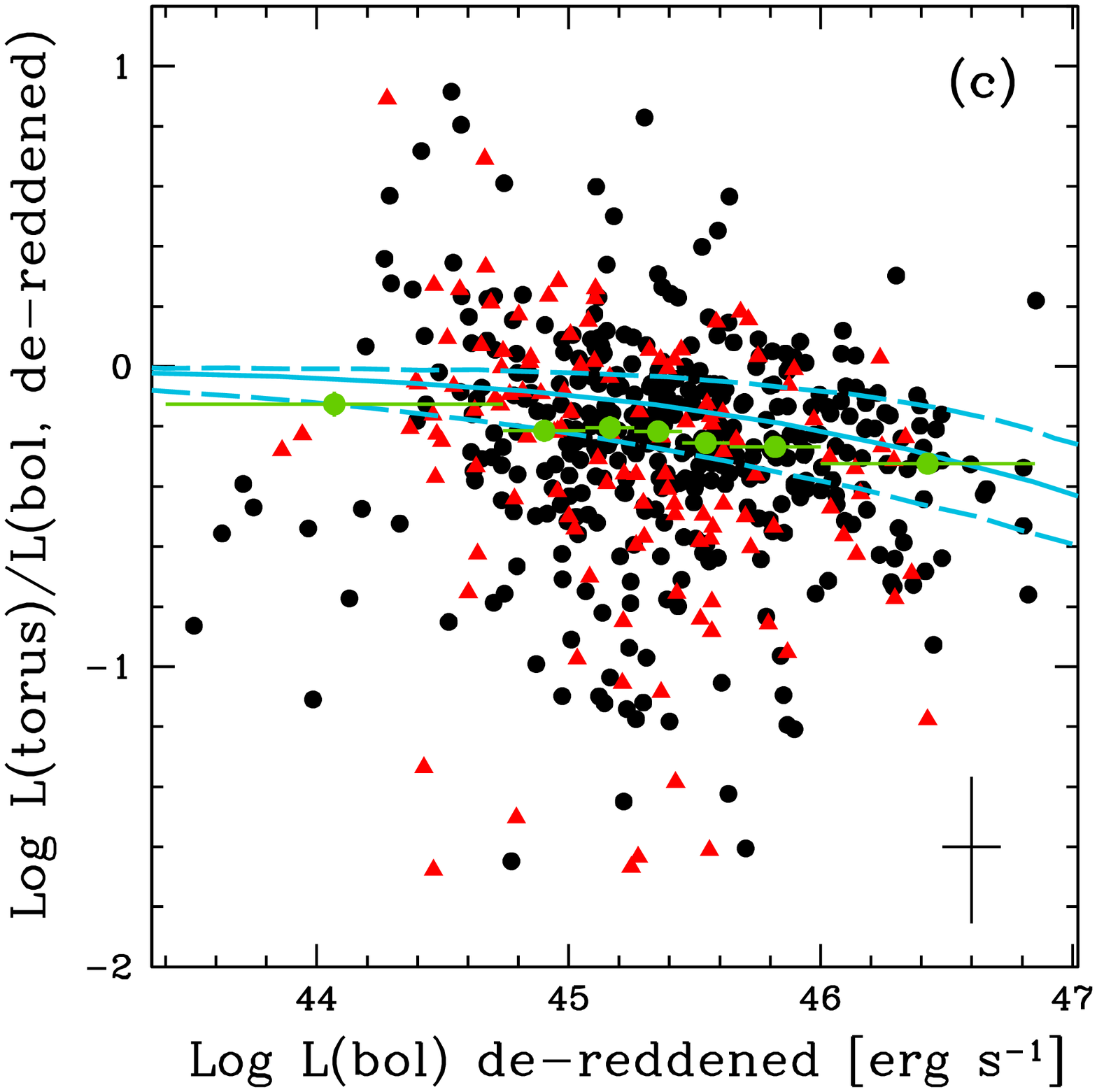}
\caption{\textit{Panel a}: $\Log \Lir/\Lopt$ as a function of $\Lopt$ computed from the observed rest-frame SED. \rev{Black points and red triangles represent the spectro-z and photo-z sample, respectively.} Green points are the median of the $\Log \Lir/\Lopt$ values in each bin (about 77 sources per bin), the bars in the y axis represent the \revs{uncertainty} on the median, while the bars in the x axis are the width of the bin. The cyan line shows the mid-infrared luminosity ratios inferred by M07. Dashed lines trace the uncertainties due to bolometric correction. \textit{Panel b}: $\Log \Ltorus/\Lbol$ as a function of host-galaxy corrected $\Lbol$ without reddening correction (about 73 sources per bin).
\textit{Panel c}: $\Log \Ltorus/\Lbol$ as a function of $\Lbol$ with both host-galaxy and reddening correction (about 73 sources per bin). The average \revs{uncertainty} on $\Lbol$ and $\Log \Ltorus/\Lbol$ are plotted in the bottom right of panel b and c.}
\label{Rlbol}
\end{figure*}

Previous analyses have found a decrease of the mid-infrared luminosity
ratios as a function of bolometric (i.e. accretion disk) luminosity
(see e.g., M07; \citealt{2008ApJ...679..140T,2009MNRAS.399.1206H}), which has been interpreted as a corresponding
decrease in the obscured fraction with luminosity, but these previous
works have \rev{one (or more) of the following} limitations: a) disk luminosities and/or mid-infrared
luminosities have been estimated using uncertain bolometric
corrections b) host-galaxy light has not been subtracted out c) the
disk has not been de-reddened in computing the bolometric luminosity.
\par 
In \S~\ref{Disk luminosities: effect of the host-galaxy and
  reddening}, we have seen that reddening can lead one to
significantly underestimate the true bolometric luminosity and hence
the luminosity ratio $R$ (the relation between the obscured fraction
and $R$ is discussed in Sect.~\ref{Discussion}).  Figure~\ref{Rlbol}
(panel a) shows our observed $\Log \Lir/\Lopt$ values as a function of $\Lopt$ where
green points are the median of $\Log \Lir/\Lopt$ in each bin (defined to have
approximately the same number of sources in each bin, about 77
objects). The luminosities have been determined by
  integrating the interpolated observed photometry (see \S.~\ref{Optical-UV and infrared luminosities from observed rest-frame SEDs}), as has the observed bolometric luminosity plotted on the x-axis. The errors on these luminosities are negligible
and so we do not show estimates here. The bars in the y--axis represent the \revs{uncertainty
on the median} estimated as the standard deviation divided by the
\revs{square root of the number of objects} in each bin ($\sigma_{\rm med}=1.4826$
MAD/$\sqrt{N}$ \rev{; Hampel 1974}; \citealt{1983ured.book.....H}; \rev{Rousseeuw \& Croux 1993})\footnote{The MAD term is the median of absolute deviation between
  data and the median of data (${\rm MAD}=\langle {\rm ABS}(d-\langle
  d\rangle)\rangle$, where $d$ are the data). \rev{The MAD value is scaled by a factor of 1.483 to become comparable with the gaussian standard deviation.}}. The bars in the x--axes are the width
of the bin. 
\rev{We decided to consider the median instead of the mean, because our measures are sometimes moderately disperse (e.g., $\Ltorus/\Lbol$ as a function of $\Lbol$ in Fig.~\ref{Rlbol}). Mean and standard deviation are heavily influenced by extreme outliers. The median and the MAD provide a measure of the core data without being significantly affected by extreme data points. 
In order to further check that the MAD is actually a robust estimator of our distribution, we have compared MAD with both bootstrap analysis and percentile. 
All uncertainties are consistent among the three different methods. 
We have then considered only one method (i.e., MAD) throughout the paper for ease of discussion.}
\par 
M07 present mid-infrared luminosity ratios for a
sample of 25 high luminous QSOs at redshift $2<z<3.5$ with
\textit{Spitzer}-IRS low resolution mid-IR spectra, combined with data
for low luminosity Type-1 AGN from archival IRS observations. Their
definition of obscuring fraction is based on the luminosity ratio at
6.7$\mu$m and 5100\AA{}, corrected by a fixed $\kbol$ ($f=0.39 ~
L_{6.7\mu {\rm m}}/L_{5100\AA{}}$, where the thermal infrared bump is
defined as $2.7 ~ L_{6.7\mu {\rm m}}$ and the accretion disk
luminosity is $7 ~ L_{5100\AA{}}$), and where the X--ray emission is
neglected.  To compare our results with M07, we have then converted
their $L_{5100\AA}$ values using a $\kbol$ of 7.   
The variation of the obscuring fraction as a
function of luminosity found by M07 is also plotted in
Fig.~\ref{Rlbol} with the solid cyan line, while the dashed
lines represent the M07 estimate of the uncertainty due to their 
adopted bolometric correction. 
\rev{Our points in Figure~\ref{Rlbol}a (ignoring host-galaxy contamination and not de-reddening the disk) show a similar trend as M07 relation. M07 has a flatter distribution, less affected by host galaxy contamination at low $\Lbol$, while our measurements show a steeper decline with luminosity. This further emphasize that proper host-galaxy and reddening correction are needed especially at low luminosities.}

The M07 analysis assumes that the luminosity ratio is
equivalent to the obscured AGN fraction (i.e., optically thin torus
regime, see Section~\ref{Dependence of obscured AGN fraction with
  bolometric luminosity}). Our median $\Lir/\Lbolobs$
value is $0.81^{+0.07}_{-0.01}$.  \rev{A Spearman rank test gives the correlation coefficient, $\rho$, of -0.51, excluding the null correlation at the level of about 14$\sigma$.}
\par About
37\% (199/539) have observed $\Lir/\Lbolobs$ values higher than one, which is not physical 
given our assumption on the optically thin torus (i.e., the energy has to be conserved).
\revs{However, this
could be due to several factors such as uncertainties in the
observational data (83 objects over 199 AGN with $\Lir/\Lbolobs>1$ do not have
spectroscopic redshift measurement), reddening
and/or host-galaxy contamination, and non-trivial torus radiative transfer effects, i.e. optically thick torus.}
\par 
The above discussion of Fig.~\ref{Rlbol}a regards the $\Log \Lir/\Lbolobs-\Log \Lbolobs$ relationship, where no host-galaxy and AGN reddening correction has been performed.
If we now use our SED-fitting to take into account and effectively subtract off
host-galaxy contamination from the $\Lbol$ estimates, without correcting for
AGN reddening, the overall effect is to shift $R$ (defined as in Eq.~[\ref{f}]) to higher values and
move $\Lbol$ to lower luminosities as shown in Figure~\ref{Rlbol}
(panel b).  This is because host-galaxy light more significantly contaminates
the near-IR to UV part of the SED more than it does in the mid-infrared, and hence
it significantly impacts our $\Lbol$ estimates. This is 
especially true for those cases where the optical-UV region is
fitted with a star-forming galaxy template (i.e., XID=16, 2072, and
53583 in Fig.~\ref{sed_outliers_ldisk}), which would lead to
unphysical results.  In fact, on average the $R$ values determined from SED
fits and where the host-galaxy is subtracted are only a factor of $\sim1.2$
higher than the $R$ determined from integrated photometry, with a median $R$ of
$0.98^{+0.02}_{-0.02}$ (\rev{$\rho=$-0.33, excluding the null correlation at the level of about $8\sigma$}).  The shift to higher $R$ values is mainly due
to the employed $\Lbol$, given that the observed and the fitted torus
luminosities are quite similar\footnote{\rev{In this case, the $\Lbol$ values (output of our code) have the AGN reddening left in. The $\Lir$ and $\Ltorus$ values are consistent within 0.16 dex, but the reddened $\Lbol$ estimates are, on average, lower than $\Lbolobs$.}} (except outliers
discussed in Appendix~\ref{appendixC}).  The resulting trend between
$R$ and $\Lbol$ is still present, although with a different
normalization.  
\par
\begin{figure*}[ht!]
     \begin{center}
        \subfigure{\label{fig:first}
            \includegraphics[width=0.4\textwidth]{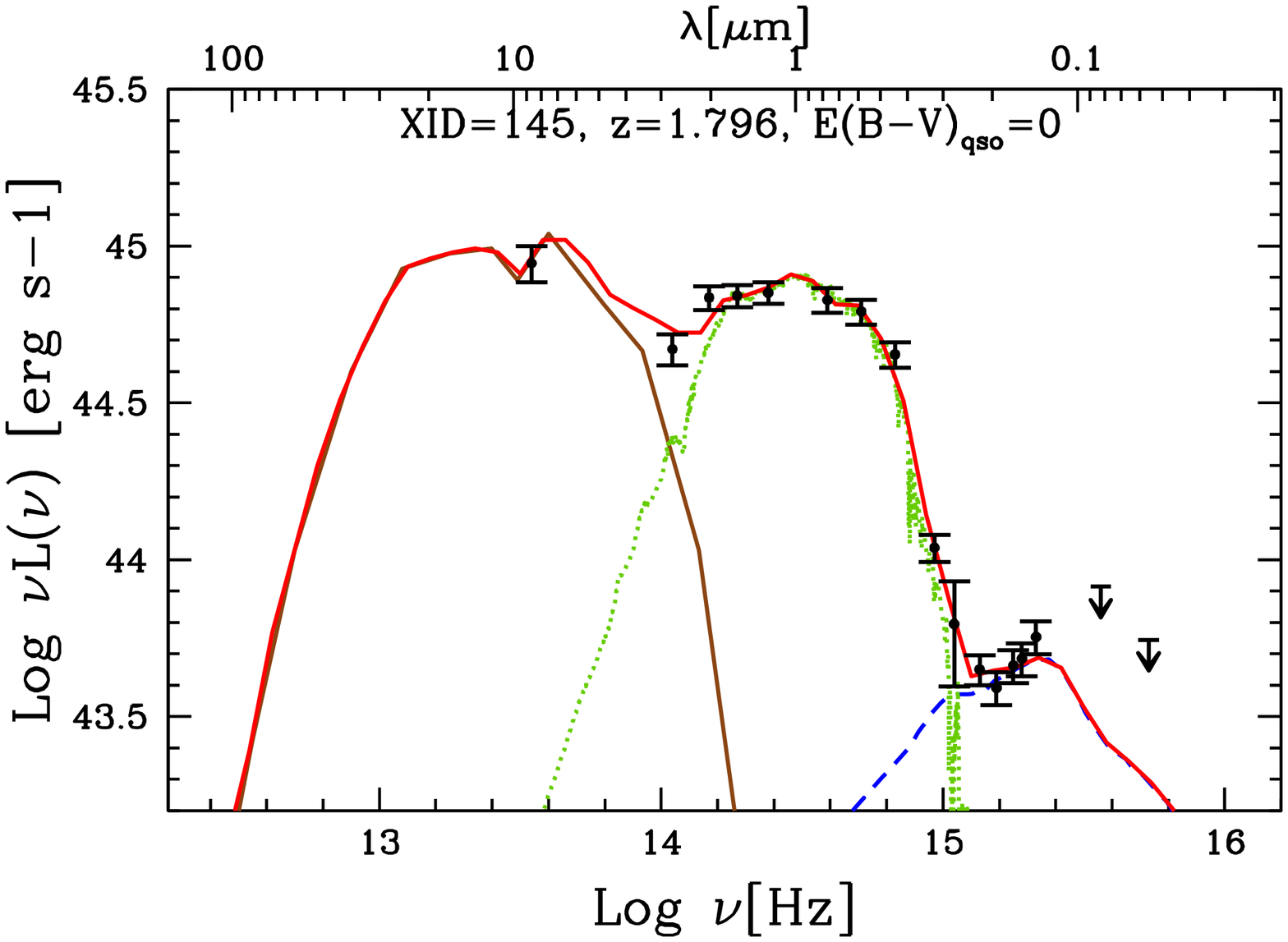}
        }
        \subfigure{\label{fig:second}
           \includegraphics[width=0.4\textwidth]{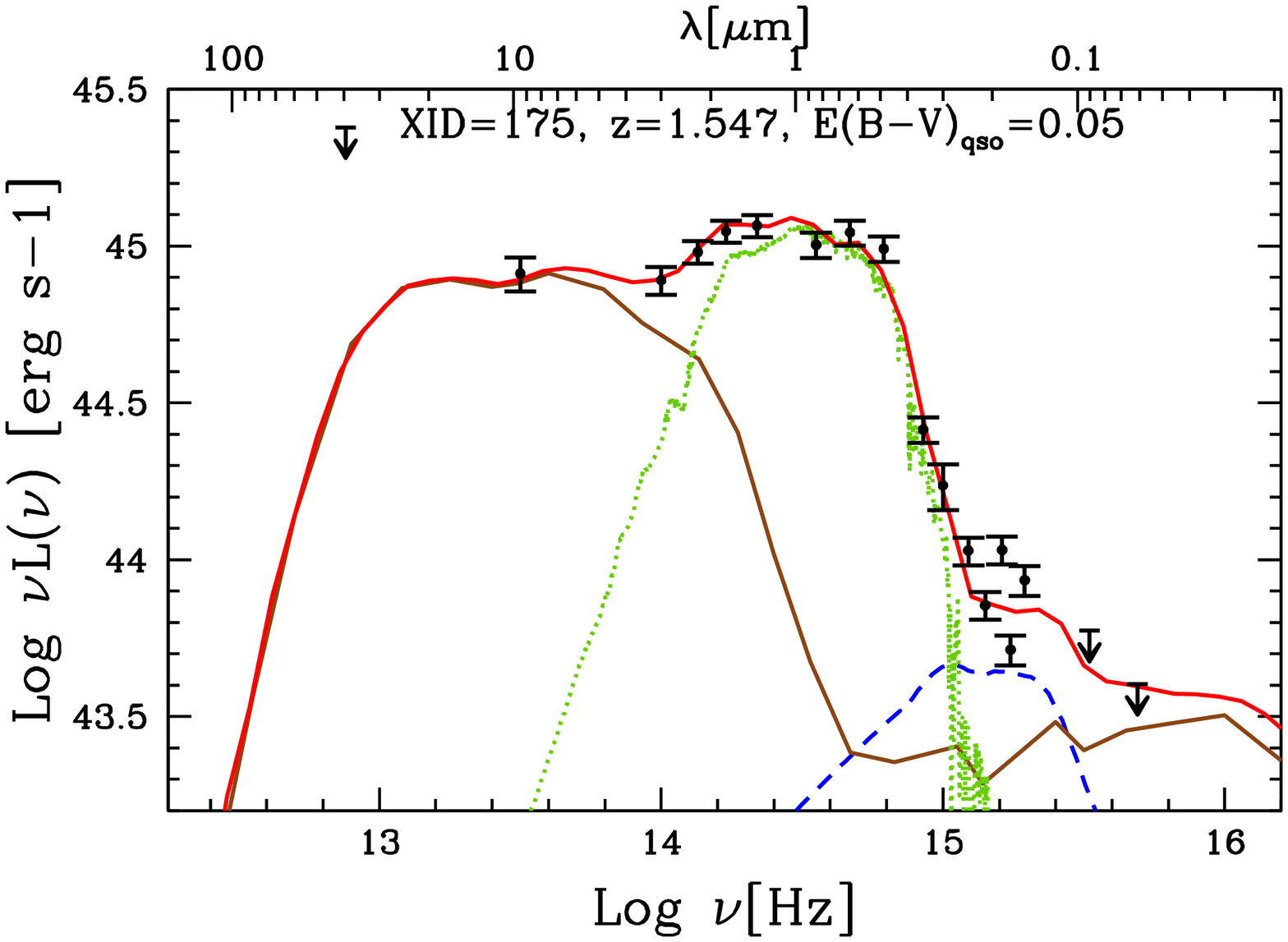}
        }\\ 
        \vspace{-2.5cm} 
        \subfigure{\label{fig:third}
            \includegraphics[width=0.4\textwidth]{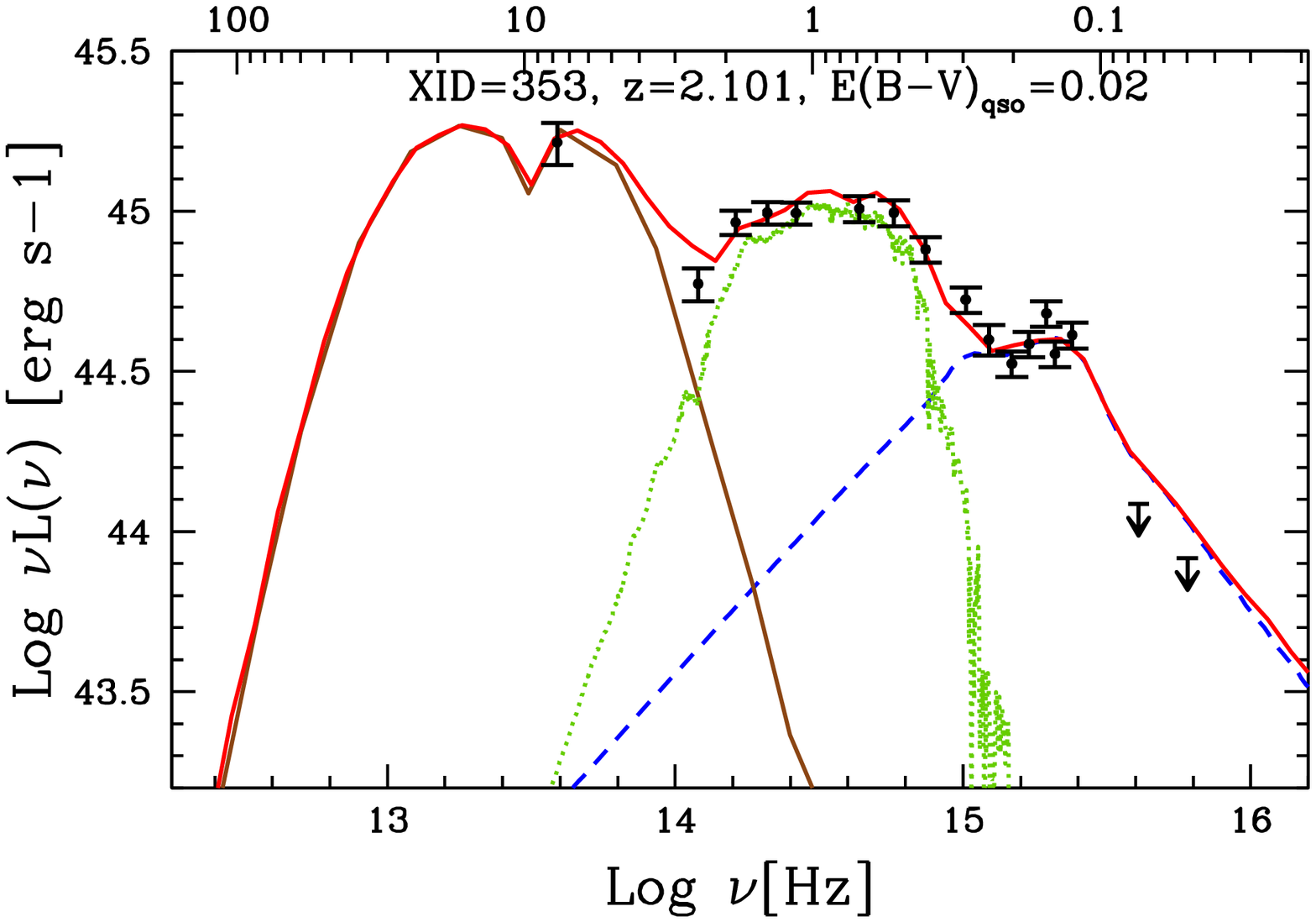}
        }
        \subfigure{\label{fig:fourth}
            \includegraphics[width=0.4\textwidth]{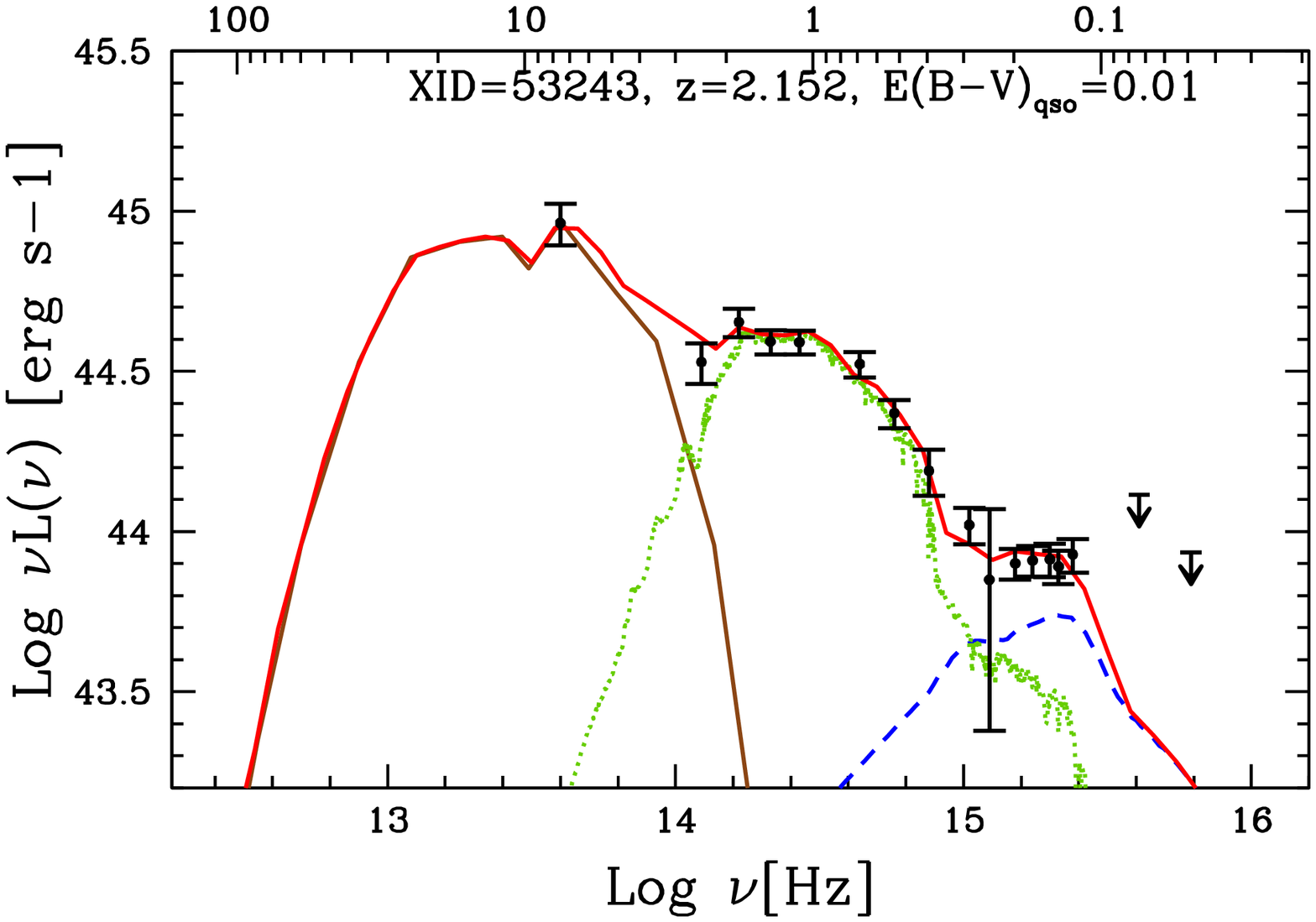}
        }
    \end{center}
    \caption{Example of SEDs with $R>1$. Keys as in Fig.~\ref{panel}.}
   \label{sed_highcf}
\end{figure*}

 
Each best-fit BBB template has been then de-reddened considering the
corresponding $\ebvq$ output of the SED-fit. Figure~\ref{Rlbol} (panel
c) shows the $R$ values as a function $\Lbol$ after employing this
correction. The median $R$ is $0.57^{+0.03}_{-0.01}$  (\rev{$\rho=$-0.19, excluding the null correlation at the level of about $4\sigma$}).
About 19\% (99/513) still have $R$ factors higher than one (70
spectro-z and 29 photo-z).  The SEDs of these objects have, on
average, a high host-galaxy contamination at low $\Lbol$ (four
examples are presented in Fig.~\ref{sed_highcf}).
\par
Summarizing, we find that the average observed $R$ value for the
Type-1 AGN sample presented here is 0.81, with a clear trend of
decreasing $R$ with increasing $\Lbolobs$.  We still observe the same
trend between $R$ and $\Lbol$ if we compute $R$ from the output
luminosities of our SED-fitting code, and correcting for the
host-galaxy contribution only, but the average $R$ is higher than the
observed one ($\langle R\rangle \simeq0.98$). We therefore need to
consider the effect of the intrinsic AGN reddening as well, whose
correction leads to a more reasonable value of $R$ ($\langle R\rangle
\simeq 0.57$), while our relation between $R$ and $\Lbol$ is shallower
than in earlier works (e.g., M07). This is presumably because,
differently from previous analyses, we have corrected for dust
reddening and subtracted off the host. Therefore, the correct average $R$ value is the one taking into account both corrections (i.e., $\langle R\rangle
\simeq 0.57$).
 \par We conclude that any
SED-based analysis needs to take into account the intrinsic AGN
reddening/host-galaxy contamination in order to properly estimate $R$ and its variation on
luminosity.  Throughout the following discussion, we consider the
intrinsic $\Lbol$ the one reddened-corrected and where the host-galaxy has been subtracted.

\section{Discussion}
\label{Discussion}

\subsection{AGN obscured fraction: optically thin versus optically thick tori}
\label{AGN obscured fraction: optically thin versus optically thick tori}

The infrared emission detected by an observer along a line of sight
that crosses the torus (Type-2) might differ from the one along a
dust-free sightline (Type-1), which is to say that infrared emission
from the torus could be \emph{anisotropic.} Following Granato \&
Danese (1994, GD94 hereafter), we define \revs{the ratio between the
integrated flux emitted by the dust in the direction of the equator and that emitted in the direction of
the pole as} $p$ (see their Eq.~[22]). This parameter, which quantifies
the anisotropy of the radiation emitted by the torus, is directly
related to the obscured fraction $\f$, and \revs{it depends} on
the optical depth (at a fixed $\f$ and \revs{for a given geometry}, see GD94 Fig.~10) of the torus to
its own mid-infrared radiation.  The basic idea is that if we assume
that the bulk of the infrared radiation is produced by an optically
thick (to its own infrared radiation) dusty torus, and the observer
has an absorption-free line of sight, $\f$ is related to the $R$
($R=\Ltorus/\Lbol$) and $p$ factors as follows
\begin{equation}
\label{fobscthick}
\f \simeq \frac{R}{1+R(1-p)}.
\end{equation}
For a torus transparent to its own radiation (optically thin) the parameter $p$ is of the order of unity (no viewing angle dependence), and therefore $\f\sim R$. If the torus is instead optically thick, the torus behaves like a black body, and the obscured fraction, for a given value of $R$, is lower than the one in the optically thin regime.
\par
GD94 studied a sample of 56 local ($z\leq0.08$) optically selected radio-quiet Seyferts, of which 16 are unobscured (Seyfert 1). In the case of unobscured AGN, 
GD94 find that the observed infrared continuum originates from an almost homogenous dust distribution, extending at least a few hundred parsec, with $\f\leq 0.6$. 
The GD94 radiative transfer models also show that optically thick and broad (extending for $\sim 1000$ pc at optical luminosities of the order of $10^{46}$ erg s$^{-1}$, but see also \citealt{2011A&A...531A..99T}) tori are able to explain the infrared continua observed in both Type-1 and Type-2 AGN. 
\par
\citet{2008ApJ...685..160N} consider clumpy torus models and showed that a number of 5-15 optically thick dusty clouds along the radial equatorial line of sight can successfully explain AGN infrared observations. 
Further observational evidence from interferometry (e.g., \citealt{2007A&A...476..713K}) and molecular emission lines (e.g., \citealt{2011ApJ...730...48P}) favor a clumpy, rather than smooth, dust distribution in AGN tori. This has stimulated additional modeling efforts by several authors (e.g.,
\citealt{2005A&A...436...47D,2010A&A...523A..27H}, and references therein).
Interestingly, another set of simulations of clumpy torus models by \citet{2011A&A...534A.121H} have found that, although the clouds are optically thick, the observed SED is dominated by emission from optically thin dust in Type-1 AGN. 
\par
Given all the conflicting views about whether the torus is optically thick or
thin to infrared radiation, we remain agnostic and consider $p$ as a
free parameter. But in order to compute $\f$, we would need a determination
of $p$ for each source in our sample, which would require information 
about the mid-infrared optical depth of each AGNs torus. 
Since such estimates are not available and, moreover, as $p$
strongly depends on assumptions about the distribution of dust in the torus and 
its geometry, we instead explore the two extreme cases
where $p=1$ ($\f=R$, optically thin torus) and $p\ll 1$ ($\f\simeq
R/(1+R)$, optically thick torus). The {\it true} obscured AGN fraction
is bounded by these two extremes.

\subsection{Dust reddened AGN population}
\label{Dust reddened AGN population}
All of our analysis on the obscured fraction is based on the parameter $R$, which is the ratio between the mid-infrared and the bolometric emission. In presence of a reddened optical--UV emission, our model fitting procedure should be able to correct for the extinction, so that we can use the de-reddened $\Lbol$. 
However, as we have mentioned at the beginning of this section, the
infrared emission along a line of sight that crosses the torus can be
different from the one along a dust-free line of sight, and this
difference increases as the torus becomes more optically thick ($p$
decreases from unity).  The difference goes in the direction that
infrared emission from an obscured line-of-sight (i.e.  equatorial
emission) is smaller than along a non-obscured line-of-sight
(i.e. polar emission). In cases for which we find that our Type-1 BBB
disk model fits require significant extinction, these lines
of sight are not strictly speaking
dust-free, and their (more equatorial) infrared emission will tend to be smaller. However, the application of Eq.~(\ref{fobscthick}) from
GD94 requires that the $\Ltorus$ in the numerator of $R$, is determined
from a (more polar) dust-free line-of-sight, i.e. it is the $\Ltorus$ of a Type-1
AGN. Thus highly extinct Type-1 AGN will have smaller $L_{\rm torus}$
values, resulting in lower value for $R$ and thus the
obscured fraction that we obtain for such objects would be
systematically smaller than it is in reality. 
\par
Such highly reddened AGN, which are broad-line sources with
nevertheless significantly reddened BBB SEDs, clearly reside in a gray
area of the AGN unification classification into only two types of AGN
(i.e. purely obscured and unobscured).  These sources are likely 
observed from intermediate viewing angles, and thus the simple
modeling of GD94 (parameterized by a Type-1 AGNs polar $L_{\rm torus}$ emission) is no
longer applicable for such objects.  We thus quantify the fraction
of such reddened AGN in our sample by considering the best-fit $\ebvq$
output of our model fitting (see Fig.~\ref{histebvq}).


In what follows, we
present our results for the main sample of 513 Type-1 AGN and for the
sub-sample of 391 objects with $\ebvq\leq0.1$.  Our choice of
$\ebvq=0.1$ is rather arbitrary, but it is effective in defining a
sub-sample of AGN with representative SEDs of the main population (see Fig.~\ref{histebvq}, but see also
Fig.~6 and related discussion in \citealt{richards03}).


\subsection{Dependence of obscured AGN fraction with bolometric luminosity}
\label{Dependence of obscured AGN fraction with bolometric luminosity}
\begin{figure*}
\epsscale{1.15}
\plottwo{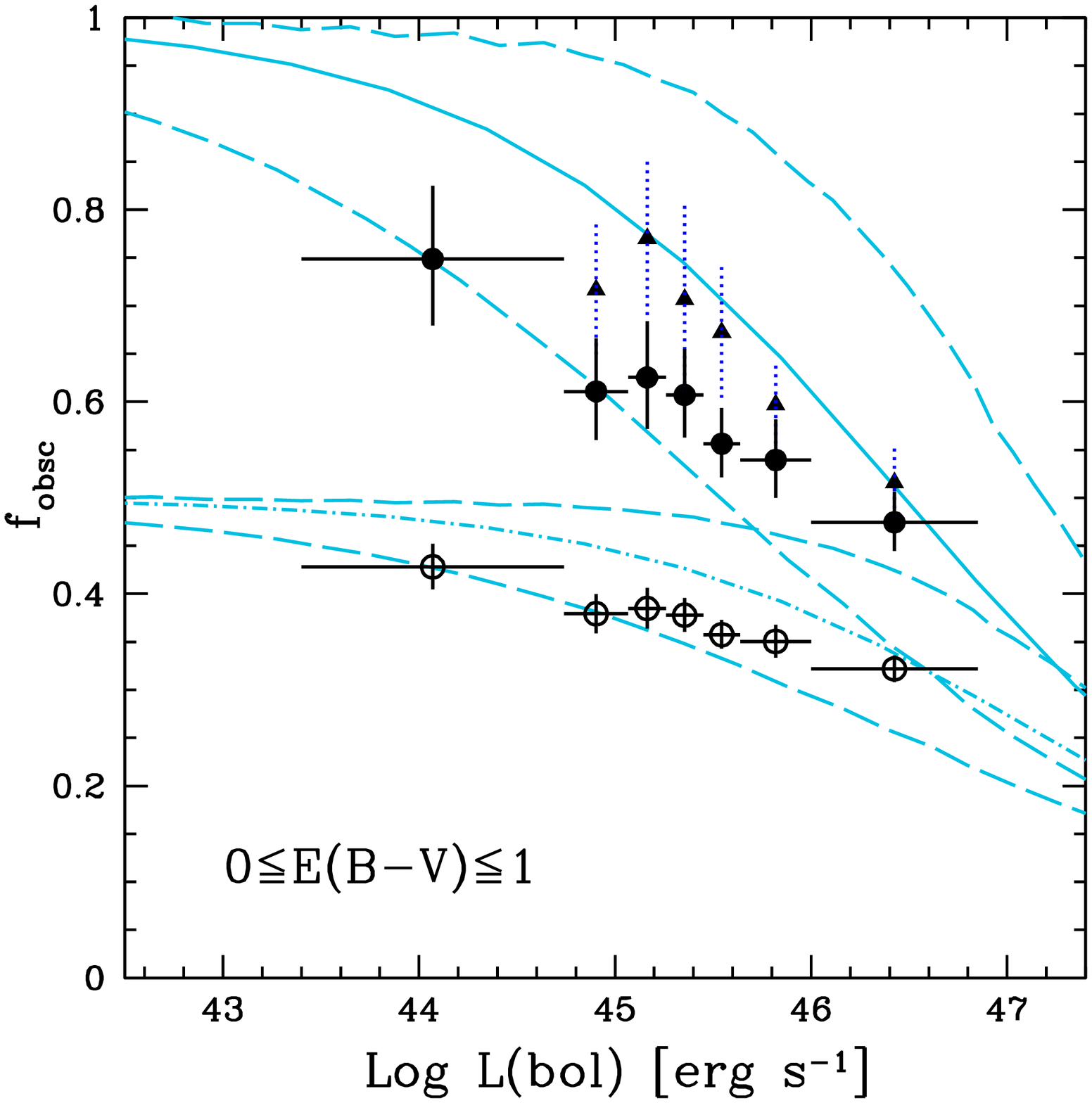}{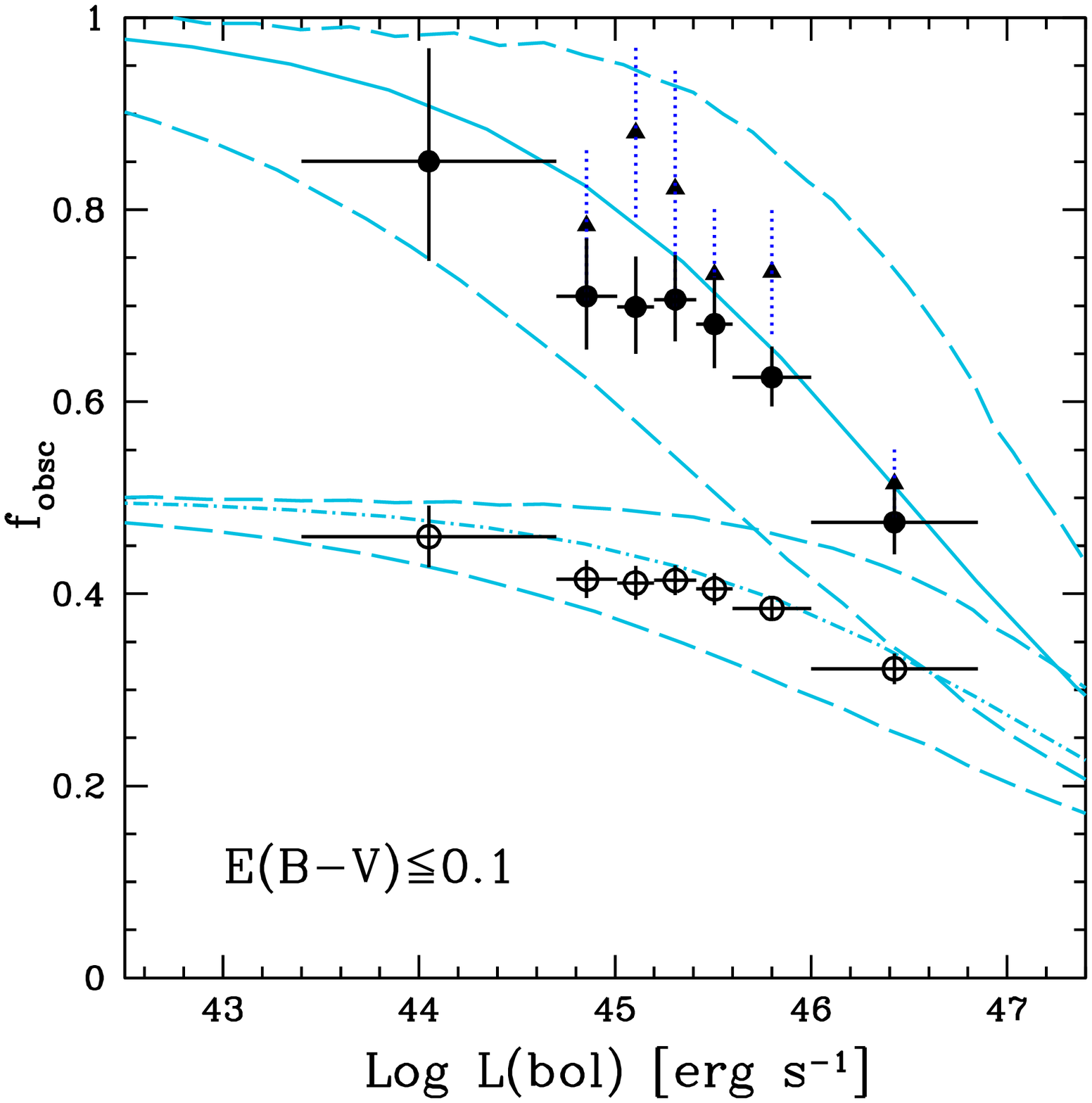}
\caption{\textit{Left panel}: Obscured AGN fraction as a function of $\Lbol$ for the main sample of 513 Type-1 AGN. Filled circles represent our median estimates of the $\f$ parameter in the optically thin torus regime ($p=1$), while open circles represent the $\f$ parameter in the optically thick torus regime ($p\ll1$). \rev{Triangles represent the mean of the obscuring fraction in each bin, while dotted lines are $1~\sigma$ error bars}. The cyan solid line is the obscured AGN fraction as originally estimated by M07 (thin case), while the cyan dot-dashed line represent the obscured AGN fraction by M07 in the optically thick torus case. Dashed lines trace the uncertainties due to bolometric correction. \textit{Right panel}: Obscured AGN fraction as a function of $\Lbol$ for the 391 Type-1 AGN with $\ebvq\leq0.1$.}
\label{fobsclbolgen}
\end{figure*}

Our estimates of the obscured AGN fraction as a function of $\Lbol$ in
the optically thin and optically thick regimes described above are presented in
Figure~\ref{fobsclbolgen}. Filled circles represent the optically thin
torus case, while open circles show the results for the thick torus
prescription.  As a comparison, we have overplotted $\f$ as found by
M07.  \rev{For completeness we also plotted the mean $\f$ in each bin, and the $1~\sigma$ error bars. The mean $\f$ of the first bin is out of scale (i.e., $\langle\f\rangle=1.24$). 
Mean $\f$ estimates are more sensitive to the data with $\f>1$ (20\% of the all sample, but mainly found at the low luminosity end), thus in this case the median is more representative of our 
obscuring fraction distribution.} 
By defining the obscured fraction $\f=R$, the M07 analysis assumes an
optically thin torus (solid cyan line in
Fig.~\ref{fobsclbolgen}). If we instead consider the optically thick
case where $p\ll 1$, the obscured AGN fraction implied by the M07 $R$
measurements is reduced as shown by the dot-dashed cyan line.  We
confirm that a decrease of $\f$ with increasing $\Lbol$ exists in both
the main sample (left panel of Fig.~\ref{fobsclbolgen}) and the sample
with $\ebvq\leq0.1$ (right panel of Fig.~\ref{fobsclbolgen}).  Our
$\f-\Lbol$ relations is within M07's 1~$\sigma$ dispersion.
Assuming the optically thin case, the obscured AGN fraction for the main sample 
changes from $\sim75\%$
at $\Lbol\simeq1.5\times10^{44}$ erg s$^{-1}$ to $\sim45\%$ at
$\Lbol\simeq2.5\times10^{46}$ erg s$^{-1}$. 
This decreasing trend is strongly suppressed in the optically thick
torus regime, where $\f$ ranges approximately from $\sim$45\% to
$\sim$30\%.  If we instead consider the low-reddening sub-sample, the
slope of the $\f-\Lbol$ relation does not change significantly, but
the normalization is shifted to higher $\f$ values. This is expected
because the
full sample includes the reddened AGN, which tend to have lower 
$L_{\rm torus}$ (see \S~\ref{Dust reddened AGN population}) and hence
lower $R$ values, implying lower obscured fractions. 
 
\begin{figure}
 \centering\includegraphics[width=9cm,clip]{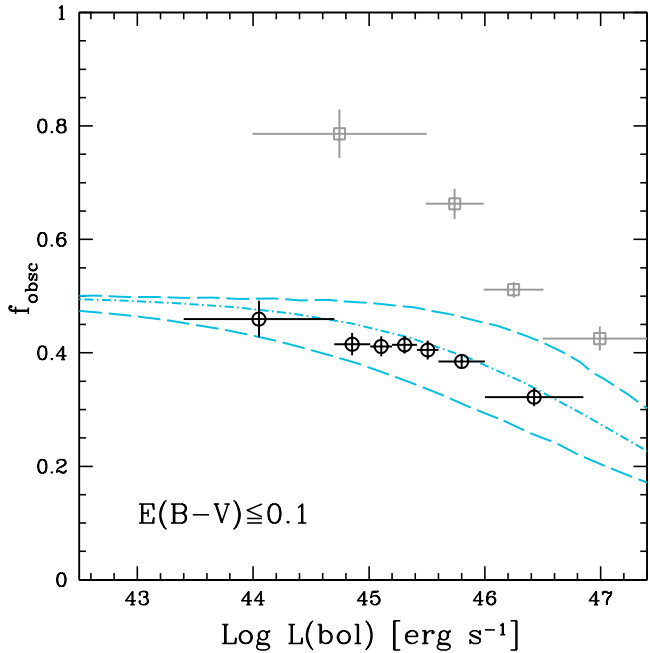}
\caption{Obscured AGN fraction as a function of $\Lbol$ in the optically thick torus regime ($p\ll1$) for the 391 Type-1 AGN with $\ebvq\leq0.1$. Open circles represent our estimates of the $\f$ parameter. Open squares are the $\f$ estimates by T08 (thick regime).}
\label{fobsclbolt}
\end{figure}

\par Another SED based approach has been presented in
\citet{2008ApJ...679..140T} (T08 hereafter). This analysis consider
230 Type-1 AGN (with spectroscopic redshifts) at $z\sim1$, selected
from several surveys (206 AGN are drawn from SDSS, 10 from GOODS, and
14 from COSMOS), with archival 24~$\mu$m MIPS photometry, and GALEX
data. This sample spans a similar range in $\Lbol$ from $10^{44}$ to
$10^{47.5}$ erg s$^{-1}$, and their $\f$ measurement are plotted in
Fig.~\ref{fobsclbolt} with grey open squares. T08 argue that their
measurements agree with M07, and the agreement is rather remarkable;
however, in the T08 analysis $\f$ is estimated by assuming an
anisotropic infrared emission coming from an optically thick torus
($p\ll 1$, see their Eq.~[1]), whereas the M07 results are derived
under the assumption of an optically thin torus $\f\sim R$. The level
of agreement between $\f$ by M07 and those evaluated by T08 is
therefore unexpected.  The obscured fraction produced by an optically
thick torus should be lower than the one originated in a thin torus at
a given $R=\Ltorus/\Lbol$ ratio.  

In order to understand this rather
confusing agreement between T08 and M07, it is worth discussing the
T08 analysis in more detail.  The obscured AGN fraction in T08 is
evaluated from the ratio between the observed luminosity at 24~$\mu$m,
corresponding approximately to the rest frame 12~$\mu$m luminosity,
and the bolometric luminosity (neglecting X--ray emission), and with
no correction for host-galaxy and reddening
contamination. Consequently, they need to compute the fraction of the
total dust-reprocessed luminosity falling within the MIPS band as a
function of opening angle ($f_{12}(\theta)$), which can be interpreted
as inverse of a bolometric correction in the infrared. They find that
$f_{12}(\theta)$ varies from 0.06 to 0.08 considering a series of
models constructed with the code described in
\citet{2005A&A...436...47D}. These $f_{12}(\theta)$ values correspond
to a bolometric correction at 24~$\mu$m of $\sim12.5-17$, which might
be responsible for higher total mid-infrared luminosity than the one we observed,
and therefore leading T08 to overestimate the obscured fraction. 
However, we note that a bolometric correction of 10 (consistent with the recent findings by
\citealt{2012MNRAS.426.2677R}) does not lead to a significantly better
agreement with the optically thick case.  Given the angular dependence
of $f_{12}(\theta)$, it is not straightforward to determine what aspect
of the T08 calculation leads to overestimated obscured fractions. 
\rev{We have also estimated the obscuring fraction for a sub-sample of AGN in the same redshift 
range explored by T08 ($0.8\leq z\leq 1.2$, 89 objects with $0\leq \ebvq\leq1$, 70 with $\ebvq\leq0.1$), but we do not find a better agreement. However, we caution that this sub-sample is significantly smaller than the one considered by T08.} 

The agreement between M07 and T08 thus remains puzzling, given that our
analysis is consistent with M07 under similar assumptions, and has
been carried out with a completely independent method, and without any
bolometric correction prescription.
\begin{figure*}
\epsscale{1.15}
\plottwo{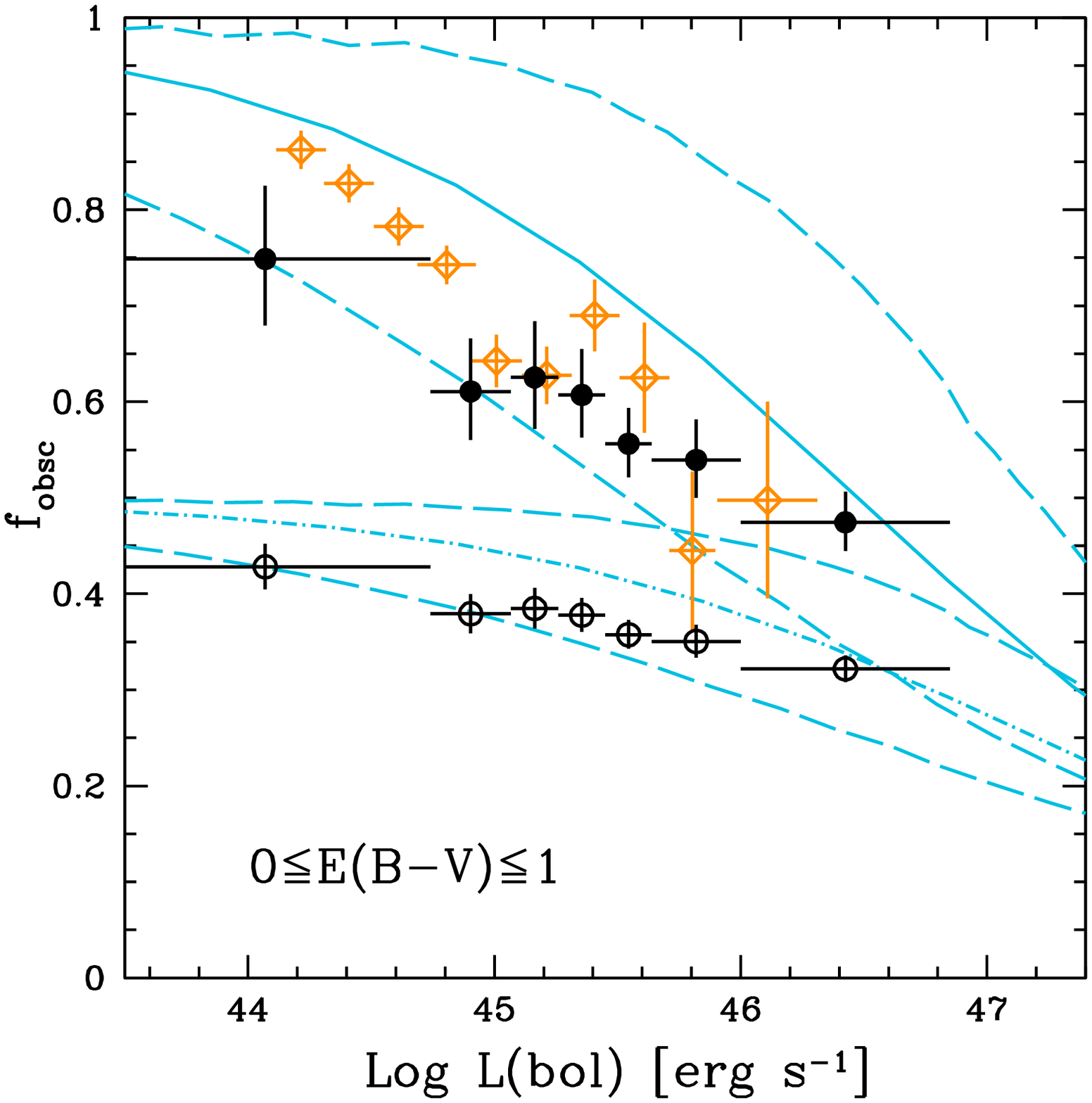}{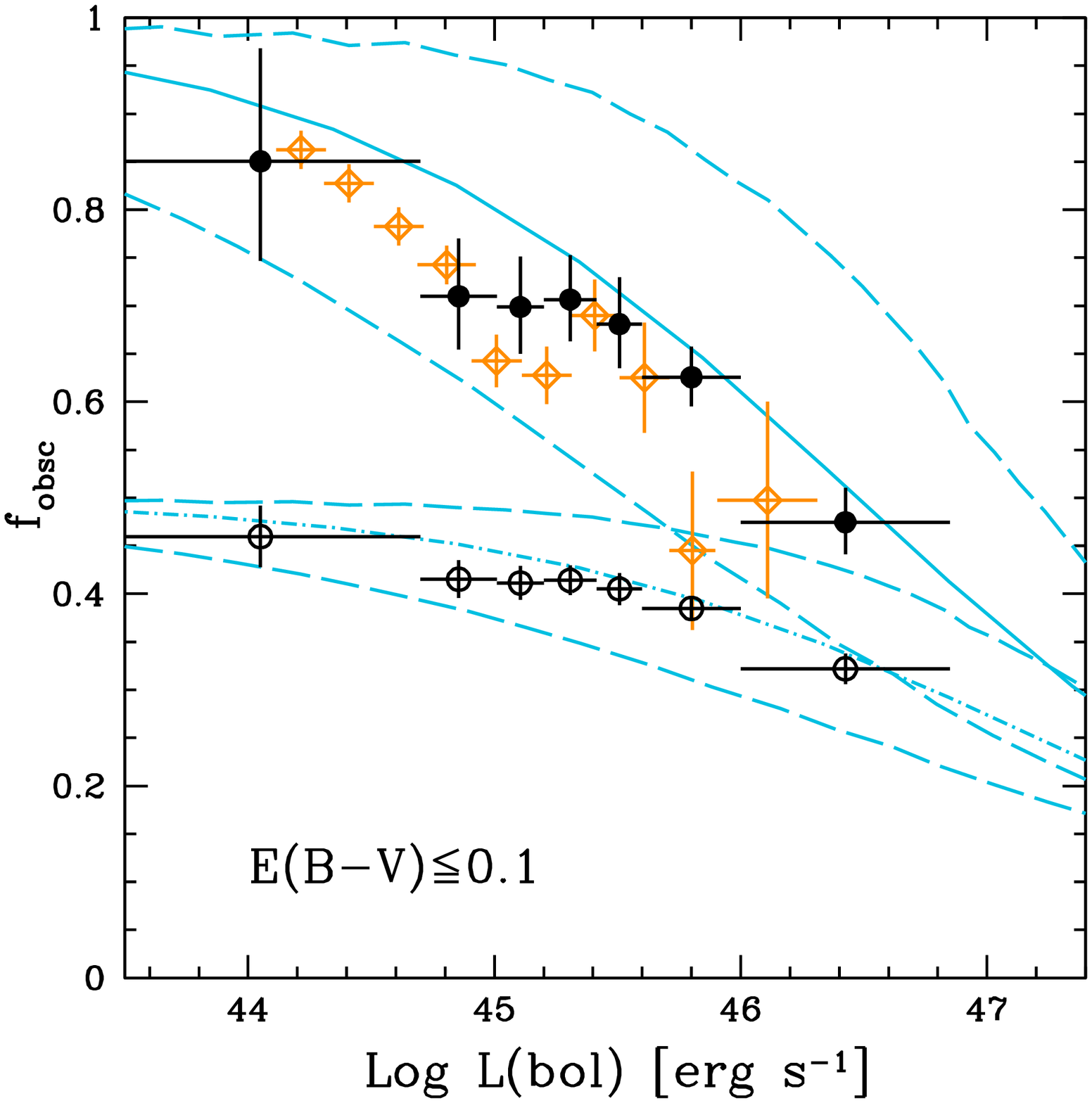}
 \caption{\textit{Left panel}: Obscured AGN fraction as a function of $\Lbol$ for the main sample of 513 Type-1 AGN. Filled circles represent our median estimates of the $\f$ parameter in the optically thin torus regime ($p=1$), while open circles represent the $\f$ parameter in the optically thick torus regime ($p\ll1$). The cyan solid line is the obscured AGN fraction as originally estimated by M07 (thin case), while the cyan dot-dashed line represent the obscured AGN fraction by M07 in the optically thick torus case. Dashed lines trace the uncertainties due to bolometric correction. \textit{Right panel}: Obscured AGN fraction as a function of $\Lbol$ for the 391 Type-1 AGN with $\ebvq\leq0.1$. Open diamonds are the $\f$ values estimated by S05.}
 \label{fobsclbol_simpson}
\end{figure*}
\subsection{Comparison to demographic based analysis.}
\par We can now compare our $\f$ estimates with demography-based
analyses (e.g., \citealt{2005AJ....129.1795H,2005MNRAS.360..565S}).
\citet{2005MNRAS.360..565S} (hereafter S05) used a magnitude-limited
AGN sample\footnote{S05 classified objects which display broad
  ($FWHM>1000$ km s$^{-1}$) \ion{H}{$\alpha$} and \ion{H}{$\beta$}
  emission lines as a Type-1 AGN, while objects showing only broad
  "wings" of the \ion{H}{$\alpha$} line have been classified
  intermediate Type-1 and grouped into the Type-2 AGN sample.} from the
SDSS to determine that the fraction of Type-2 AGN relative to the
total (i.e. the obscured fraction) decreases with the luminosity of
the [\ion{O}{iii}] narrow emission line, where it has been assumed
that the [\ion{O}{iii}] luminosity is a good proxy for the bolometric
AGN emission and, crucially, that the Type-2 AGN sample is complete.
\par
A comparison between our measurement of the obscured AGN fraction
with S05 is presented in Figure~\ref{fobsclbol_simpson} for the total
and the low-reddening AGN sample. We converted the [\ion{O}{iii}] luminosities
to bolometric using  $\kbol$ of $\sim3200$
\citep{2011ApJS..194...45S}.  We find that the obscured fraction
estimated by S05 is fully consistent with the optically thin torus
regime.  Given that the $\f$ values from S05 are computed using a
completely different and independent method, this may be an indication that the
reprocessed infrared emission in AGN occurs in the optically thin regime 
(we will address this issue in Sect.~\ref{models}).  Assuming a constant $k_{[\rm
    {O~III}]}$ for $L_{[\rm {O~III}]}$ is a rather crude
approximation, as it has been found that an anti-correlation exists between
the equivalent widths of emission lines and the continuum luminosity
of AGNs, i.e. the so-called {\it Baldwin effect}, may also exist in
narrow emission lines such as [\ion{O}{iii}] line (e.g.,
\citealt{2002ApJ...581..912D,2004ApJ...614..558N,2012arXiv1211.1113Z};
but see also \citealt{2002MNRAS.337..275C} for a different result).
If the [\ion{O}{iii}] luminosity can be considered a good proxy for $\Lbol$ (e.g., \citealt{2004ApJ...613..109H}), there may be the possibility that $k_{[\rm {O~III}]}$ value is very different from what we have considered and may not be constant with $\Lbol$. We have then applied the relation between $\Lbol$ and [\ion{O}{iii}] luminosity as found recently by \citet{2012MNRAS.426.2703S} (see their Eq.~[4]), and the data are still fully consistent with an optically thin torus.
 
\par 
\citet{2008AJ....136.2373R} present the obscured AGN fraction (i.e., the ratio of Type-2 to total (Type-1 + Type-2) quasar number densities.) for a large sample of optically selected Type-2 AGN from SDSS with redshifts $z<0.83$. 
They carefully take into account selection
effects and biases in their estimate of the obscured fraction, 
giving reliable lower limits
to this parameter (in agreement with S05 results). In their analysis
it is clearly pointed out that [\ion{O}{iii}] luminosity is not
a perfect tracer of $\Lbol$, and there is considerable scatter between
these two luminosities for Type-1 AGN (see their Fig. 9; see also
\citealt{2006A&A...453..525N} for similar results).  They have also
found indications that [\ion{O}{iii}] line is slightly more extincted
in Type-2 AGN than in Type-1.  These findings thus conclude that AGN
samples selected through this line might be biased toward Type-1
objects and this would artificially reduce the obscured AGN fraction
derived in demography-based studies.  
\par 
Summarizing, we confirm
that a correlation exists between $\f$ and the $\Lbol$ in the
optically thin regime, while the correlation is very weak in the thick case.  The slope
of $\f-\Lbol$ relation does not vary significantly considering the
total and the low-reddening AGN sample, but the overall relation for
the low-reddening AGN sample is shifted to higher $\f$ than the one
for total AGN sample.  Finally, a comparison of our SED-based obscured
fraction results to demography-based determinations seems to favor
the optically thin regime.

\subsection{Dependence of obscured AGN fraction with hard X--ray luminosity}
\label{Dependence of obscured AGN fraction with hard X--ray luminosity}
\begin{figure*}
\epsscale{1.1}
 \plottwo{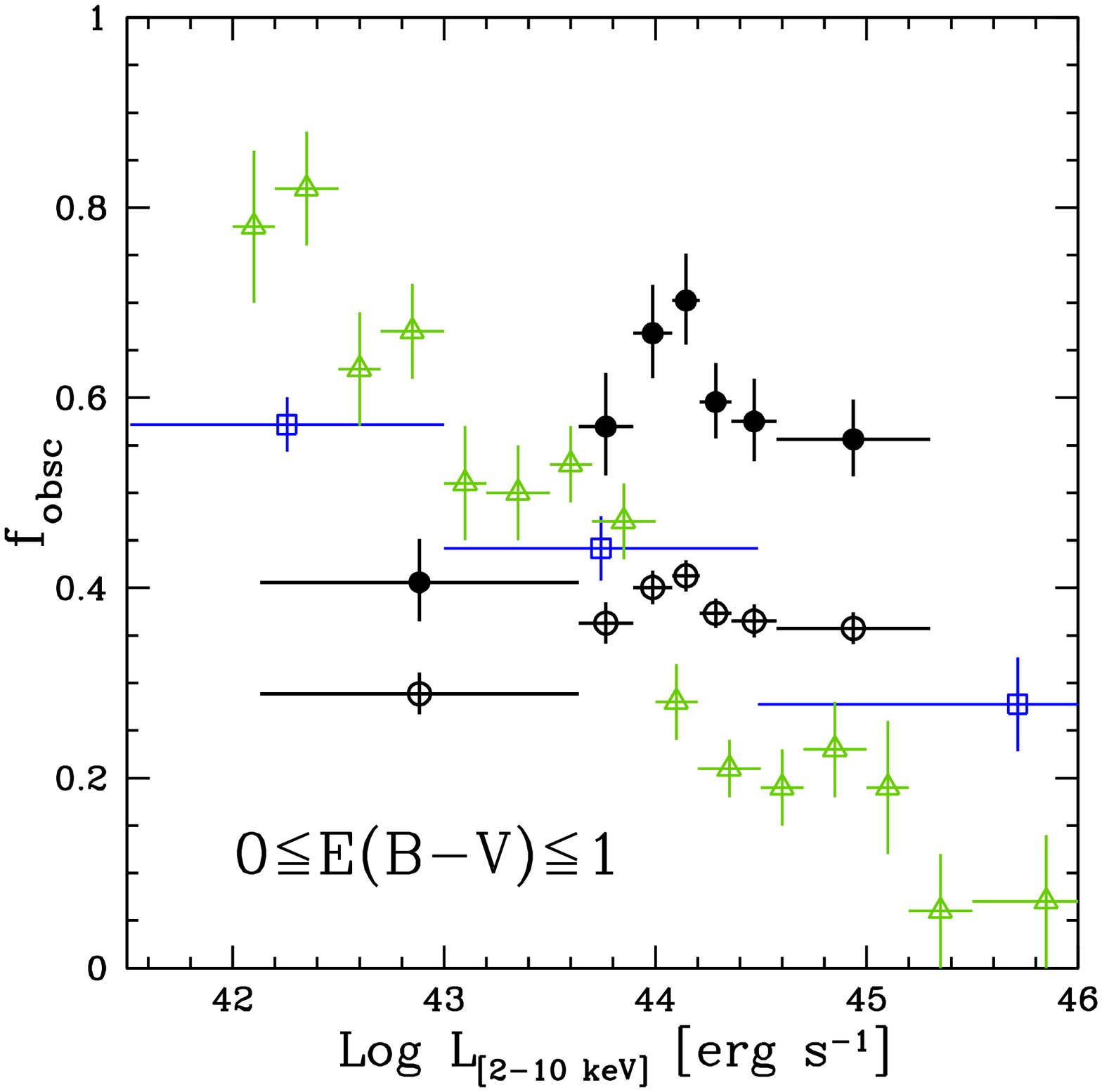}{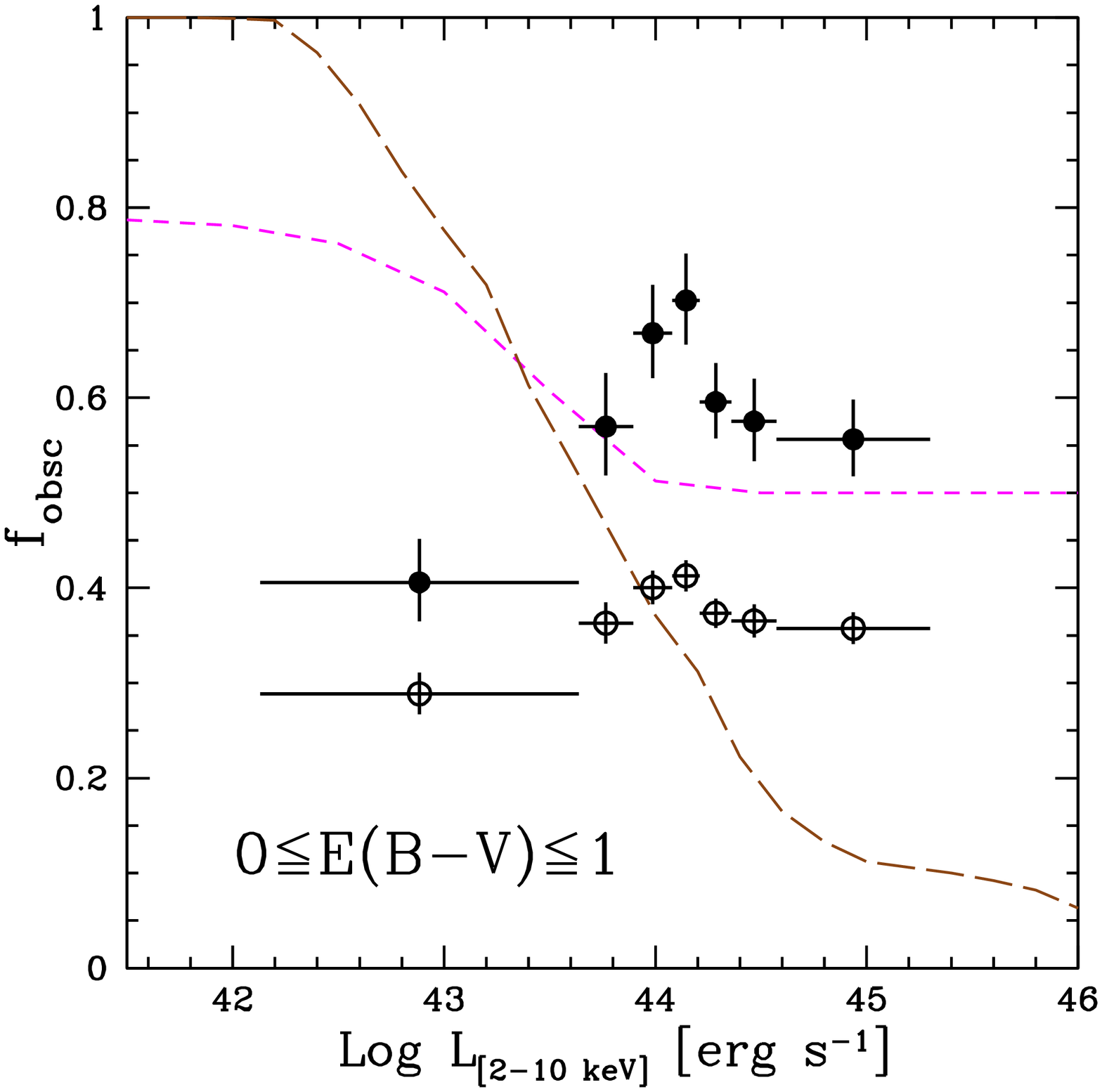}\\
\epsscale{1.1}
 \plottwo{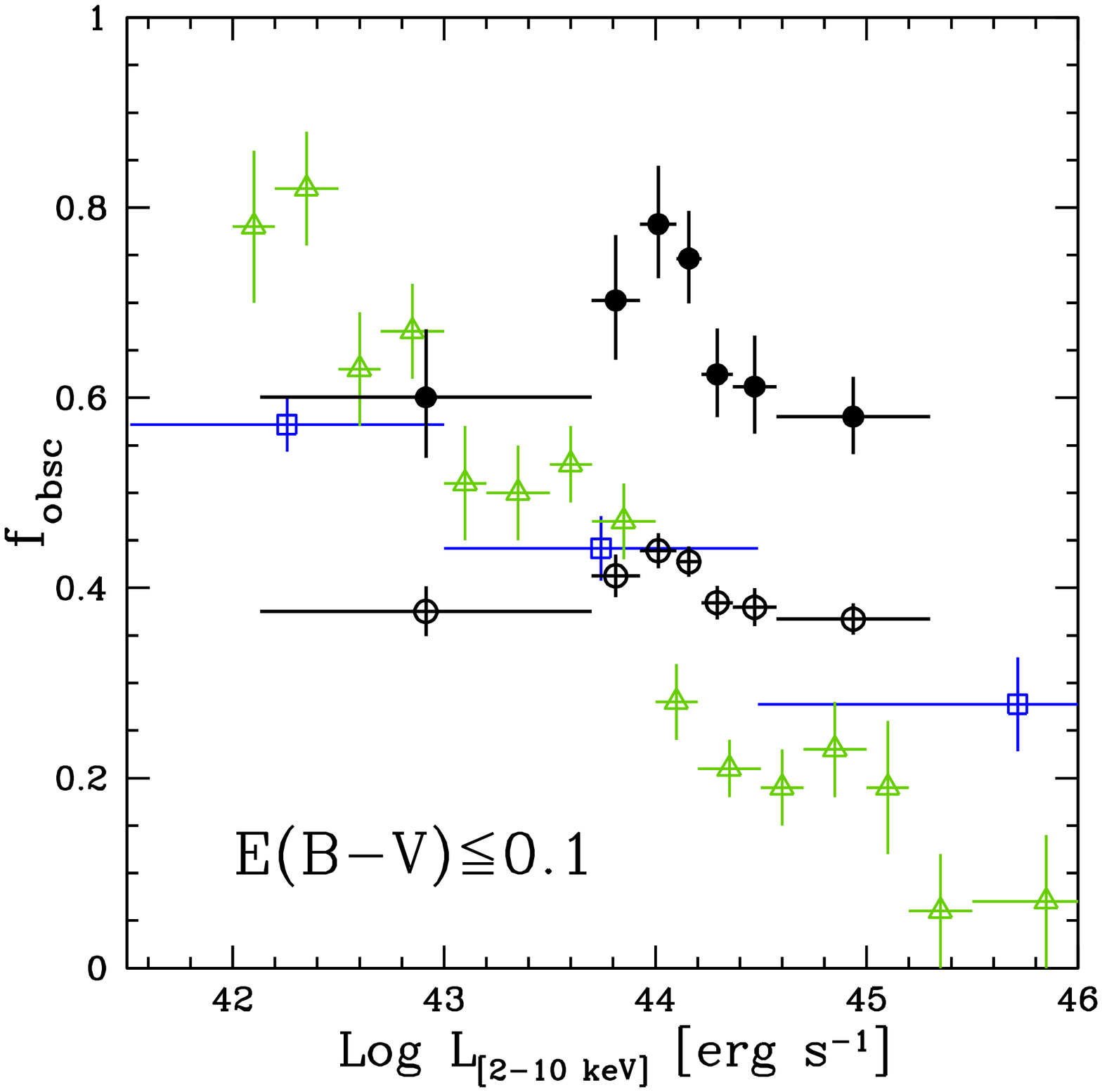}{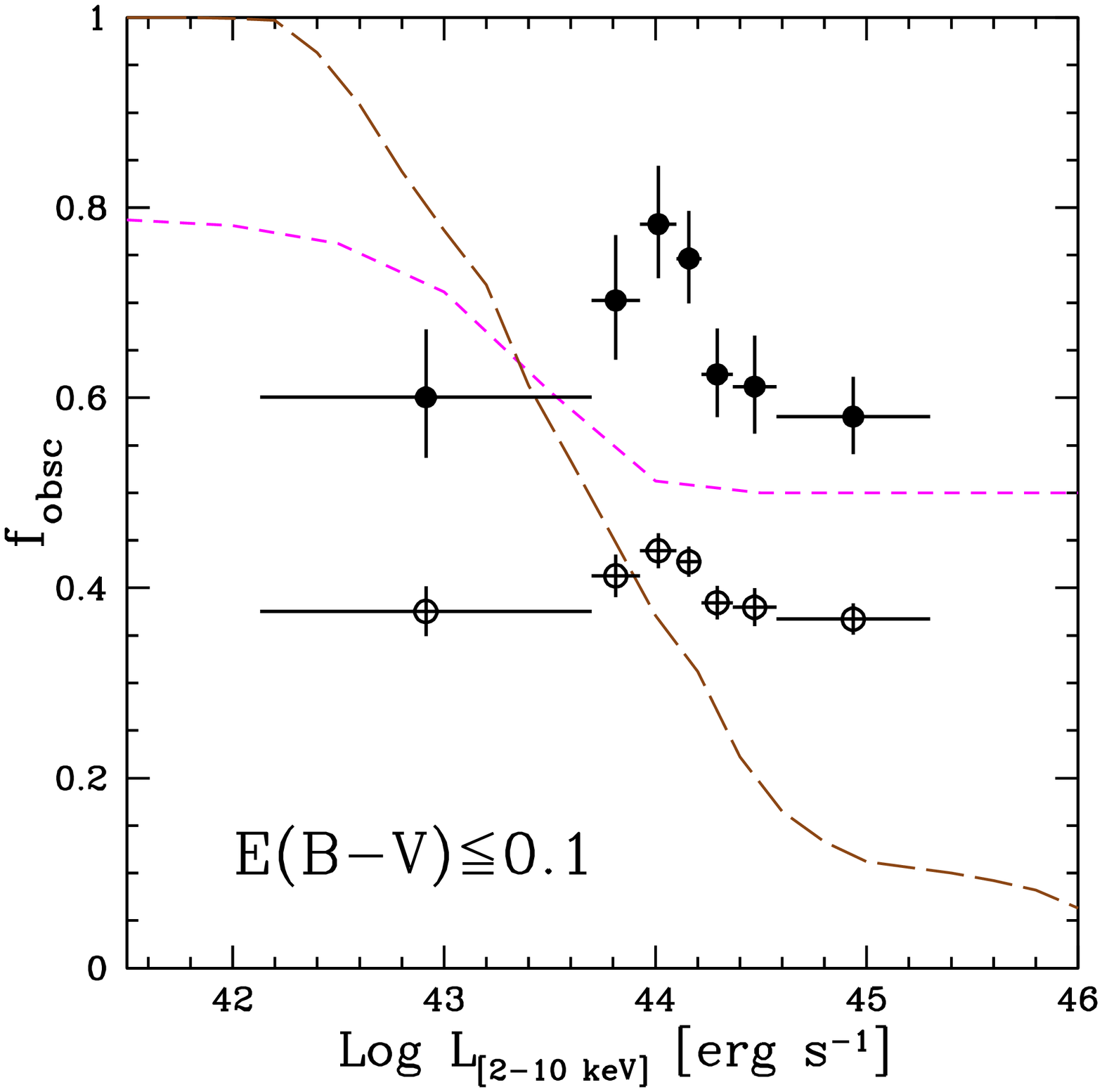}
\caption{\textit{Top row:} Obscured AGN fraction as a function of the X--ray luminosity for the total AGN sample assuming $p=1$ (filled black circles) and $p\ll1$ (open black circles), Ueda et al. (2003, blue open squares), and Hasinger et al. (2008, green open triangles). Magenta dashed line represents the ratio of Compton-thin absorbed AGN to all Compton-thin AGN assumed in the Gilli et al. (2007) population synthesis model. The long-dashed brown line represents the $\f-\Lhard$ relation found by Treister et al. (2009). \textit{bottom row:} Obscured AGN fraction as a function of the X--ray luminosity for the low-reddening AGN sample. Keys as in the top row.}
\label{fobsclx}
\end{figure*}

\citet{2003ApJ...598..886U} were the first to find a significant (almost linear) decrease of the obscured AGN fraction with increasing hard X--ray luminosity ($\Lhard$, see also \citealt{2003ApJ...596L..23S}), considering a combination of hard X--ray surveys, namely the High Energy Astronomy Observatory 1 ({\it HEAO 1}), \textit{ASCA}, and {\it Chandra} for a total of 247 AGNs in the $2-10$ keV luminosity range of $10^{41.6-46.5}$ erg s$^{-1}$ and redshift up to 3. After that work, a series of almost independent analyses of X--ray selected samples have been performed confirming this trend (e.g., \citealt{2003ApJ...596L..23S,2005AJ....129..578B,2005ApJ...635..864L,2008A&A...490..905H}). 
All these works have computed the obscured AGN fraction considering the ratio of Type-2 AGN over the total AGN population.
\par
Figure~\ref{fobsclx} shows the obscured fraction as a function of $\Lhard$ in the optically thin ($p=1$, filled circles) and thick ($p\ll1$, open squares) regime for the main sample ($0\leq\ebvq\leq1$, top row), and for the low-reddening AGN one  ($\ebvq\leq0.1$, bottom row). 
As a comparison, we have over-plotted $\f$ measurement from \citet{2003ApJ...598..886U} (open squares), \citet{2008A&A...490..905H} (open triangles), and the $\f-\Lhard$ relation found by \citet[brown dashed curve, T09 hereafter]{2009ApJ...693.1713T}, utilizing a demographic approach, for 339 X-ray AGN in the Extended Chandra Deep Field South.
There are several interesting points to note.
First, we find that $\f$ does not show a clear trend with $\Lhard$, but it has a peak at $\sim2\times10^{44}$ erg s$^{-1}$, and decreases towards low $\Lhard$.
Second, the variation of $\f$ with $\Lhard$ in the optically thick regime is extremely weak.
Third, X-ray demography-based samples find an obscured fraction lower than our estimates (SED-based) by a factor of $\sim2$ in the optically thin regime, and by a factor of $\sim1.3$ in the optically thick one at high $\Lhard$ ($>10^{44}$ erg s$^{-1}$). 
Evidence for a higher $\f$ than X--ray surveys has already been obtained by \citet{2008AJ....136.2373R}. 
They find a lower limit of the obscured fraction significantly higher than that derived from X--ray surveys (especially Hasinger et al. 2008, and T09) at [\ion{O}{iii}] luminosities higher than $10^{43}$ erg s$^{-1}$.
\par
Our SED-based obscured fraction determination provides an independent confirmation that X-ray demographic analyses are systematically missing obscured and highly obscured 
AGN, which are likely to be the Compton thick AGN.
\par
In Fig.~\ref{fobsclx} we also show the relation between absorbed (i.e., $21.5<\Log\NH\,{\rm [cm^{-2}]}<23.5$) Compton-thin AGN to all Compton-thin AGN assumed in the Gilli et al. (2007, G07 hereafter) X-ray background population synthesis model (a luminosity-dependent $\f$ parameter has been assumed). The G07 model predicts a relatively large fraction of obscured AGNs, $\sim$50\%, at high luminosities, while the observed value by T09 is only $\sim$20\%.
Our $\f$ estimates show a different behavior in both the optically thick and thin regimes than the $\f-\Lhard$ relation in G07.
However, at $\Lhard>10^{44}$ erg s$^{-1}$ the $\f$ values we observe are in better agreements with the $\f$ values in G07 if we consider the optically thin case.
\par
\begin{figure}
 \centering\includegraphics[width=9cm,clip]{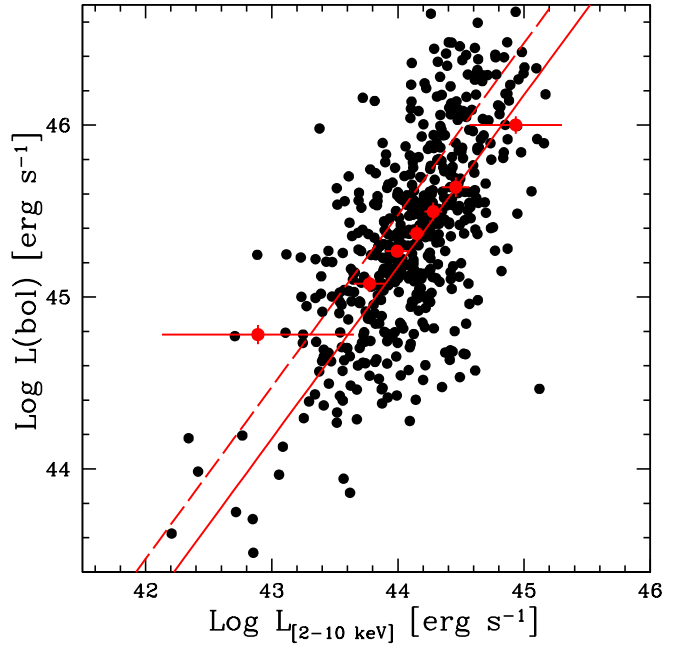}
 \caption{Bolometric luminosity as a function of $\Lhard$. Red points are the median of the $\Lbol$ values in each bin (about 73 sources per bin), the bars in the y axis represent the \revs{uncertainty} on the median (1.4826 MAD$/\sqrt{N}$, see \S~\ref{Mid-infrared to bolometric luminosity ratio versus bolometric luminosity}), while the bars in the x axis are the width of the bin. The red solid line represents \revs{an average hard X--ray bolometric correction of $\sim15$ ($\Lbol/\Lhard$, see Fig.9 in L12 for the average $\Lbol$ of our sample of $\sim2.5\times10^{45}$ erg s$^{-1}$)}. The red dashed line represents \revs{$\Lbol/\Lhard\sim30$ (see Fig.~3 in \citealt{marconi04} for $\Lbol\sim2.5\times10^{45}$ erg s$^{-1}$)}. }
 \label{lxlbol}
\end{figure}

While in \S~\ref{Dependence of obscured AGN fraction with bolometric luminosity} we observed an essentially monotonically linear
decrease of $\f$ with $\log\Lbol$, we find a non-monotonic dependence
of $\f$ on $\log\Lhard$ and thus no compelling evidence for a
decreasing trend with hard X-ray luminosity. The differing
behavior of $\f$ with these luminosities could be explained if $\Lbol$ and $\Lhard$ are not monotonically related, and/or if their relationship has a large scatter. 
In Figure~\ref{lxlbol} $\Lbol$ is plotted as a function of $\Lhard$. 
The relation seems to be almost linear, although the scatter is large 
($\sim0.44$ dex).
\par

Another possible explanation is that Type-1 AGN with high obscured
fractions ($\f\sim 0.6-0.8$) have, for some reasons, been systematically 
excluded from our sample at low X--ray luminosities. For example, if these faint AGN were 
erroneously misclassified as obscured (Type-2) AGN, in a way which is not random, 
but rather dependent on their dust covering factor, then our low-luminosity 
Type-1 sample would be biased towards lower covering factors.

\subsection{Evolution with redshift?}
\label{Evolution with redshift}
\begin{figure}
 \centering\includegraphics[width=9cm,clip]{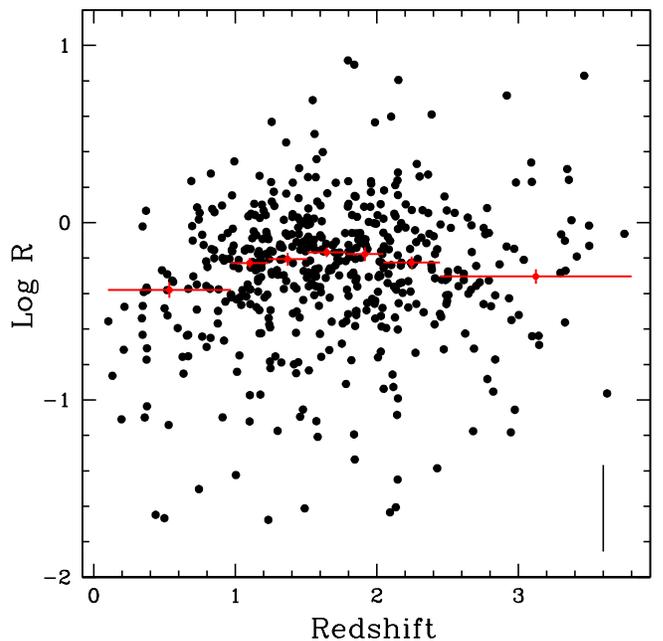}
 \caption{$R$ as a function of redshift considering the all Type-1 AGN sample. The average error bar on $R$ measurements is plotted in the bottom right.}
 \label{cfzdered}
\end{figure}
\citet{2003ApJ...598..886U} did not find clear evidence of $\f-z$ dependence (see also \citealt{2007A&A...463...79G}), while  \citet{2006ApJ...652L..79T} have found a significant increase of the obscured fraction with redshift (as $(1+z)^\alpha$, with $\alpha=0.4\pm0.1$ using a demography-based approach; see also \citealt{2005ApJ...635..864L,2006ApJ...653.1070B,2008A&A...490..905H}) combining seven wide and deep surveys, for a total sample of 2341 objects. To investigate if an evolution of $R$ (and therefore of $\f$) with redshift exists using our data, we have binned the whole sample in the $\Log R-z$ plane considering the same number of objects in each bin ($\sim73$ sources per bin). The distribution is presented in Figure~\ref{cfzdered}. 
To first order, there is no clear evolution between $R$ with redshift \revs{($\rho=$0.11 consistent with no correlation)}. However, this approach can hide possible trends due to the fact that we have considered a flux limited sample, and hence the range of luminosities probed in each redshift bin is not the same. 
We have therefore investigated any possible dependence of $\f$ with both $\Lbol$ and $\Lhard$ by selecting two complete samples in the $\Lbol-z$ and $\Lhard-z$ plane.
\begin{figure*}
\epsscale{1.15}
\plottwo{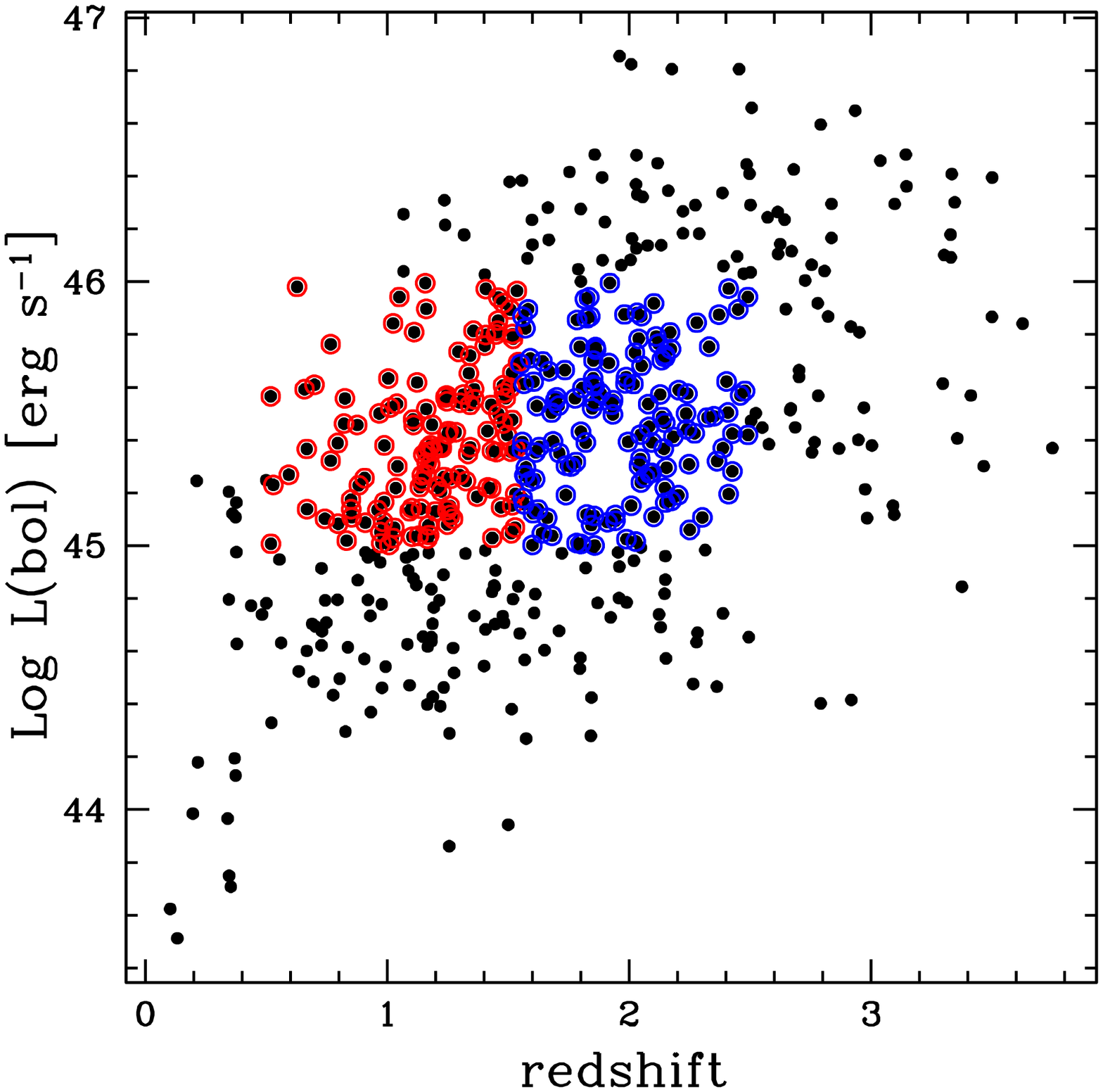}{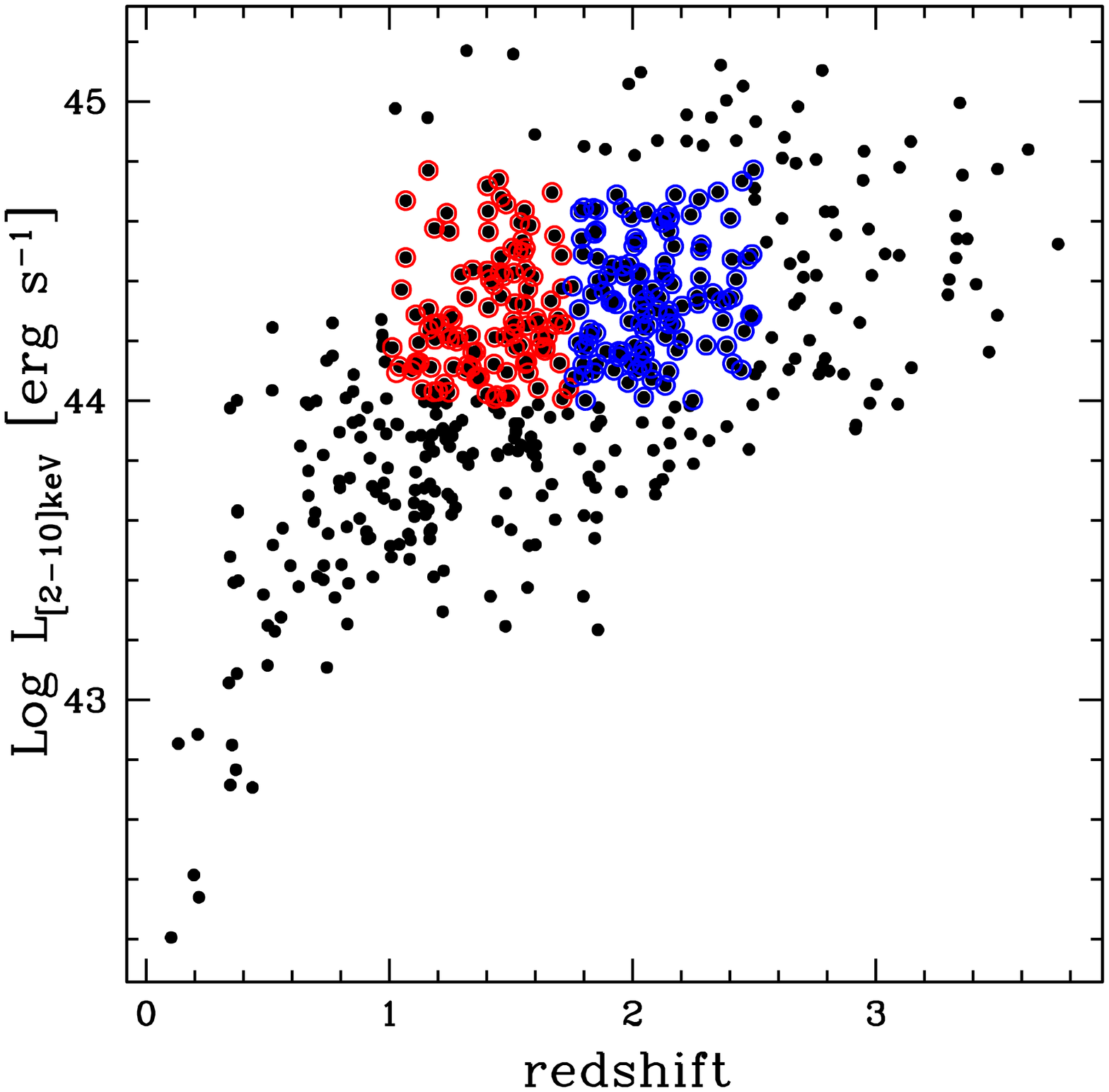}
\caption{\textit{Left panel}: Distribution of $\Lbol$ as a function of redshift. Red and blue open circles mark the two subsamples at $\Log \Lbol=45-46$ \revs{[erg s$^{-1}$]} with redshift $0.5\leq z\leq1.55$ (136 objects) and $1.55< z\leq2.5$ (139 objects), respectively. \textit{Right panel}: Distribution of $\Lhard$ as a function of redshift. Red and blue open squares mark the two subsamples at $\Log \Lhard=44-44.8$ \revs{[erg s$^{-1}$]} with redshift $1.0\leq z\leq1.75$ (120 objects) and $1.75< z\leq2.5$ (112 objects), respectively.}
\label{lum}
\end{figure*}

\subsubsection{Evolution with redshift of the $R-\Lbol$ relationship}
\label{Evolution with redshift of the R-Lbol relationship}
\begin{figure*}
  \centering
  {\includegraphics[width=0.6\textwidth]{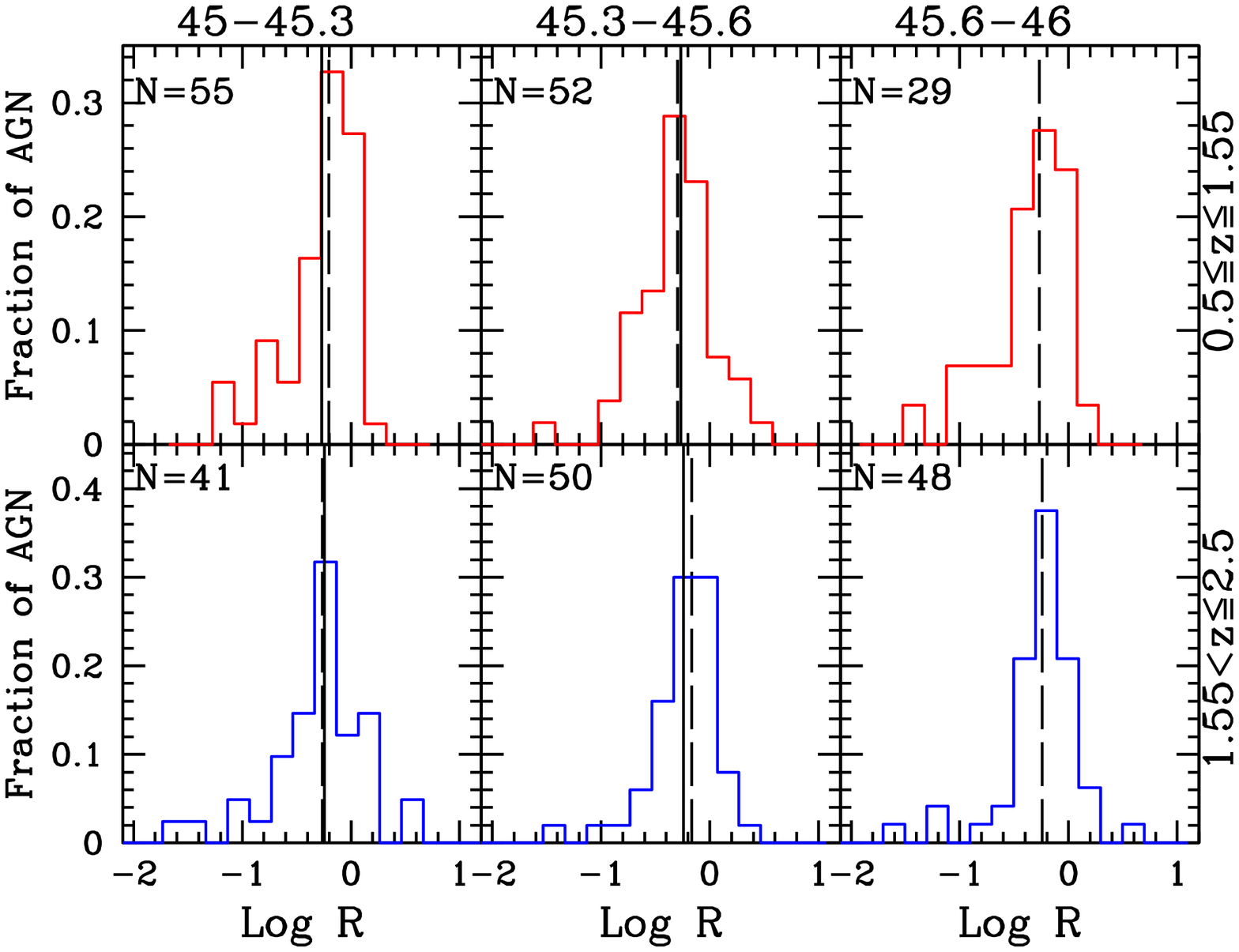}}
  {\includegraphics[width=0.4\textwidth]{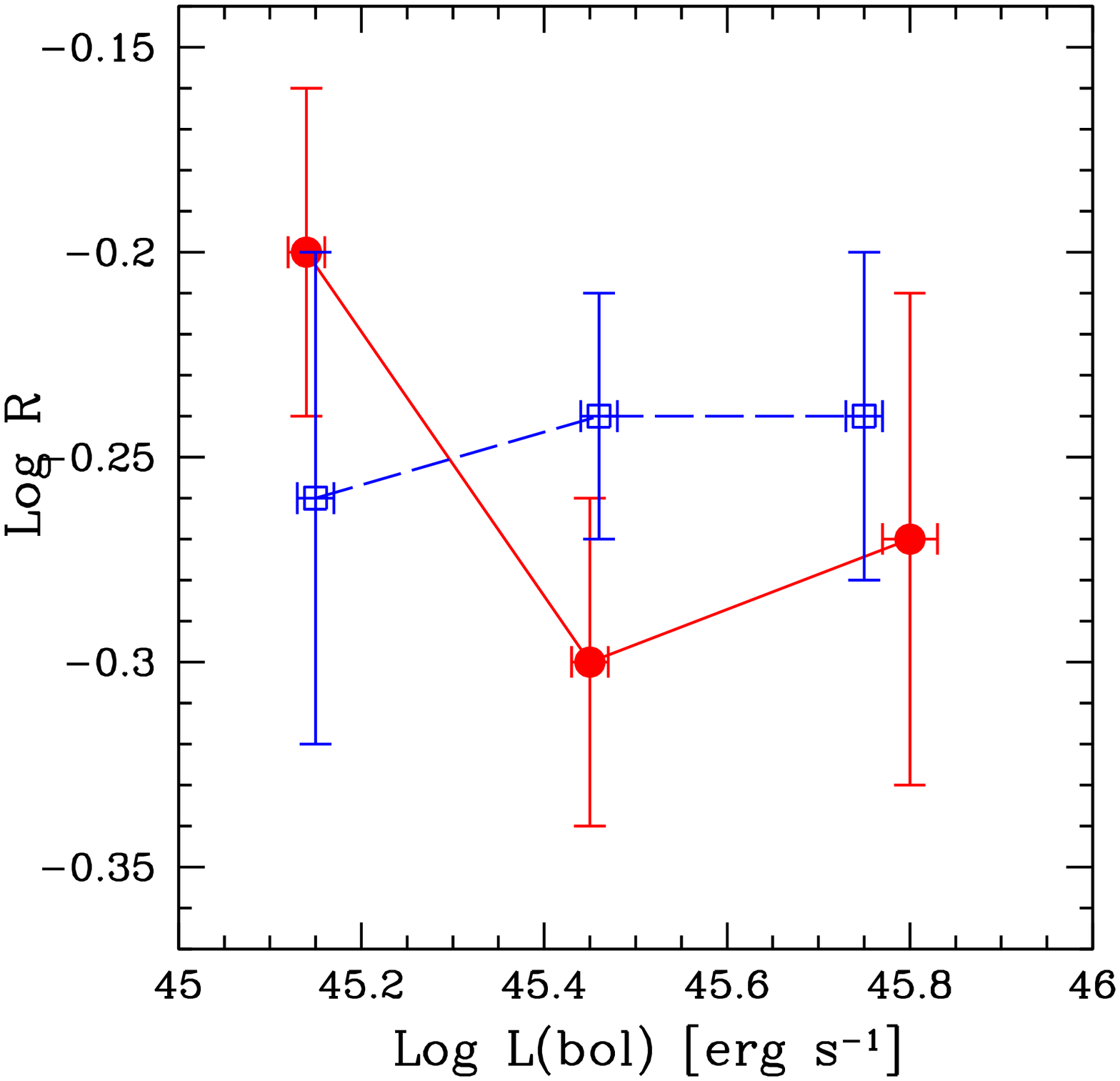}}                        
  \caption{\textit{Upper panel}: Distributions of $\Log R$ in bins of $\Lbol$ and redshift. The panels are divided between $0.5\leq z \leq 1.55$ (top three panels) and $1.55< z \leq 2.5$ (lower three panels), and bolometric luminosities increases from left to right ($\Log\Lbol$ [erg s$^{-1}$] intervals and the number of objects are reported on top of each panel). Histograms show the observed $\Log R$ distributions. The dashed lines are the median values, while the solid black lines in the first and in the second bins are plotted at the $\Log R$ value corresponding to the median in the highest luminosity bin. \textit{Lower panel}: Median $\Log R$ as a function of the median $\Lbol$. Filled circles and open squares represent the median $\Log R$ for $0.5\leq z \leq 1.55$ and $1.55< z \leq 2.5$, respectively. Solid lines connect low redshift bins, while dashed ones connect high redshift bins. Error bars on the median are estimated considering the \rev{MAD} divided by the square root of the number of observed AGN.}
  \label{cflbzevol}
\end{figure*}

The possibility of a dependency of the $R-\Lbol$ relationship with redshift has been explored by binning in $z$ and $\Lbol$. The sample is divided in two redshift bins and three $\Lbol$ bins. The redshift bins are $0.5\leq z \leq 1.55$ and $1.55< z \leq 2.5$ in the $\Lbol$ range of $10^{45-46}$ erg s$^{-1}$, while the luminosity cuts in each redshift bin are chosen in order to explore almost the same luminosity range ($\Log\Lbol$\revs{[erg s$^{-1}$]}$=45-45.3$, $45.3-45.6$ , and $45.6-46$).  
In Fig.~\ref{lum} (left panel) the $\Lbol$ distribution as a function of redshift is presented. Red and blue open circles mark the two complete subsamples selected with the criteria above. There are 136 AGN in the low redshift bin with a median $\Lbol=10^{45.37}$ erg s$^{-1}$, while there are 139 objects in the high-redshift bin with a median $\Lbol=10^{45.52}$ erg s$^{-1}$. The two median $\Lbol$ differ only by a factor of $\sim1.4$. 
Given the absence of a clear trend in Fig.~\ref{cfzdered}, a factor 1.4 difference results in a small change in the mid-infrared to bolometric luminosity ratio. 
We have performed a (two-sided) Kolmogorov-Smirnov test in order to further test whether these two AGN subsamples are consistent with having the same distribution of $\Lbol$ ($D$ value of 0.17 and probability of 0.042). 
The histograms in Fig.~\ref{cflbzevol} (upper panel) show the $\Log R$ distributions, while the dashed lines represent the median $\Log R$ in each bin. 
In Fig.~\ref{cflbzevol} (lower panel) the median $\Log R$ is plotted against the median $\Lbol$. Different symbols for low redshift (filled circles) and high redshift (open squares) are introduced.
\rev{All bins are consistent within the errors.
\revs{Moreover, a Spearman rank test between $\Log R$ and $\Lbol$ indicates no correlation for both high and low redshift bins.}}
Given that with our data-set we can investigate only a narrow range of $\Lbol$, we cannot claim whether a significant evolution of $R$ as a function of $\Lbol$ is present on a wider $\Lbol$ range.

\subsubsection{Evolution with redshift of the $R-\Lhard$ relationship}
\label{Evolution with redshift of the R-Lhard relationship}
\begin{figure*}
  \centering
  {\includegraphics[width=0.6\textwidth]{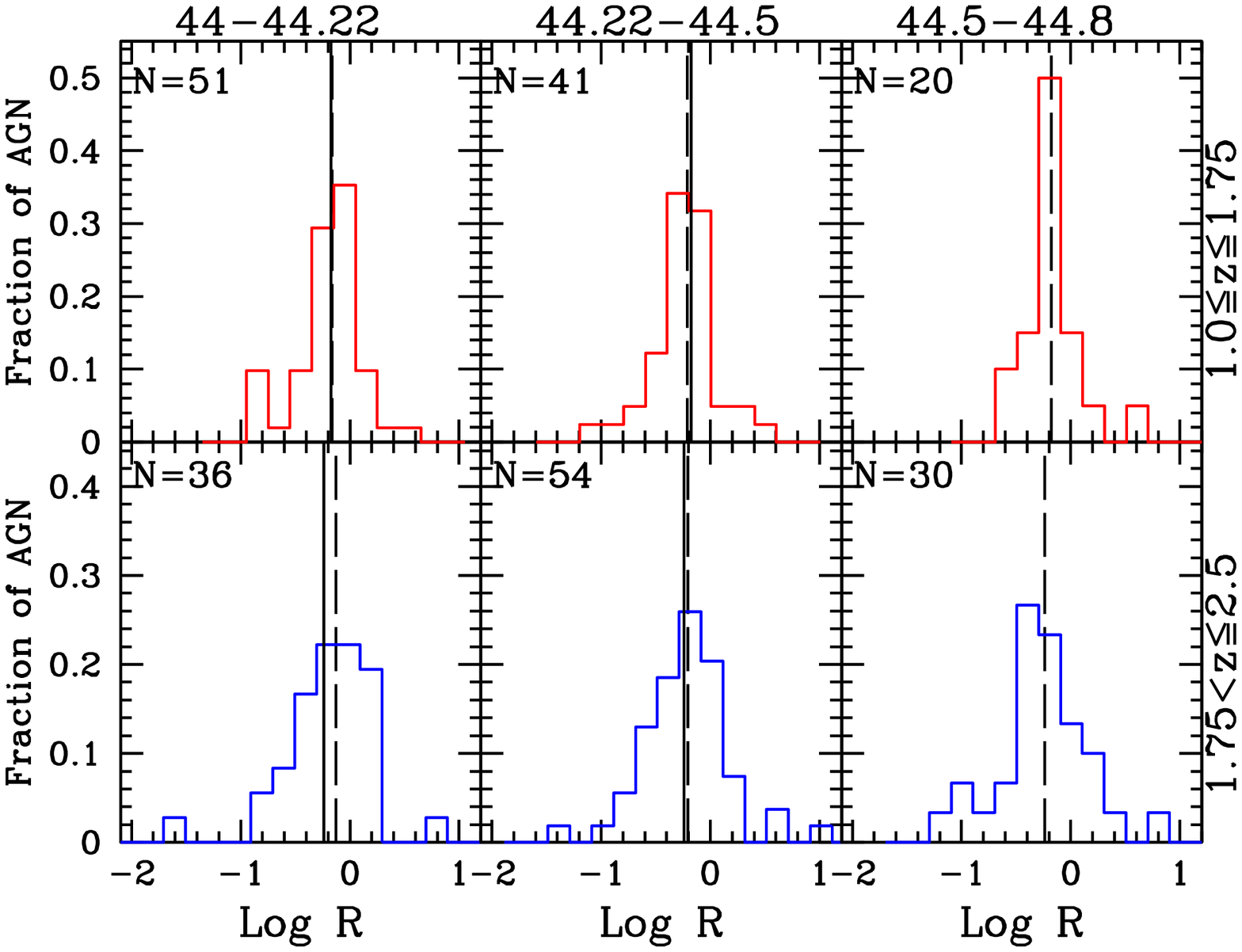}}
  {\includegraphics[width=0.4\textwidth]{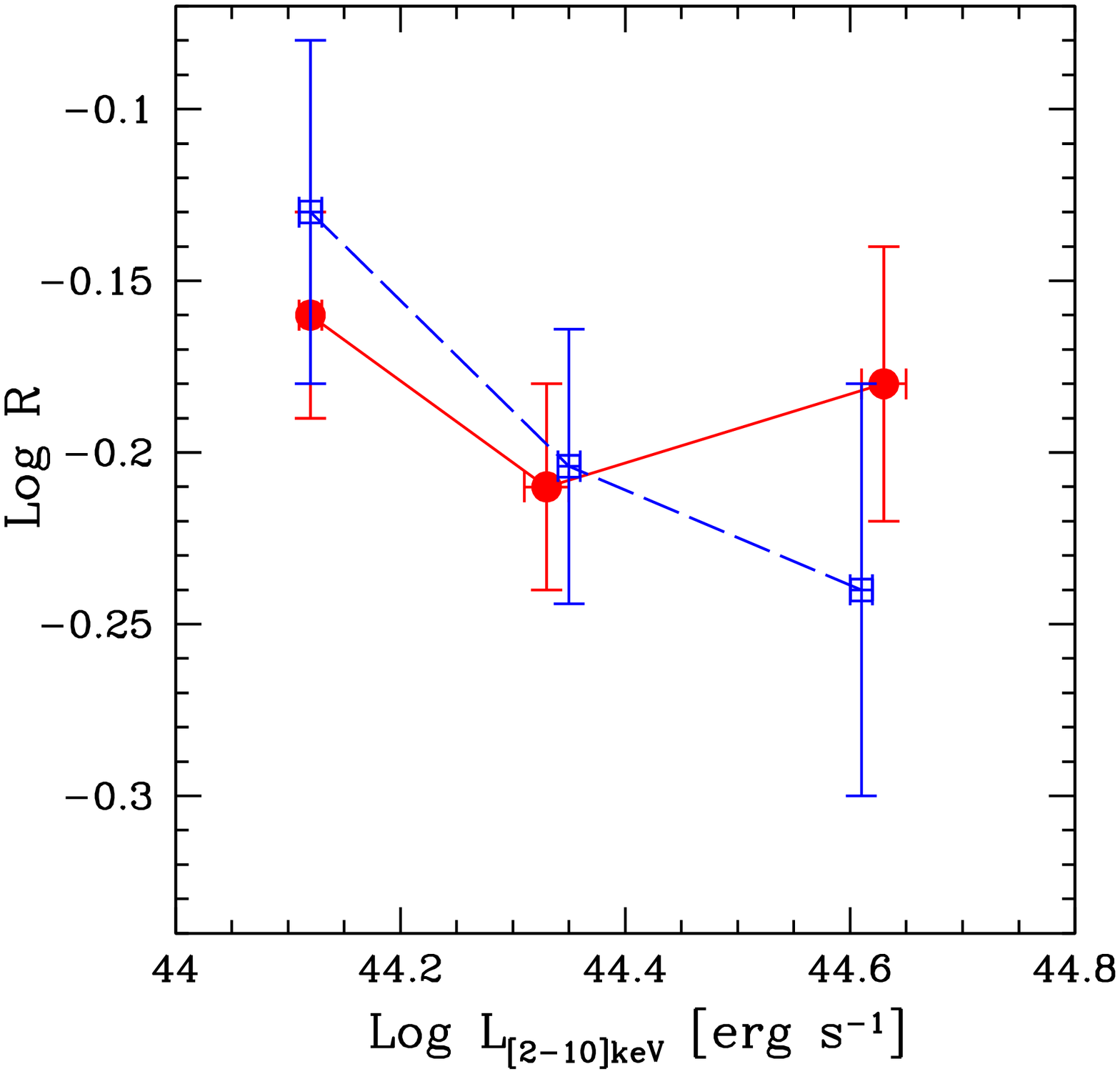}}                        
  \caption{\textit{Upper panel}: Distributions of $\Log R$ in bins of $\Lhard$ and redshift. The panels are divided between $1.0\leq z \leq 1.75$ (top three panels) and $1.75< z \leq 2.5$ (lower three panels), and bolometric luminosities increases from left to right ($\Log\Lbol$ [erg s$^{-1}$] intervals and the number of objects are reported on top of each panel). Histograms show the observed $\Log R$ distributions. The dashed lines are the median values, while the solid black lines in the first and in the second bins are plotted at the $\Log R$ value corresponding to the median in the highest luminosity bin. \textit{Lower panel}: Median $\Log R$ as a function of the median $\Lbol$. Filled circles and open squares represent the median $\Log R$ for $1.0\leq z \leq 1.75$ and $1.75< z \leq 2.5$, respectively. Solid lines connect low redshift bins, while dashed ones connect high redshift bins. Error bars on the median are estimated considering the \rev{MAD} divided by the square root of the number of observed AGN.}
  \label{cflxzevol}
\end{figure*}

The same analysis of the previous Section has been applied at the $R-\Lhard$ relationship by binning in $z$ and $\Lhard$. The redshift bins are $1.0\leq z \leq 1.75$ and $1.75< z \leq 2.5$ in the $\Lhard$ range of $10^{44-44.8}$ erg s$^{-1}$, while the luminosity cuts in each redshift bin are $\Log\Lbol$\revs{[erg s$^{-1}$]}$=44-44.22$, $44.22-44.5$, and $44.5-44.8$.  
In Fig.~\ref{lum} (right panel) the $\Lhard$ distribution as a function of redshift is presented. Red and blue open circles mark the two complete subsamples selected with the criteria above. There are 113 AGN in the low redshift bin with a median $\Lbol=10^{44.25}$ erg s$^{-1}$, while there are 120 objects in the high-redshift bin with a median $\Lbol=10^{44.34}$ erg s$^{-1}$. There is a factor of $\sim1.2$ difference and a (two-sided) Kolmogorov-Smirnov test gives a $D$ value of 0.19 and probability of 0.014. 
The histograms in Fig.~\ref{cflxzevol} (upper panel) show the $\Log R$ distributions, while the dashed lines represent the median $\Log R$ in each bin. 
In Fig.~\ref{cflxzevol} (lower panel) the median $\Log R$ is plotted against the median $\Lhard$. 
The two redshift bins do not show significantly different trends, and the median is consistent within the errors.
The narrow range of $\Lhard$ does not allow us any claim whether a significant evolution of $R$ as a function of $\Lhard$ is present.

\subsection{Torus models}
\label{models}

The presented analysis of the obscured AGN fraction over about four decades of $\Lbol$ is consistent with what has been found by demographics (i.e., S05), yet considering a completely different and independent method.
This result favors the "receding torus" scenario where re-emission occurs in the optically thin torus regime. 
Interestingly, \citet{2011A&A...534A.121H}, studying the infrared emission of Type-1 AGN through a set of simulations of clumpy torus models, have found that, although the clouds are optically thick, the visible AGN SED is dominated by optically thin dust.
Increasing AGN radiation pressure may cause large thick dust clouds to be driven out of the torus at higher luminosities, and this effect can be also explained by optically thin layers of dust illuminated by the nuclear source \citep{2010A&A...523A..27H}.
\par
\begin{figure*}
\epsscale{1.15}
\plottwo{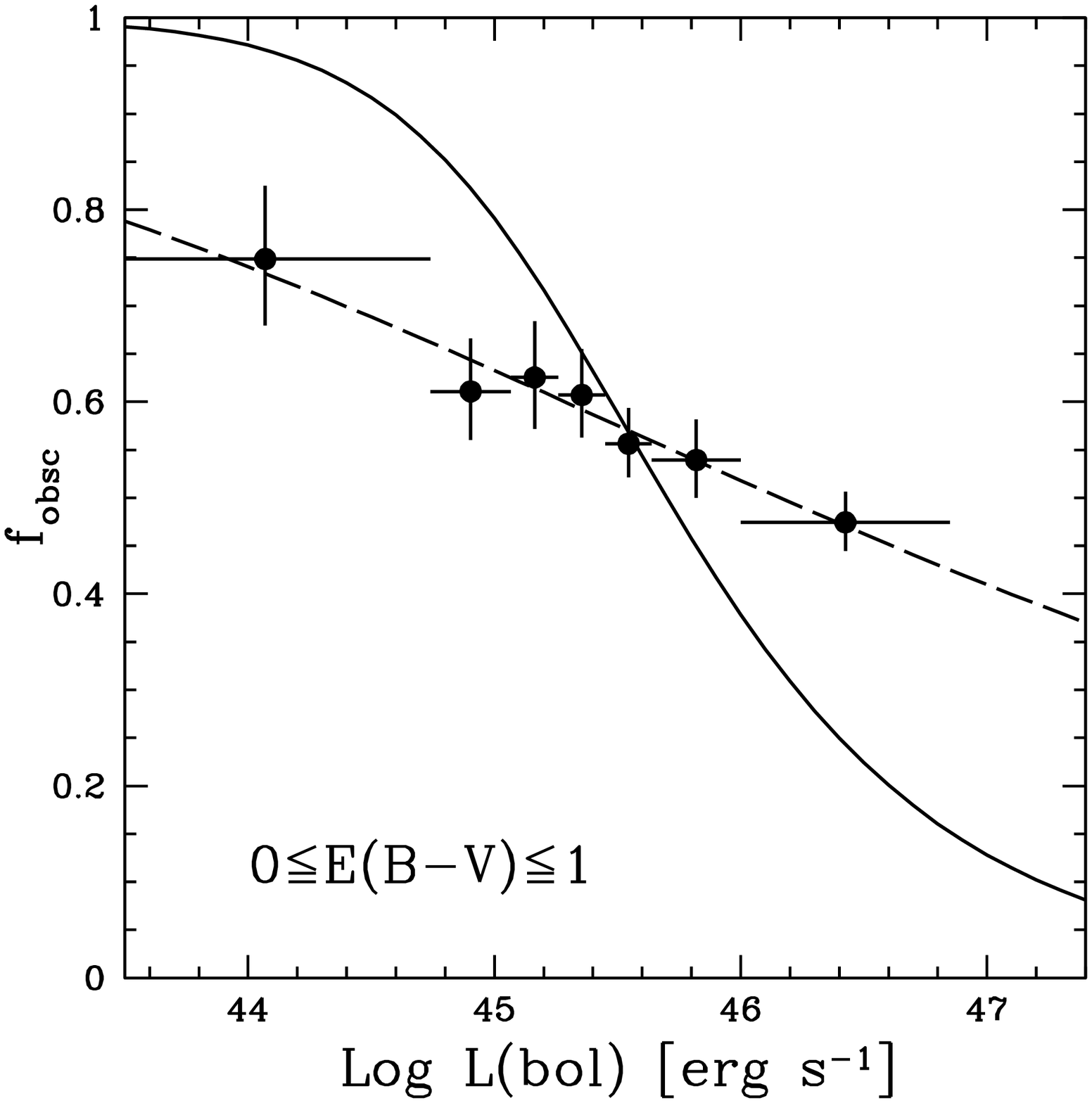}{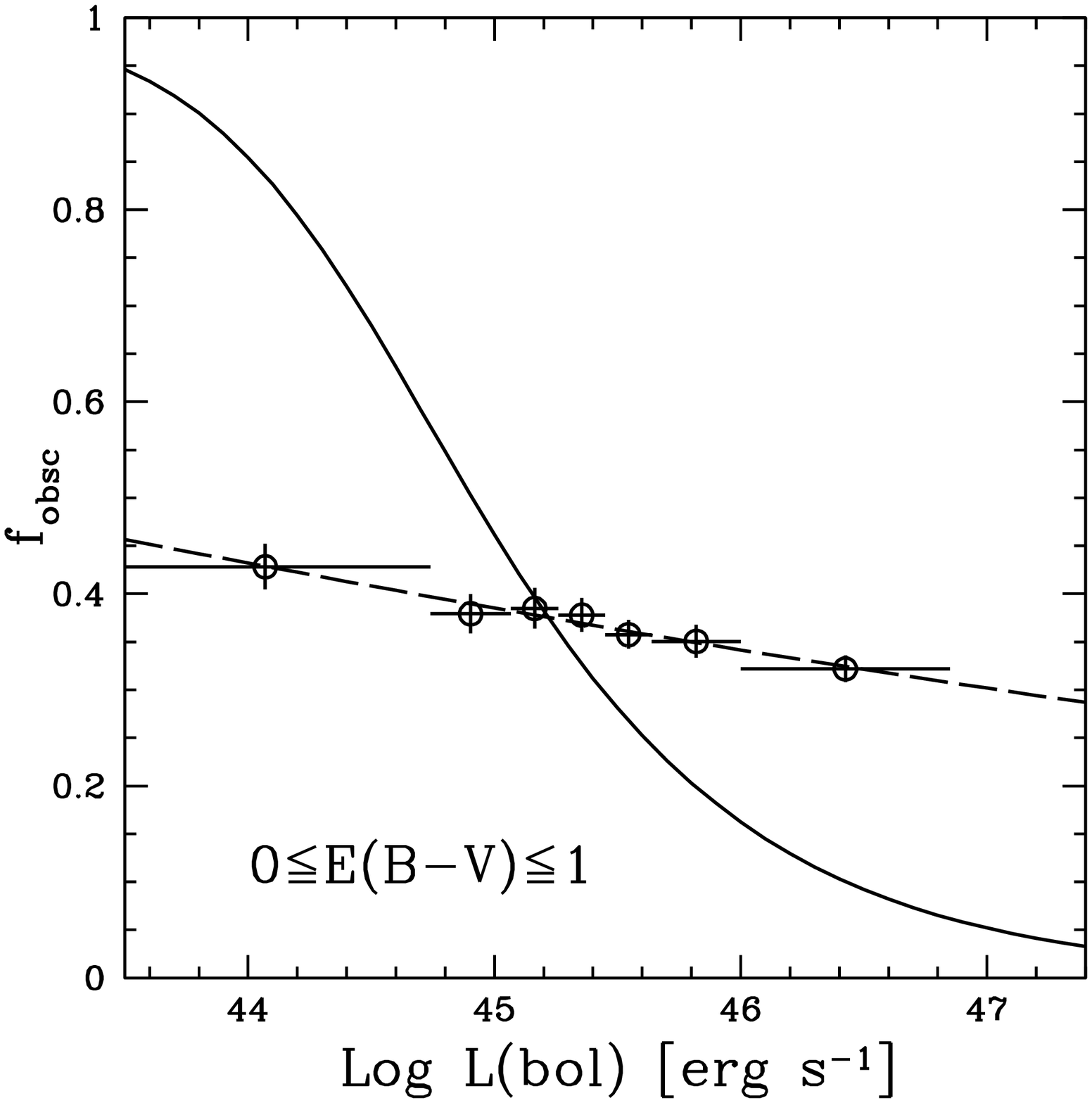}
\caption{Obscured AGN fraction as a function of $\Lbol$ for the main AGN sample. \textit{Left panel}: Optically thin torus models. The solid line shows the best-fit ``standard" receding torus model (Eq.~[\ref{fstdsimpson98}]), while the long dashed line represents the best-fit model where the torus height varies (Eq.~[\ref{fmodsimpson98}]). \textit{Right panel}: Optically thick torus models. Keys as the in left panel.}
\label{figmodelsall}
\end{figure*}
\begin{figure*}
\epsscale{1.15}
\plottwo{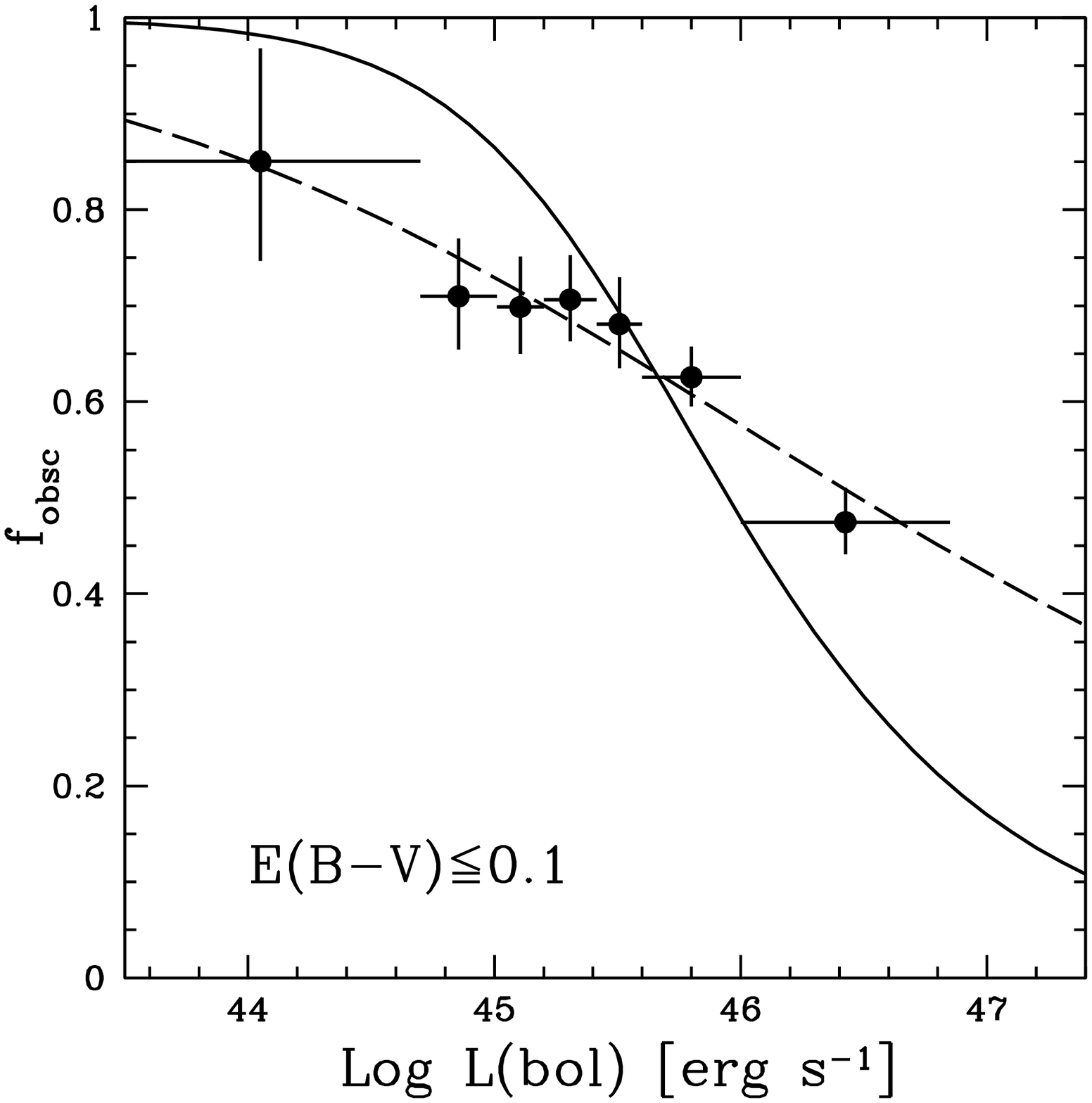}{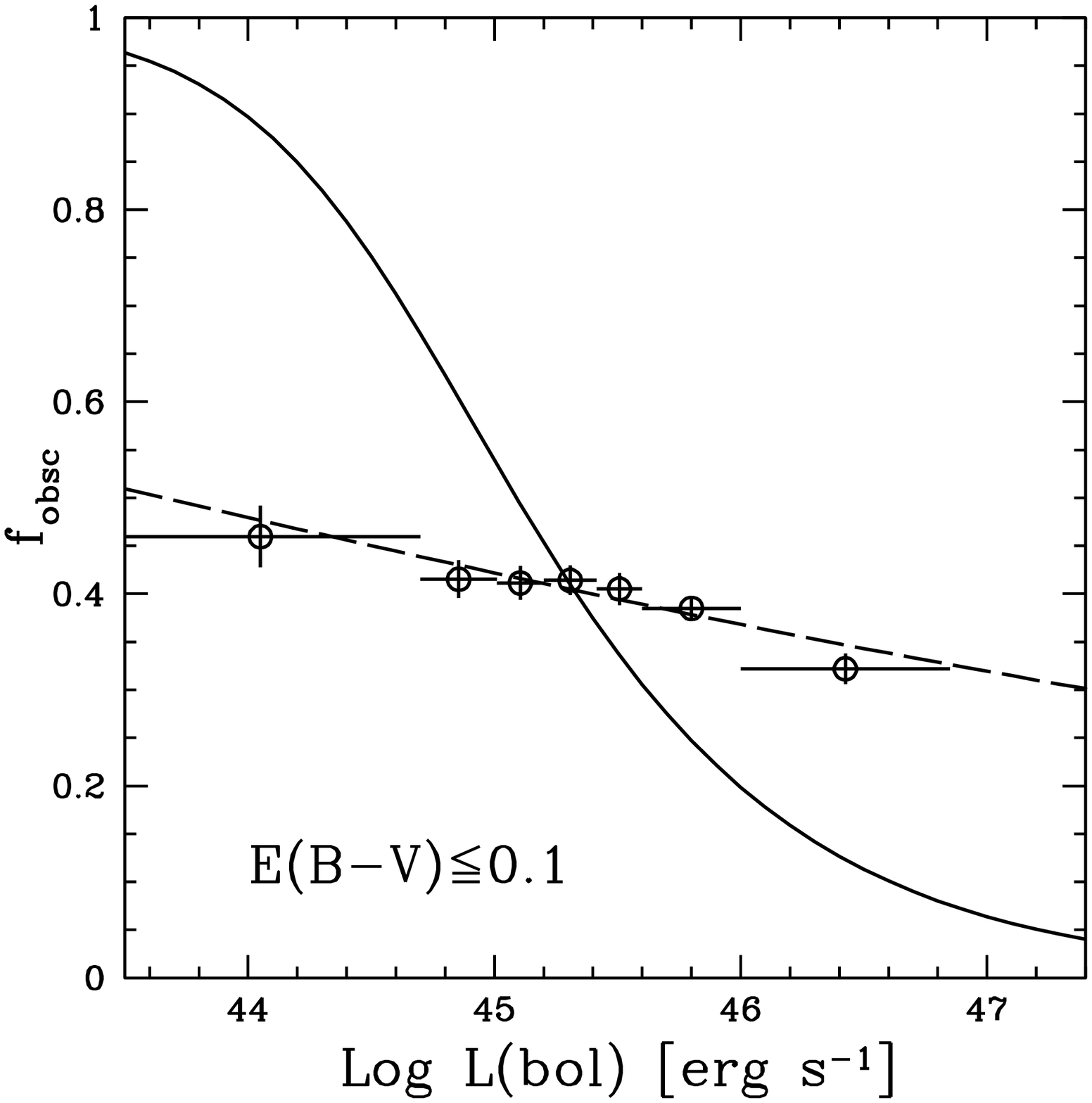}
\caption{Obscured AGN fraction as a function of $\Lbol$ for the low-reddening AGN sample. \textit{Left panel}: Optically thin torus models. The solid line shows the best-fit ``standard" receding torus model (Eq.~[\ref{fstdsimpson98}]), while the long dashed line represents the best-fit model where the torus height varies (Eq.~[\ref{fmodsimpson98}]). \textit{Right panel}: Optically thick torus models. Keys as the in left panel.}
\label{figmodelslowred}
\end{figure*}

In the context of the receding torus, the obscured fraction can be defined as (see \citealt{1998MNRAS.297L..39S,2005MNRAS.360..565S})
\begin{equation}
\label{fstdsimpson98}
\f = (1+3 \Lbol/\mathcal{L}_0)^{-0.5},
\end{equation}
where $\mathcal{L}_0$ is the luminosity for an opening angle of
$60^\circ$ (i.e. equal number of Type-1 and 2 AGN), under the
assumption that the height of the torus is constant with luminosity.
This model for the main and low-reddening AGN sample are shown as solid lines in Figs.~\ref{figmodelsall} and
\ref{figmodelslowred} for the optically thin (left panel) and thick
(right panel) torus regime.  For both cases our data are not well fit
by this model. Considering the main sample of 513 Type-1 AGN (i.e., $0\leq\ebvq\leq1$), the
best-fit for the optically thin case has a reduced (6 degrees of
freedom) $\chi^2$ of 142 for $\mathcal{L}_0 = 10^{45.71\pm0.04}$ erg
s$^{-1}$, while the optically thick case has a reduced $\chi^2$ of 200
for $\mathcal{L}_0 = 10^{44.92\pm0.04}$ erg s$^{-1}$ (see
Fig.~\ref{figmodelsall}).  If we instead consider the sub-sample of
391 Type-1 AGN (i.e., $\ebvq\leq0.1$), the best-fit for the optically thin case has a reduced
(6 degrees of freedom) $\chi^2$ of 90 for $\mathcal{L}_0 = 10^{45.95\pm0.03}$
erg s$^{-1}$, while the optically thick case has a reduced $\chi^2$ of
177 for $\mathcal{L}_0 = 10^{45.10\pm0.03}$ erg s$^{-1}$ (see
Fig.~\ref{figmodelslowred}).
\par
We instead adopt the modified receding torus model presented in
\citet{2005MNRAS.360..565S} where the height of the torus is allowed
to vary ($h\propto L^\xi$), and hence Eq.~(\ref{fstdsimpson98})
becomes
\begin{equation}
\label{fmodsimpson98}
\f = (1+3 \Lbol/\mathcal{L}_0)^{1-2 \xi}.
\end{equation}
This model produces a very good fit of our data in both torus regimes. 
Considering the main AGN sample, the best-fit for the optically thin case has a reduced (5 degrees of freedom) $\chi^2$ of 2 for $\mathcal{L}_0 = 10^{46.16\pm0.16}$ erg s$^{-1}$ and $\xi=0.37\pm0.02$, while the optically thick case has a similar $\chi^2$ for $\mathcal{L}_0 = 10^{42.65\pm1.24}$ erg s$^{-1}$ and $\xi=0.44\pm0.03$ (see Fig.~\ref{figmodelsall}). For the low reddening AGN sample we found that the best-fit for the optically thin case has a reduced (5 degrees of freedom) $\chi^2$ of 5 for $\mathcal{L}_0 = 10^{46.48\pm0.16}$ erg s$^{-1}$ and $\xi=0.32\pm0.03$, while the optically thick case has a reduced $\chi^2$ of 1.2 for $\mathcal{L}_0 = 10^{43.66\pm0.72}$ erg s$^{-1}$ and $\xi=0.43\pm0.03$ (see Fig.~\ref{figmodelsall}).
In the light of this, the model employing a luminosity-dependent torus height is preferred.
\par
As a further support to our analysis, we note that \citet[C05 hereafter]{2005ApJ...619...86C}, considering a sample of 64 Palomar-Green (PG) QSOs with infrared SEDs ($3-150~\mu$m) observed by the {\it Infrared Space Observatory}, have estimated 
how the torus height varies with luminosities considering an SED-based approach\footnote{i.e., through the $\f-\Lbol$ relation with $\f=\Lir/\Lbol$. The host-galaxy/reddening contribution is neglected given that these objects are bright PG QSOs with $\Lbol\sim10^{45-47}$ erg s$^{-1}$. C05 sample contains also two QSOs with $\Lbol\sim10^{48}$ erg s$^{-1}$.}.
C05 have found that the height of the torus scales with $\Lbol$ (estimated from the optical continuum luminosity at 5100\AA{}  and a $\kbol=9$) with a slope $\xi$ of $0.37\pm0.05$. 
Fifty-four PG QSOs in C05 sample have FWHM of the \ion{H}{$\beta$} line larger than 2000 km s$^{-1}$ (broad-line QSO sample). For this sub-sample, C05 has found a slope $\xi=0.34\pm0.04$.
The fact that our best-fit $\xi$ value agrees with the the C05 provides a further (independent) evidence that a torus optically thin to its own radiation is the preferred solution.
\par
Summarizing, our data favor the optically thin solution with the height of torus varying with the bolometric luminosity with a slope $\xi=0.32-0.37$ considering the low-reddening and the main AGN sample, respectively.



\subsection{Possible biases}
\citet{2008MNRAS.386.1252H} pointed out that SED-fitting techniques are both powerful and limited. Powerful not only because samples with photometry are always orders of magnitudes larger than the spectroscopic ones, but also because photometry allows us to explore a wider range of wavelength than a single spectrum. 
The limitations are, for examples, in the quality of the photometric data and in the templates employed to explore the parameter space (i.e., degeneracies in the model parameters). 
\par
In our analysis we are employing only one BBB template which does not vary on the x-axes (i.e., we have assumed a fixed disk temperature). This might
bias our optical-UV luminosity estimates if the shape of the BBB
changed with luminosity and/or redshift, and if there is a range of disk temperatures.  However, all work in the literature on AGN SED have found that the shape of the average AGN SED does not change with redshift and luminosity \citep{1989ApJ...347...29S,1994ApJS...95....1E,2006ApJS..166..470R,2011ApJS..196....2S,2012A&A...539A..48M,2012ApJ...759....6E}. 
\par 
For what concerns the disk temperature, our estimate of disk
luminosities might be underestimated for those sources where we are
not sampling the peak of the BBB, and therefore luminosity ratios might
be overestimated.  In order to evaluate the wavelength to which BBB
peaks in the optical-UV, we have considered the relation between disk
temperature, $\Mbh$ and accretion rate given by
\citet{1997iagn.book.....P} (see Eq.~[3.20]). Assuming an average
$\Mbh$ of $3\times10^{8}~M_\odot$, an accretion rate onto the SMBH
(normalized to the Eddington one) of 0.1 \citep{2012MNRAS.425..623L},
and a scale radius of 3 Schwarzschild radius (i.e. the innermost
stable orbit), the emission of the inner part of the disk is maximum
at the frequency of $\sim7\times10^{15}$ Hz, which corresponds to a
wavelength of $\sim$430\AA{}. The energy peak of the BBB lies in the
unobserved extreme ultraviolet (e.g., 100--912\AA{}) region (e.g., see
\citealt{1987ApJ...323..456M}), therefore to determine the shape and
the peak of the BBB is of particular importance.

\section{Summary and Conclusions}
\label{Summary and Conclusions}

We have presented a homogeneous and comprehensive analysis of the
obscured fraction for a sample of 513 Type-1 AGN from the XMM--COSMOS
survey, which has the best available multi wavelength coverage
provided in the literature, over a wide range of redshifts ($0.10 \leq
z \leq 3.75$) and bolometric luminosities ($43.47\leq \Log\Lbol
\,[\rm erg \,s^{-1}]\leq 46.84$). 
The main goals of the present study are \textit{(i)} to measure the
AGN infrared luminosity from the torus and the optical-UV luminosity
from the accretion disc, and \textit{(ii)} to use these measurements
to determine the fraction of obscured AGN, and its dependence on luminosity and redshift. To achieve these goals, we have employed an
upgraded version of the SED-fitting code already presented in
\citet{2011A&A...534A.110L} and \citet{2012MNRAS.425..623L}, which
models simultaneously four components of the AGN SED; i.e., cold-dust
from star-forming region, hot-dust from the torus, optical-UV emission
from the evolving stellar population, and optical-UV from the
accretion disk.

The AGN obscured fraction has been obtained without assuming any bolometric
correction, and we have determined its dependence on 
$\Lbol$ and $\Lhard$ independently. Our SED-fitting approach allows
us to correct for both the
effect of intrinsic AGN reddening and subtract off the contamination of the
host-galaxy to isolate the AGN emission.  Moreover, we have explored two
distinct regimes bracketing the range of physical properties of the torus, 
one where the dust is optically thin ($\f=R$) and the other
where it is optically thick ($\f=R/(1+R)$) to its own infrared radiation.  
The {\it true} AGN obscured fraction lies between these two cases \rev{(under the assumption 
of a toroidal obscuring region)}.

Given that highly extincted Type-1 AGN ($\ebvq>0.1$) will have smaller $L_{\rm torus}$
values (resulting in lower values for $R$), the
obscured fraction that we obtain for such objects would be
systematically smaller than it is in reality. 
We have therefore analyzed the obscured fraction for the main sample of 513 Type-1 AGN and for the
sub-sample of 391 objects with $\ebvq\leq0.1$.

The most important results obtained in the present study can be summarized as follows:
\begin{enumerate}
 \item We confirm previous studies that found a decrease of the obscured fraction of Type-1 AGN with increasing bolometric luminosity. In particular, for the main AGN sample, $\f$ ranges in the optically thin case from about 0.45 to 0.75, while in the optically thick case the trend between $\f$ and $\Lbol$ is much flatter ranging from 0.30 to 0.45.  For the low reddening sample, $\f$ ranges in the optically thin case from about 0.45 to 0.85, while in the optically thick case the trend between $\f$ and $\Lbol$ is much flatter ranging from 0.35 to 0.45. This decrease with $\Lbol$
can be interpreted in the context of the receding torus model where the covering factor of the dust is reduced at high $\Lbol$.

 \item We favor a scenario where the torus is optically thin, with the torus height varying with bolometric luminosity ($h\propto \Lbol^{\quad0.32-0.37}$). This result is supported by the agreement between the $\f-\Lbol$ relation estimated with our SED-based approach and S05 which has used a different sample (i.e., optically selected AGN) and a completely independent method (i.e., demographics).

\item The obscured fraction does not vary monotonically with X--ray
  luminosity. The $\f-\Lhard$ relationship in the optically thick case
  is almost flat. X--ray demography-based studies found an obscured
  fraction lower by a factor $\sim2$ in the optically thin regime, and
  by a factor of $\sim1.3$ in the thick one. We argue that X--ray
  studies miss a large fraction of the highly obscured-Compton-thick AGN at $\Lhard>10^{44}$ erg
  s$^{-1}$.
  
 \item We do not find any clear evidence of evolution with redshift of the mid-infrared to bolometric luminosity ratio, and hence of the obscured fraction, as a function of both $\Lbol$ and $\Lhard$.
\end{enumerate}

We conclude that the major driver of the $\f-$luminosity relationship is the bolometric luminosity, rather than X--ray luminosity. This is expected in the receding torus scenario, i.e. the X-ray emission is not providing most of the heat that sublimates dust and regulates the torus distance.

Our $\f$ measurements could be used in the context of future demographics analyses in order to check whether AGN surveys are missing highly obscured-Compton thick AGN.
Moreover, by comparing any unbiased demographic sample to our results one could obtain deeper insights into the structure of the torus. Indeed, our comparison with S05 suggests that the torus is optically thin to its own radiation.

This analysis also provide a fitting formula (Eq.~[\ref{fmodsimpson98}], see \S\ref{models}) which can be used in all future
bolometric luminosity function papers (e.g., \citealt{hopkins07}), to determine the contribution of the
obscured accretion as a function of bolometric luminosity and to study the growth history 
of SMBHs.


\section*{Acknowledgements}
\label{sec_acknow}
We thank the anonymous reviewer for thoroughly reading the paper and providing valuable comments.
We acknowledge financial contribution from the agreement ASI-INAF
I/009/10/0 and from the INAF-PRIN-2011.  EL is indebted to Micol
Bolzonella for providing very helpful and constructive comments on the
code presented here.  EL gratefully thanks Andrea Merloni, J$\rm\ddot{o}$rg-Uwe Pott, Elisabete de Cuhna, Brent Groves, and Angela Bongiorno for useful discussions. We also thank the members of the ENIGMA group\footnote{http://www.mpia-hd.mpg.de/ENIGMA/} at the
Max Planck Institute for Astronomy (MPIA) for helpful discussions.
JFH acknowledges generous support from the Alexander von Humboldt
foundation in the context of the Sofja Kovalevskaja Award. 
GTR acknowledges the generous support of a research fellowship from the Alexander von Humboldt Foundation at the Max-Planck-Institut f$\rm\ddot{u}$r Astronomie and is grateful for the hospitality of the Astronomisches Rechen-Institut.

\appendix
\section*{Appendix}
\label{appendix}
We summarize below the main properties of the SEDs of the outliers marked with orange open squares in Figs.~\ref{ldiskcomphostcorrected}, \ref{ldderedcomp}, and \ref{lircomp}.


\begin{figure}[ht!]
     \begin{center}
        \subfigure{\label{fig:third}
            \includegraphics[width=0.4\textwidth]{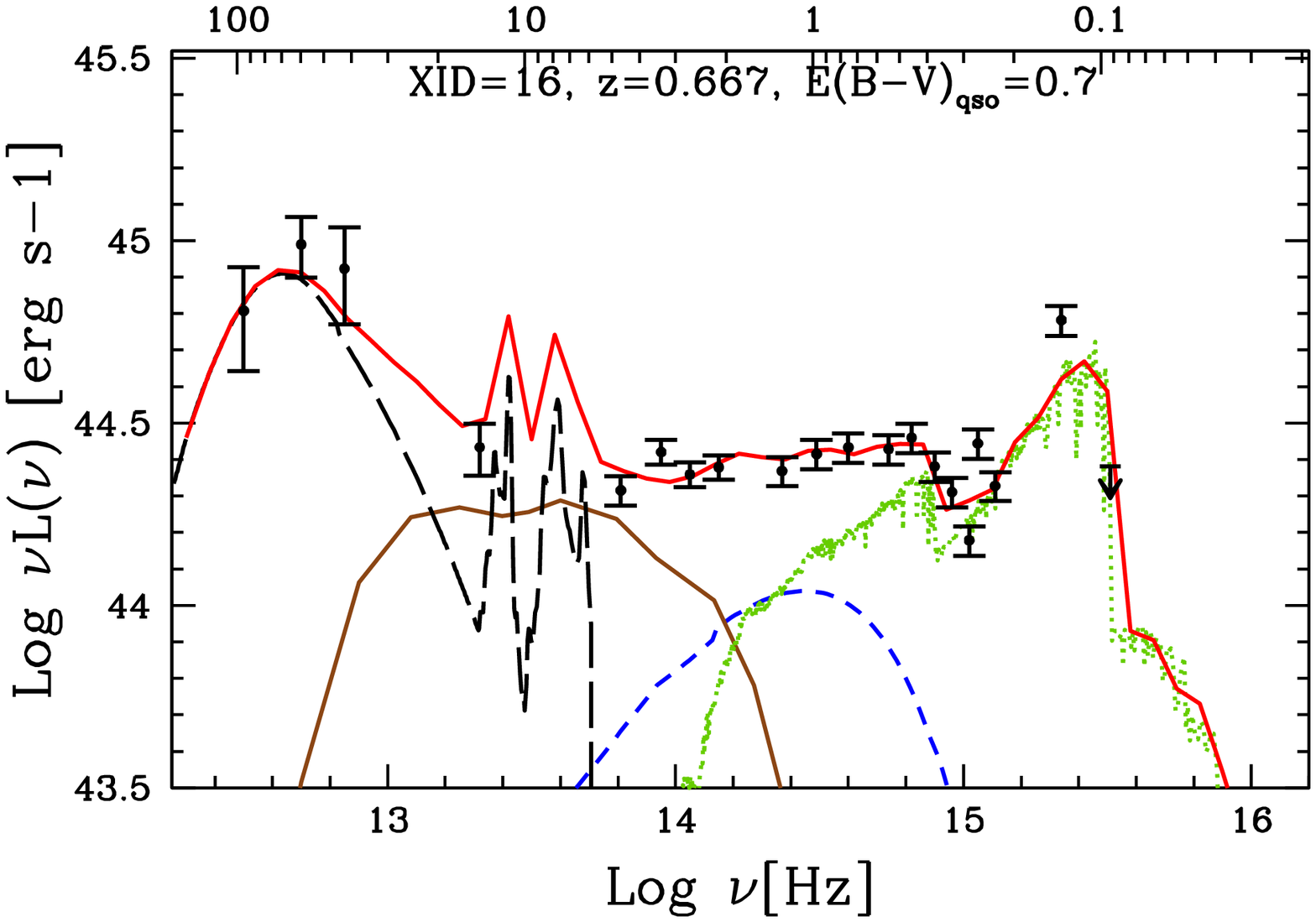}
        }
        \subfigure{\label{fig:second}
           \includegraphics[width=0.4\textwidth]{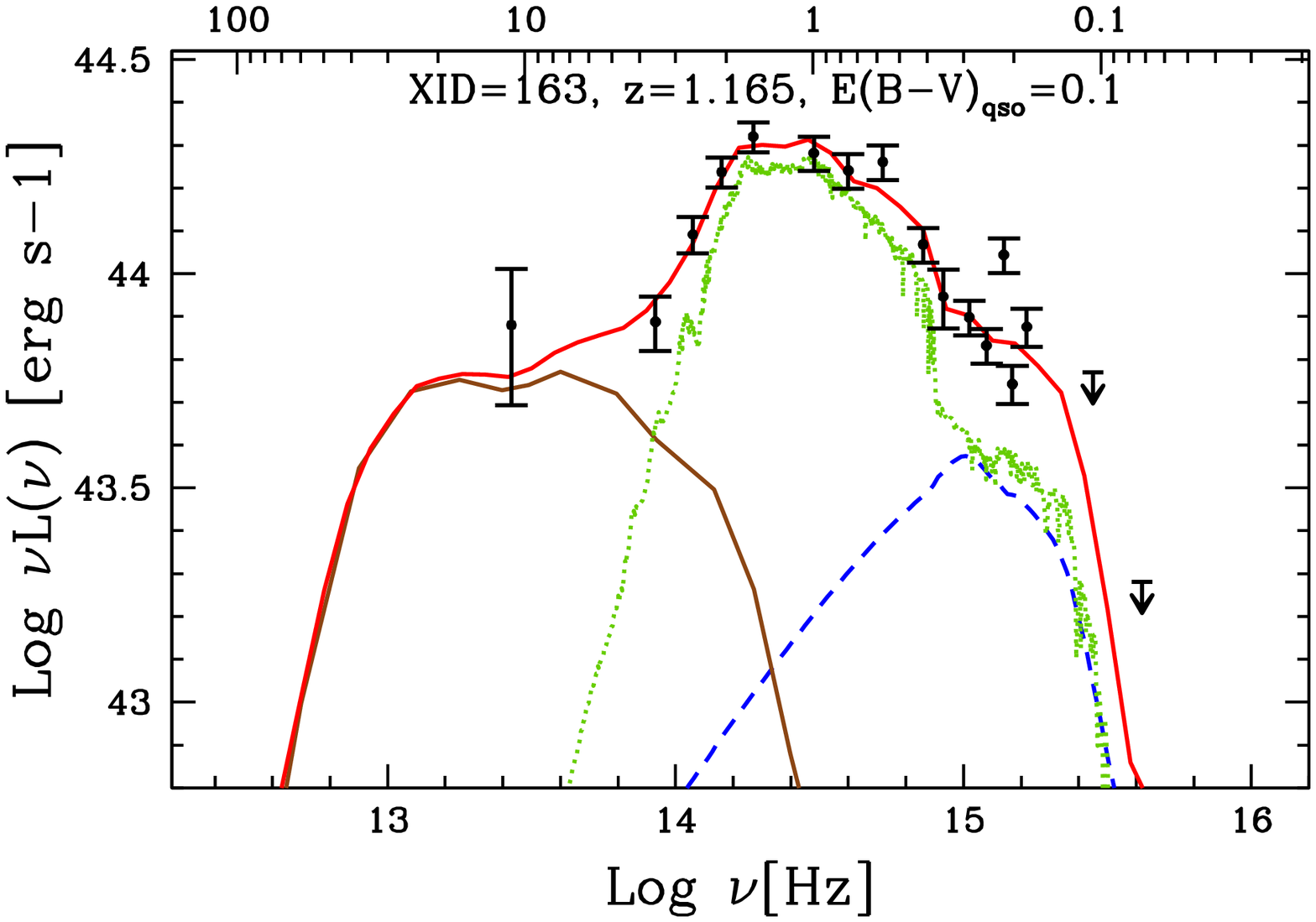}
        }\\ 
        \vspace{-2.2cm} 
        \subfigure{\label{fig:first}
            \includegraphics[width=0.4\textwidth]{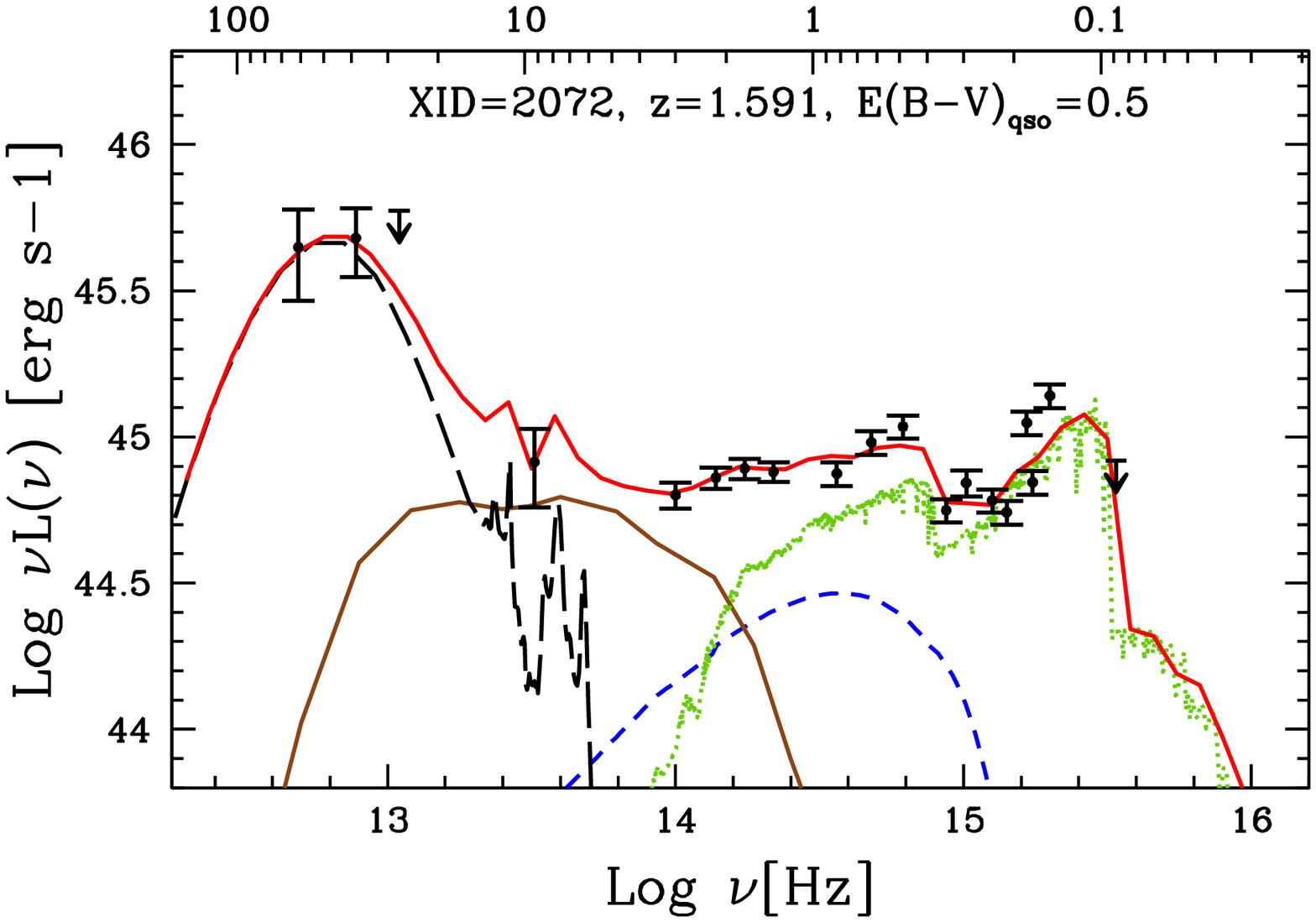}
        }
        \subfigure{\label{fig:fourth}
            \includegraphics[width=0.4\textwidth]{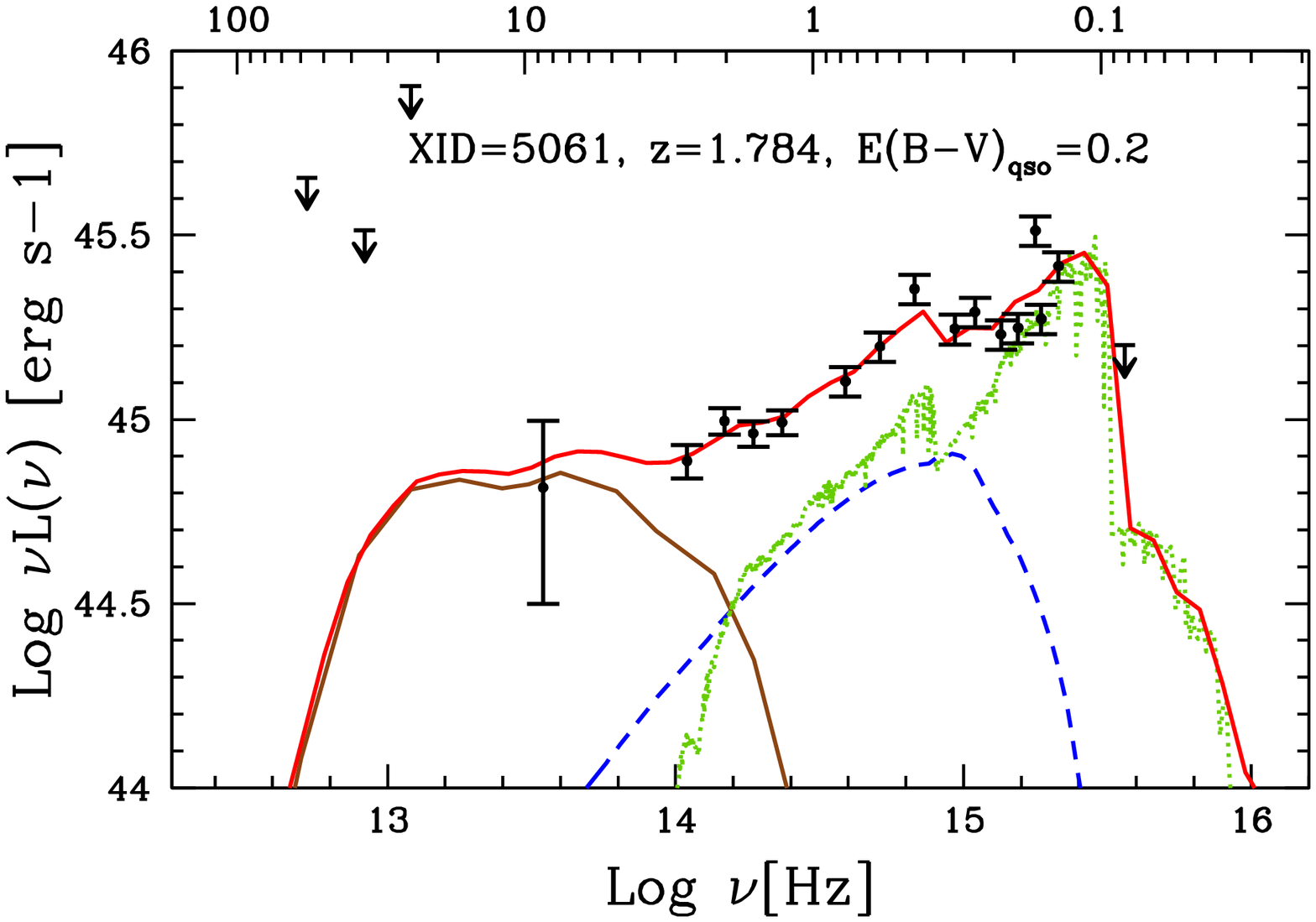}
        }\\ 
        \vspace{-2.2cm} 
        \subfigure{\label{fig:third}
            \includegraphics[width=0.4\textwidth]{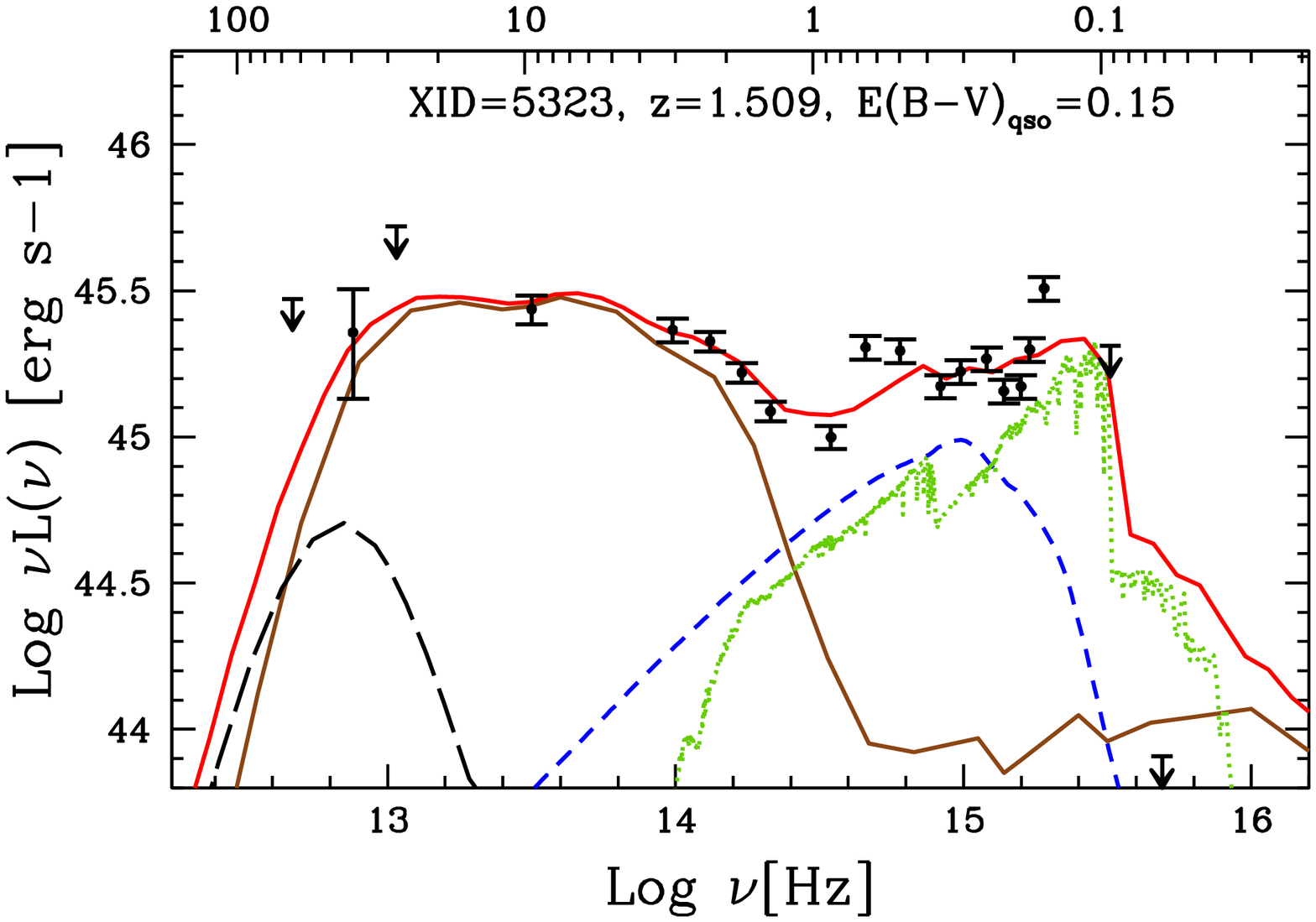}
        }
        \subfigure{\label{fig:fourth}
            \includegraphics[width=0.4\textwidth]{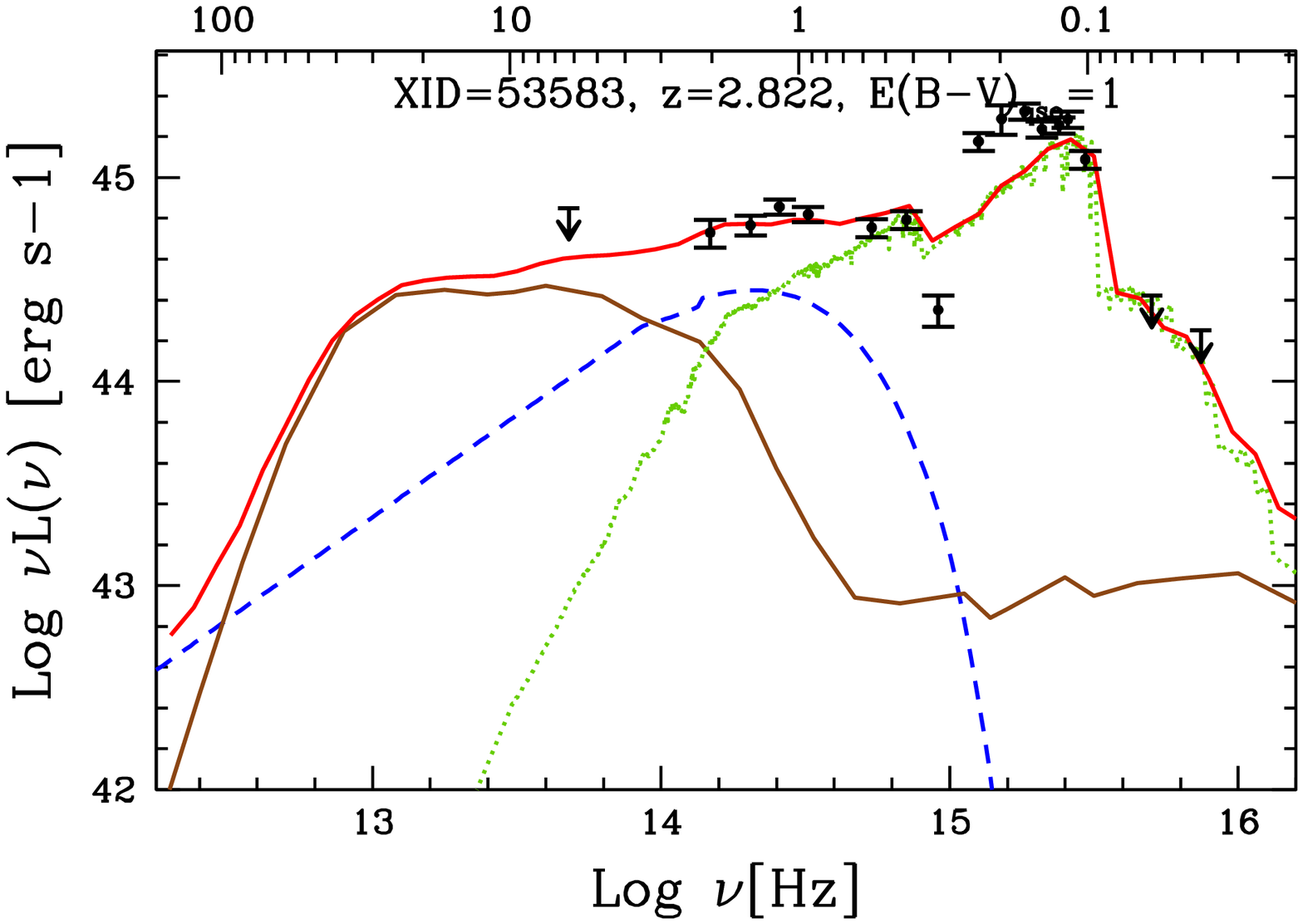}
        }\\ 
        \vspace{-2.2cm} 
        \subfigure{\label{fig:third}
            \includegraphics[width=0.4\textwidth]{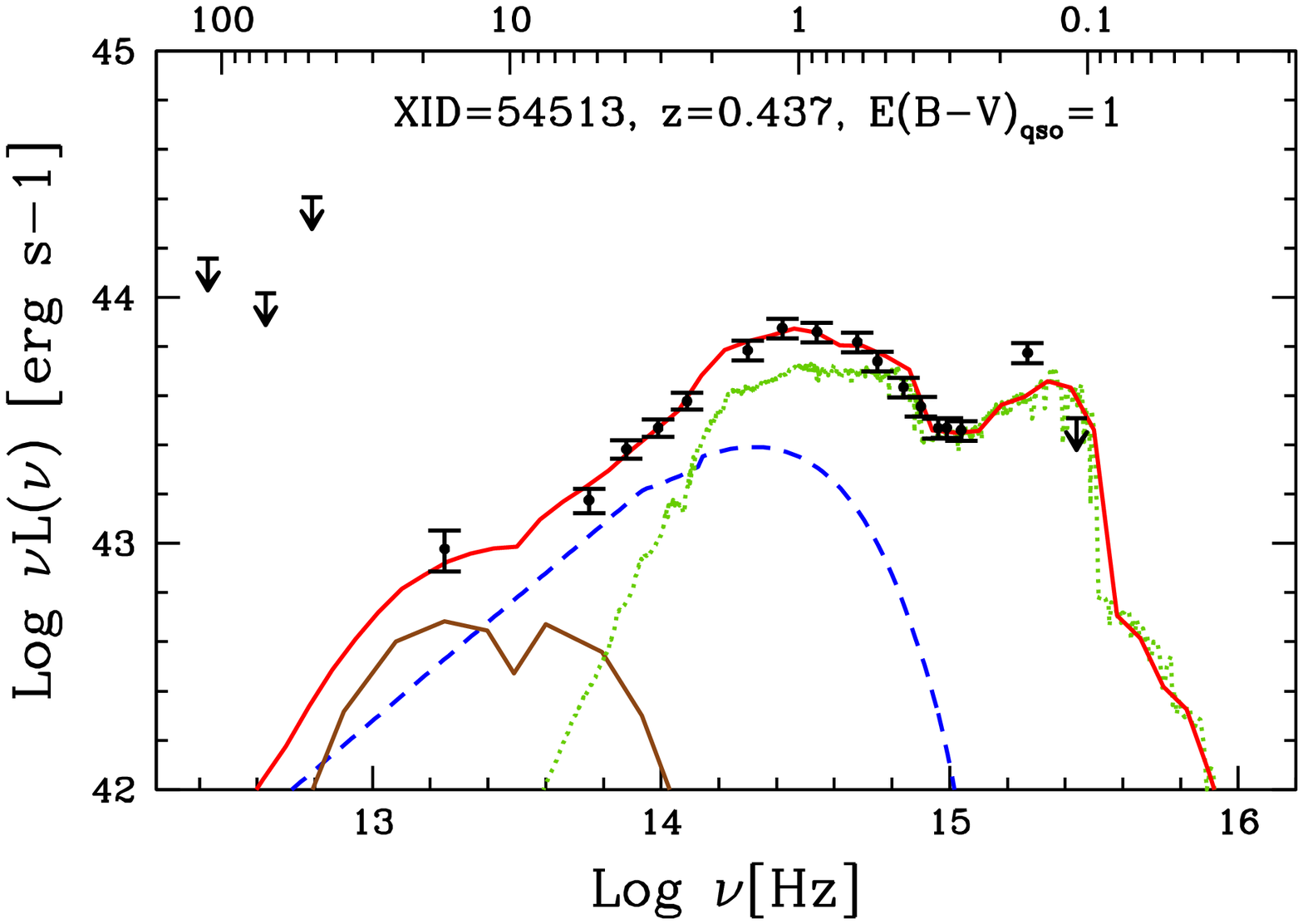}
        }
        \subfigure{\label{fig:fourth}
            \includegraphics[width=0.4\textwidth]{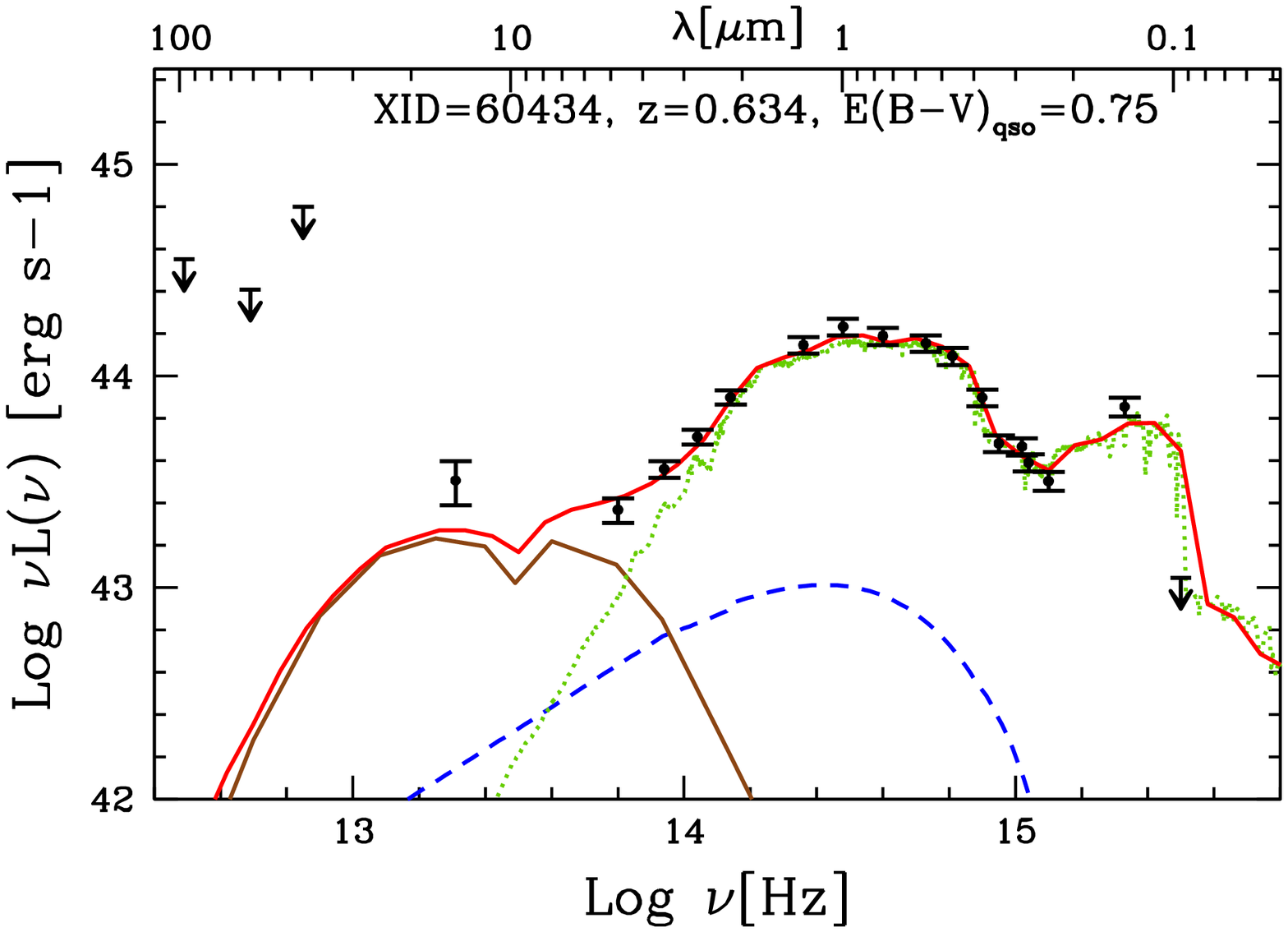}
        }
    \end{center}
    \caption{Examples of outliers in SED fitting shown in Fig.~\ref{ldiskcomphostcorrected}. Keys as in Fig.~\ref{panel}.}
   \label{sed_outliers_ldisk}
\end{figure}
\begin{figure}[ht!]
     \begin{center}
        \subfigure{\label{fig:first}
            \includegraphics[width=0.45\textwidth]{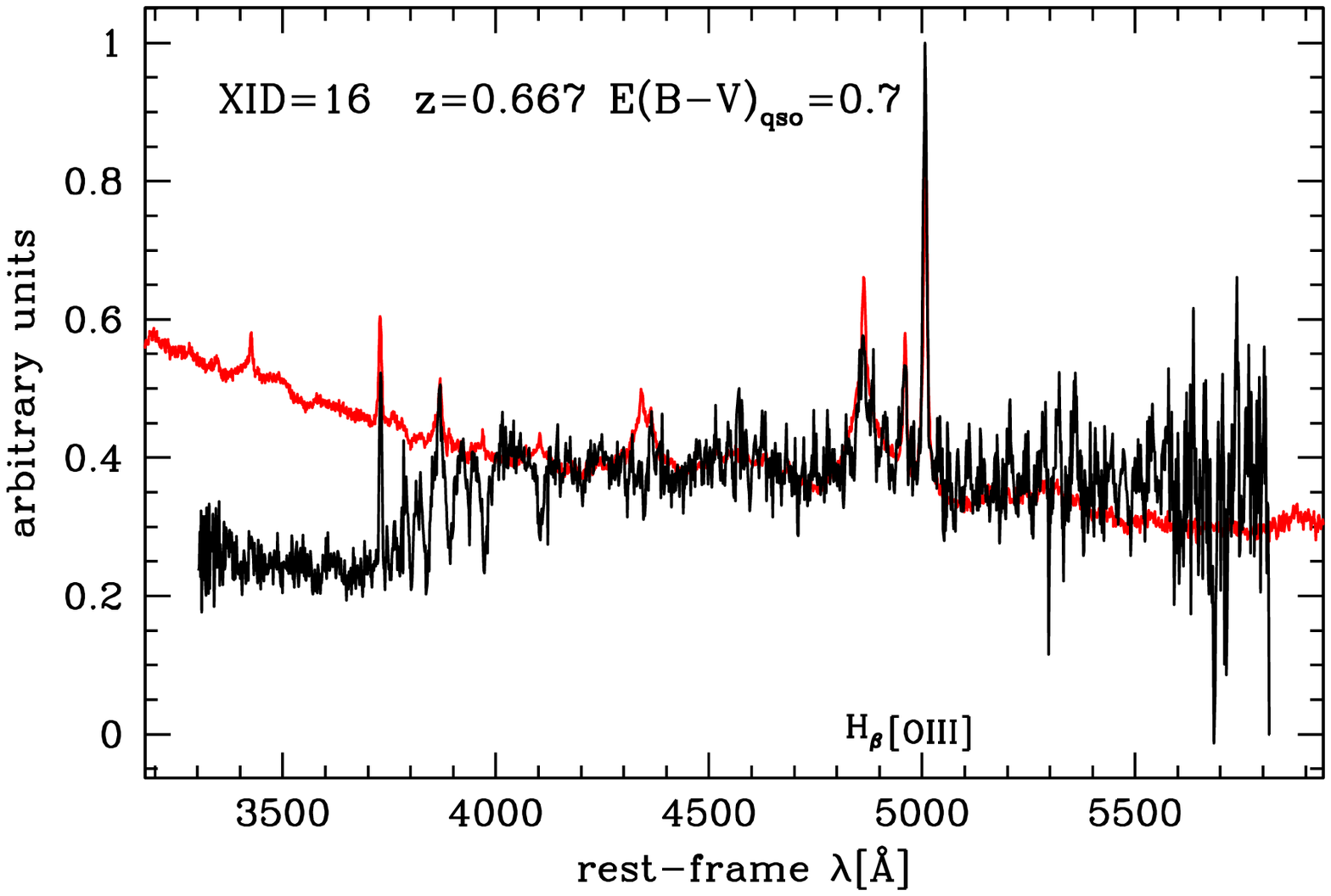}
        }
        \subfigure{\label{fig:second}
           \includegraphics[width=0.45\textwidth]{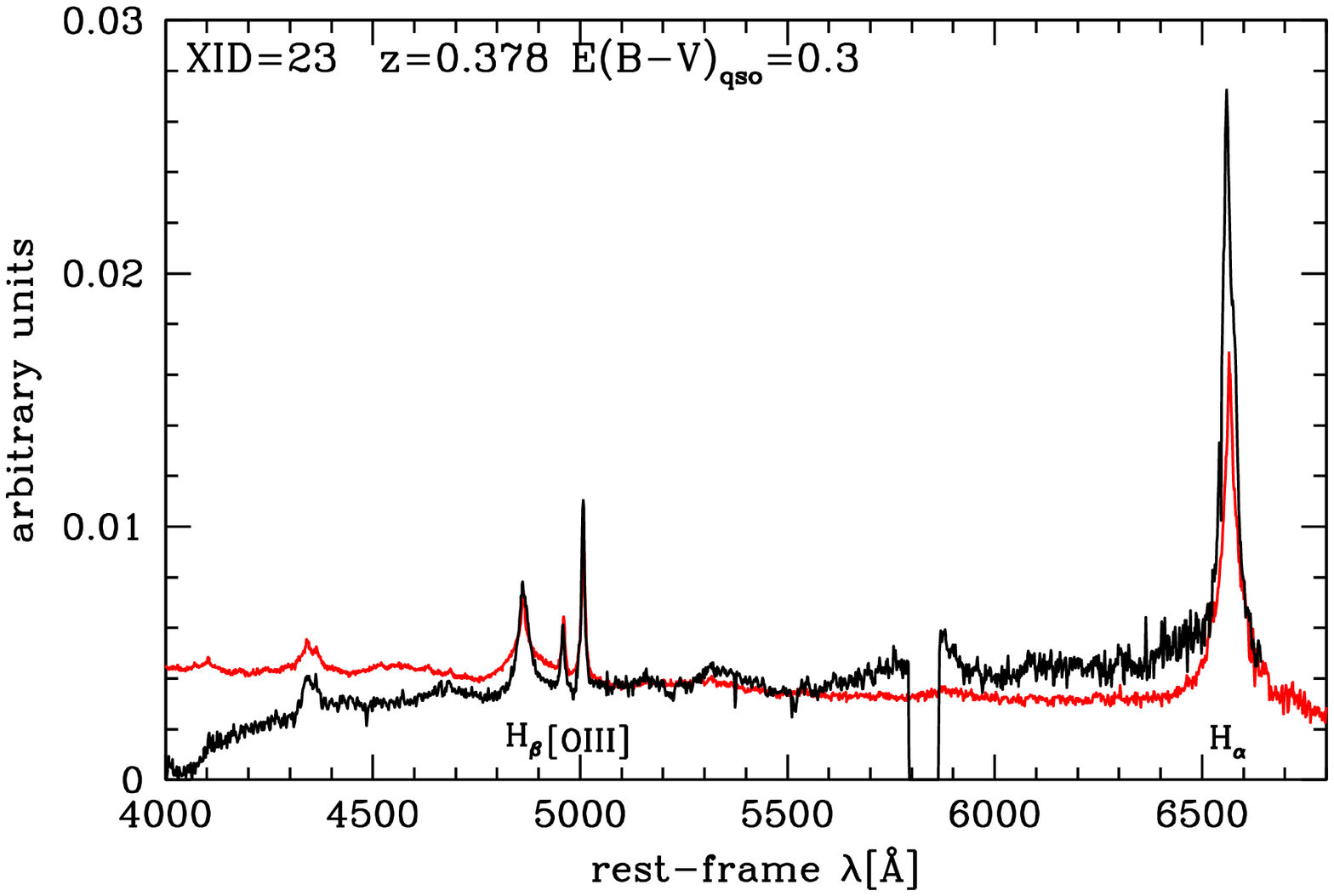}
        }\\ 
        \vspace{-2.3cm} 
        \subfigure{\label{fig:third}
            \includegraphics[width=0.45\textwidth]{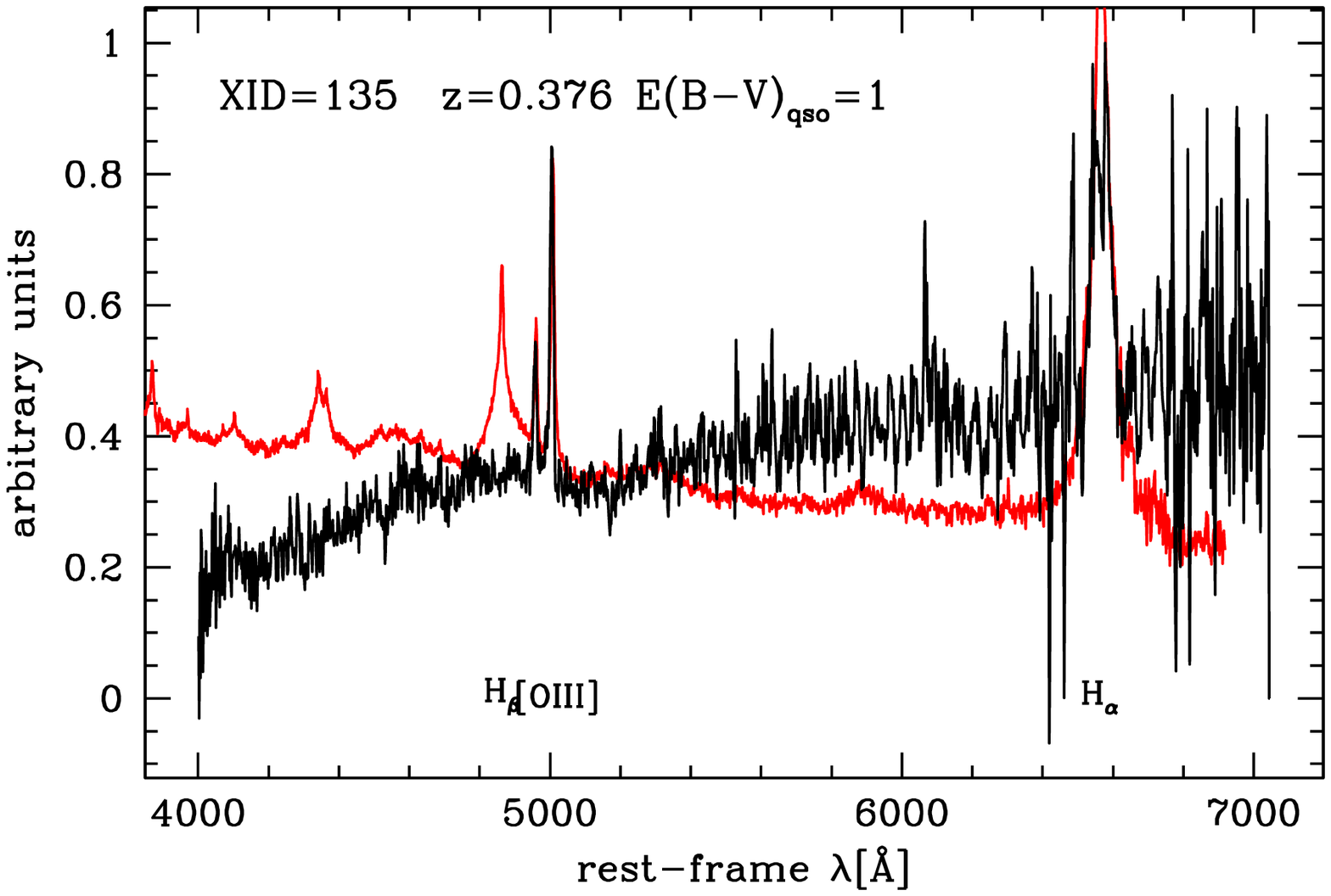}
        }
        \subfigure{\label{fig:fourth}
            \includegraphics[width=0.45\textwidth]{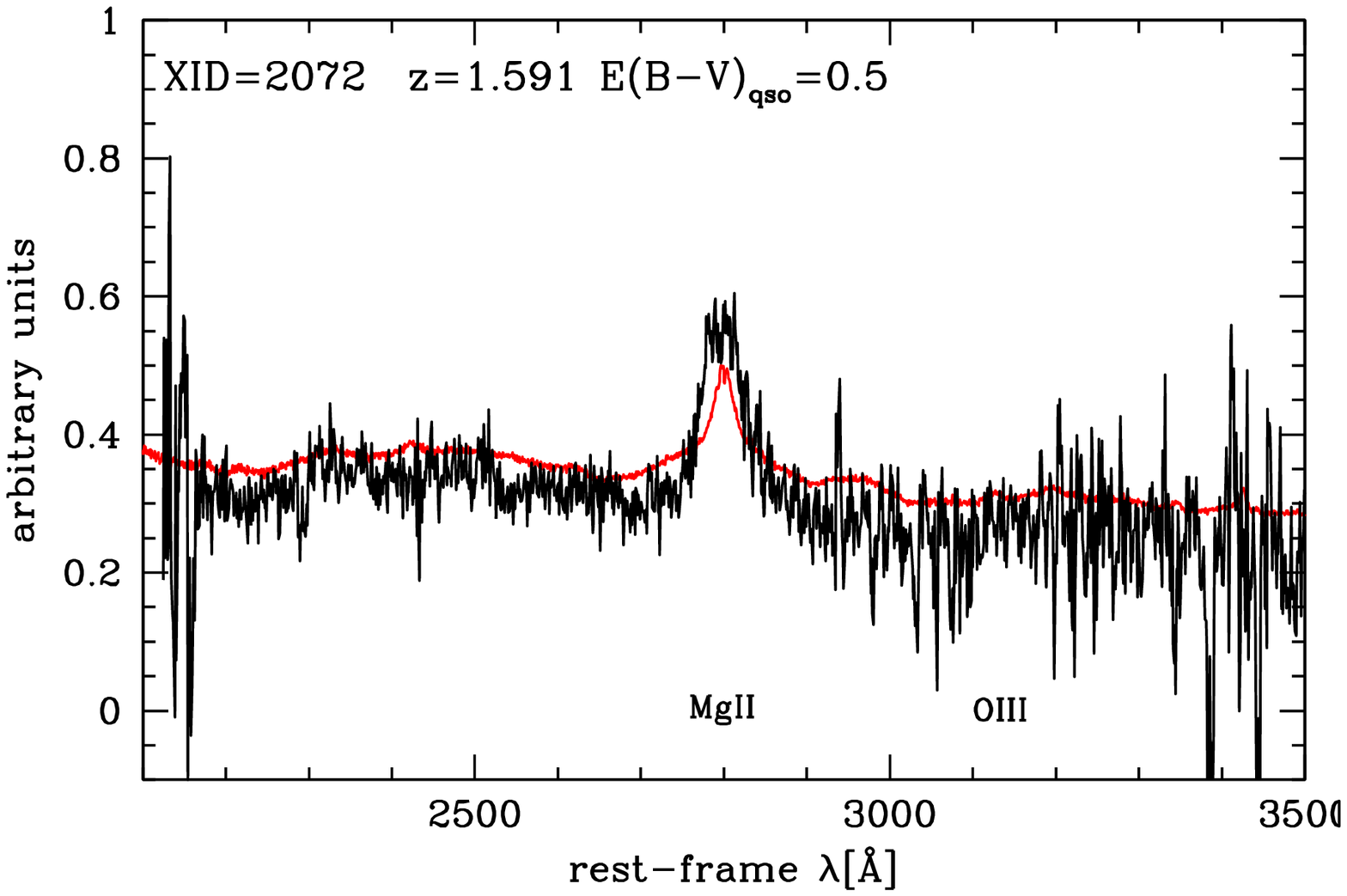}
        }\\ 
        \vspace{-2.3cm} 
        \subfigure{\label{fig:third}
            \includegraphics[width=0.45\textwidth]{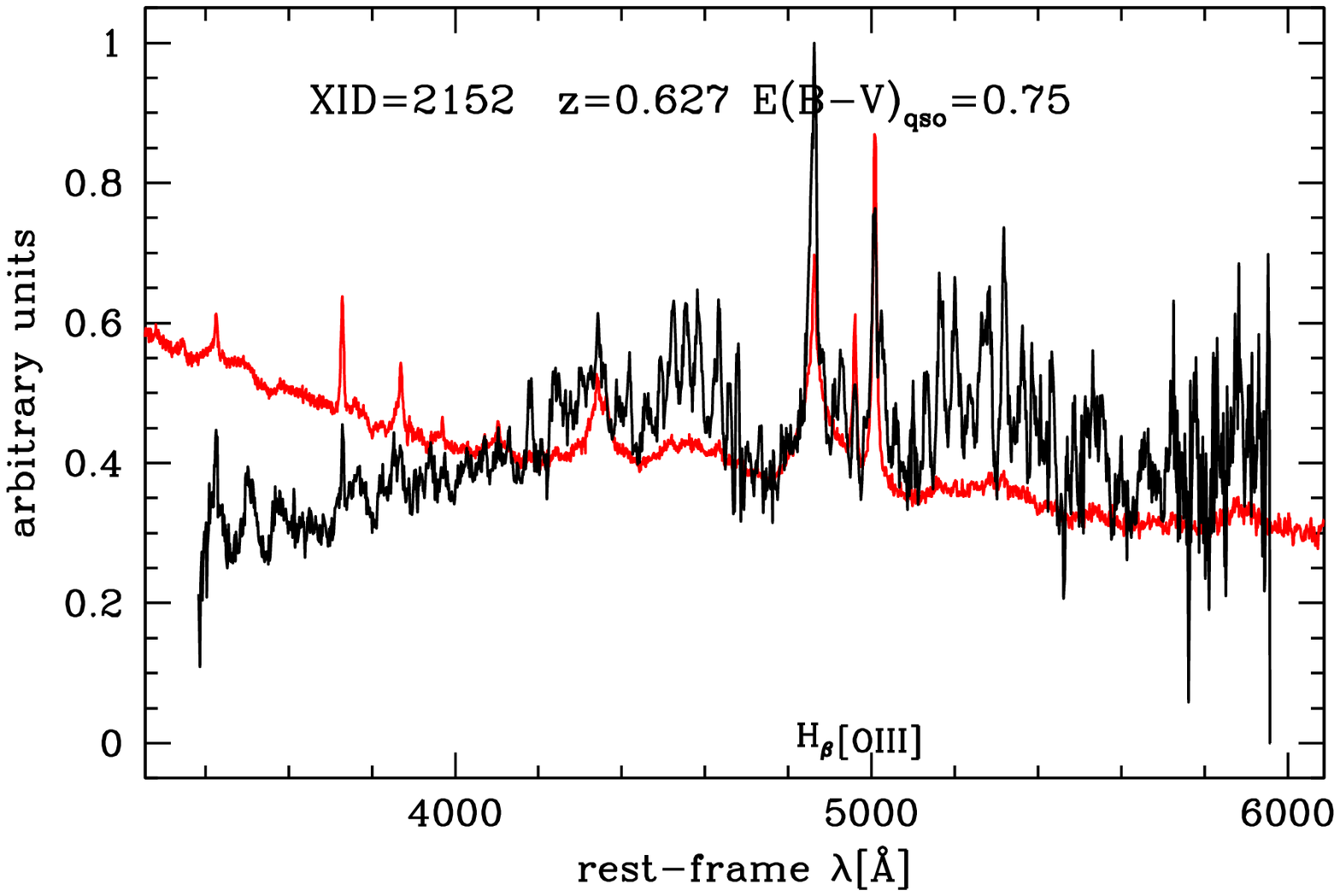}
        }
        \subfigure{\label{fig:fourth}
            \includegraphics[width=0.45\textwidth]{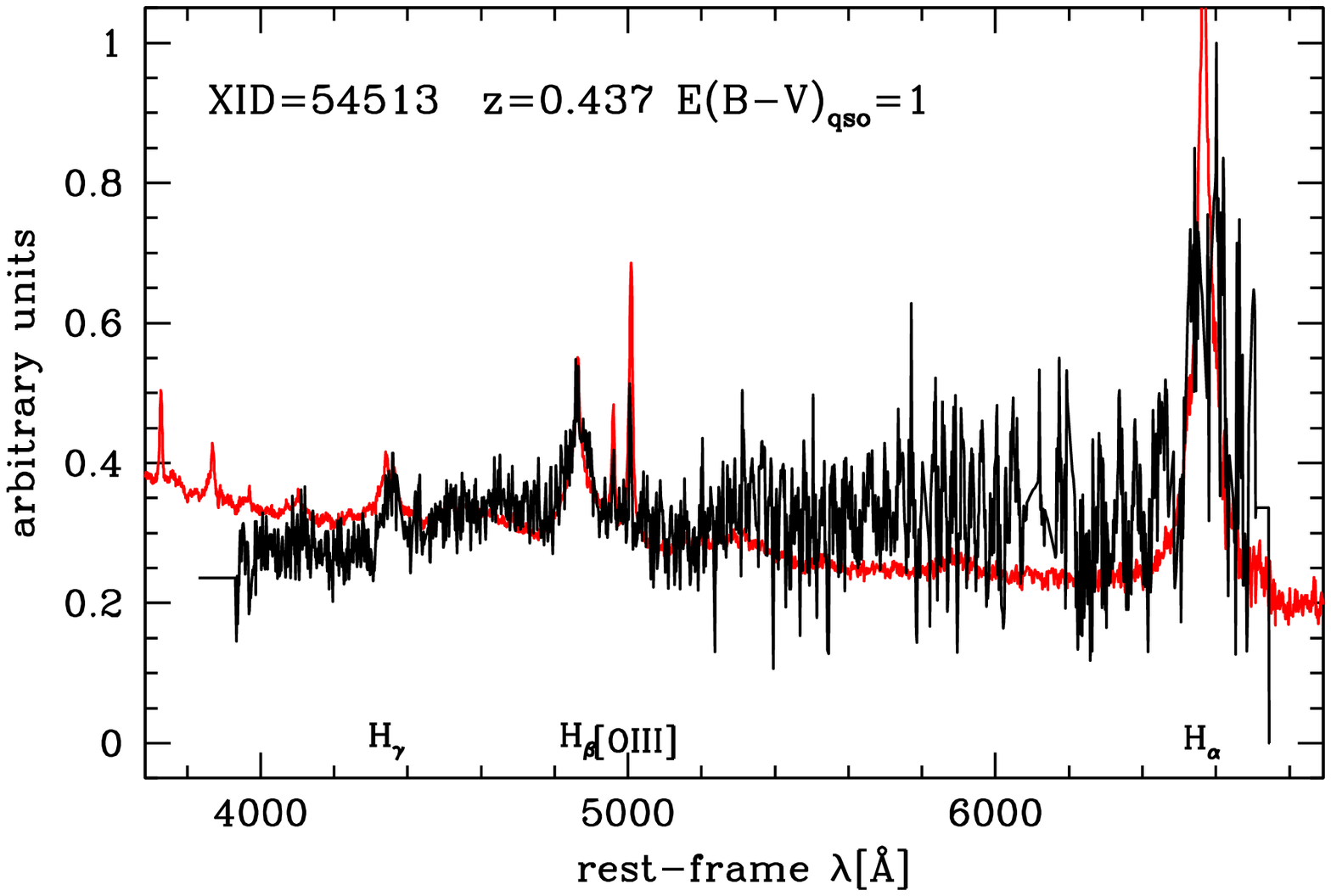}
        }
    \end{center}
    \caption{Examples of spectra of heavily reddened Type-1 AGN (zCOSMOS 20k-Bright, Lilly et al. 2007, 2013 in preparation) in the analyzed sample with identification of the main emission lines. The flux is per unit wavelength ($F_\lambda$), and normalization is arbitrary. The red line is the SDSS reddened quasar composite \citep{richards03}.}
   \label{spectra}
\end{figure}

\section{Examples of outliers in SED fitting shown in Fig.~5}
\label{appendixA}
Representative examples of SEDs of outliers plotted in the left panel of Fig.~\ref{ldiskcomphostcorrected} are shown in Fig.~\ref{sed_outliers_ldisk}. The majority of them are fitted with a reddened BBB, while the optical-UV emission is mainly coming from a young stellar population. 
From Fig.~\ref{templates} it is clear that a (partial) modeling degeneracy between reddened disks and star-forming galaxies is present in the optical-UV.
To break this degeneracy an independent measurement of the host-galaxy emission is needed, but it is not straightforward to estimate for unobscured AGN. We have then tried to verify whether the best-fit SED and the optical spectrum are qualitatively in agreement. 
The optical spectra of these outliers are therefore fitted considering a set of SDSS composite spectra from \cite{richards03}, representative of the quasar emission, and a grid of 39 theoretical galaxy template spectra from \citet[hereafter BC03]{2003MNRAS.344.1000B}, spanning a wide range in age and metallicity, to account for the stellar component. 
Five spectra, mainly at low $\Lbolobs$, have poor signal-to-noise, and therefore it is not possible to obtain a reasonable good fit for them.  
Four sources can be fitted only if, along with an SDSS reddened quasar composite, a significant host galaxy component is also included. 
For three objects (see the SED of XID=54513, 2072, and 5323) there is no detectable host galaxy component, and the spectrum is well fitted with an SDSS quasar composite spectrum alone, although one of the reddest composite spectra. \revs{The SEDs are presented in Fig.~\ref{sed_outliers_ldisk}, and some spectra example from zCOSMOS 20k-Bright (Lilly et al. 2007, 2013 in preparation) in Fig.~\ref{spectra} where we show the SDSS reddened quasar composite spectrum for clarity}. 
The best-fit SEDs of these objects are not in agreement with the spectral fit.
\par
As a further sanity check we have fitted the whole sample without host-galaxy component. The scatter in the $\Log \Lbolobs/\Lbol$ distribution is obviously smaller with a median $\Log \Lbolobs/\Lbol$ of 0.20 . 
The average obscuring fraction and the results discussed in our analysis are not significantly affected. 

\section{Examples of outliers in SED fitting shown in Fig.~6}
\label{appendixB}
\begin{figure}[ht!]
     \begin{center}
        \subfigure{\label{fig:first}
            \includegraphics[width=0.4\textwidth]{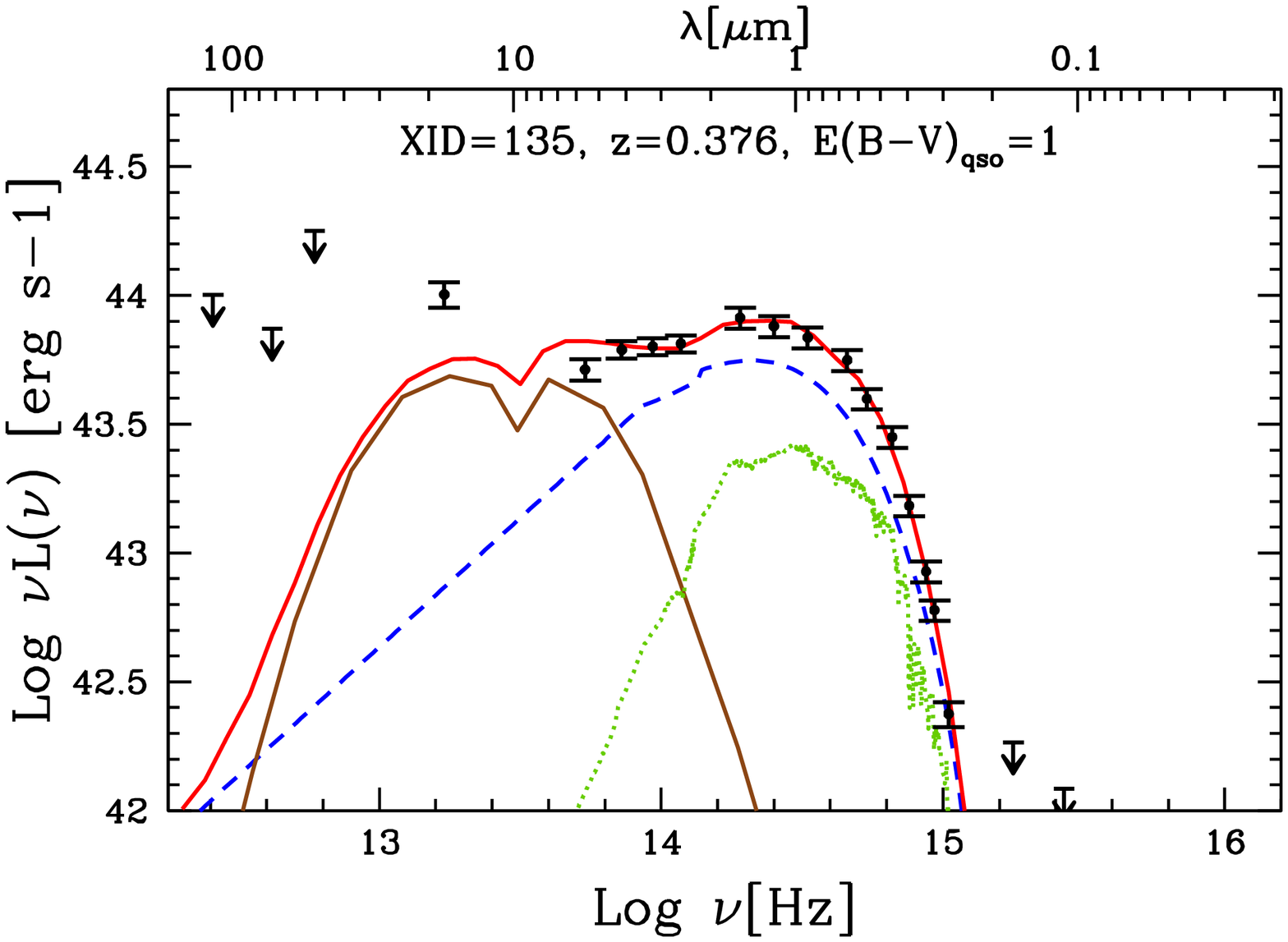}
        }
        \subfigure{\label{fig:second}
           \includegraphics[width=0.4\textwidth]{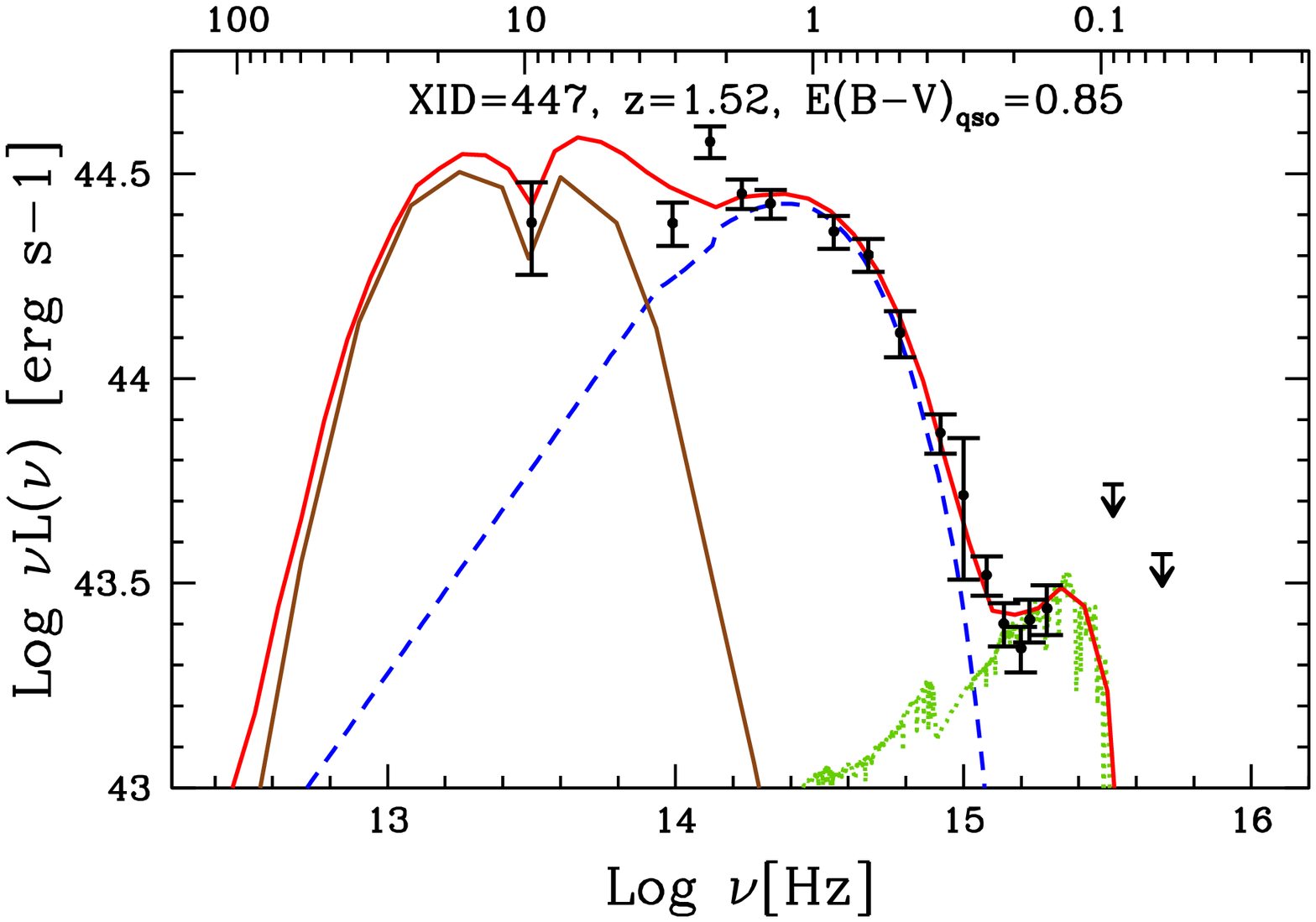}
        }\\ 
        \vspace{-2.2cm} 
        \subfigure{\label{fig:third}
            \includegraphics[width=0.4\textwidth]{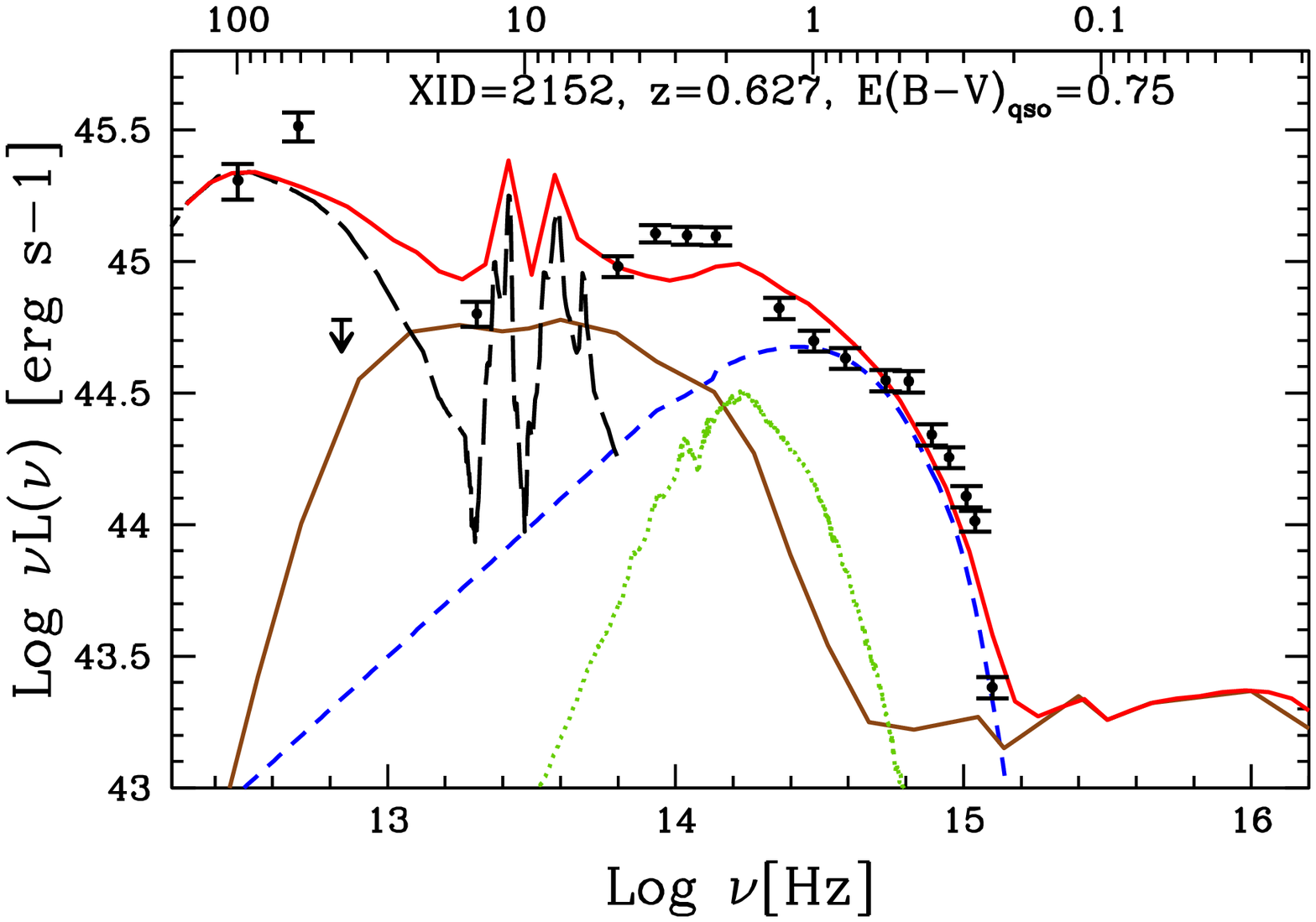}
        }
        \subfigure{\label{fig:fourth}
            \includegraphics[width=0.4\textwidth]{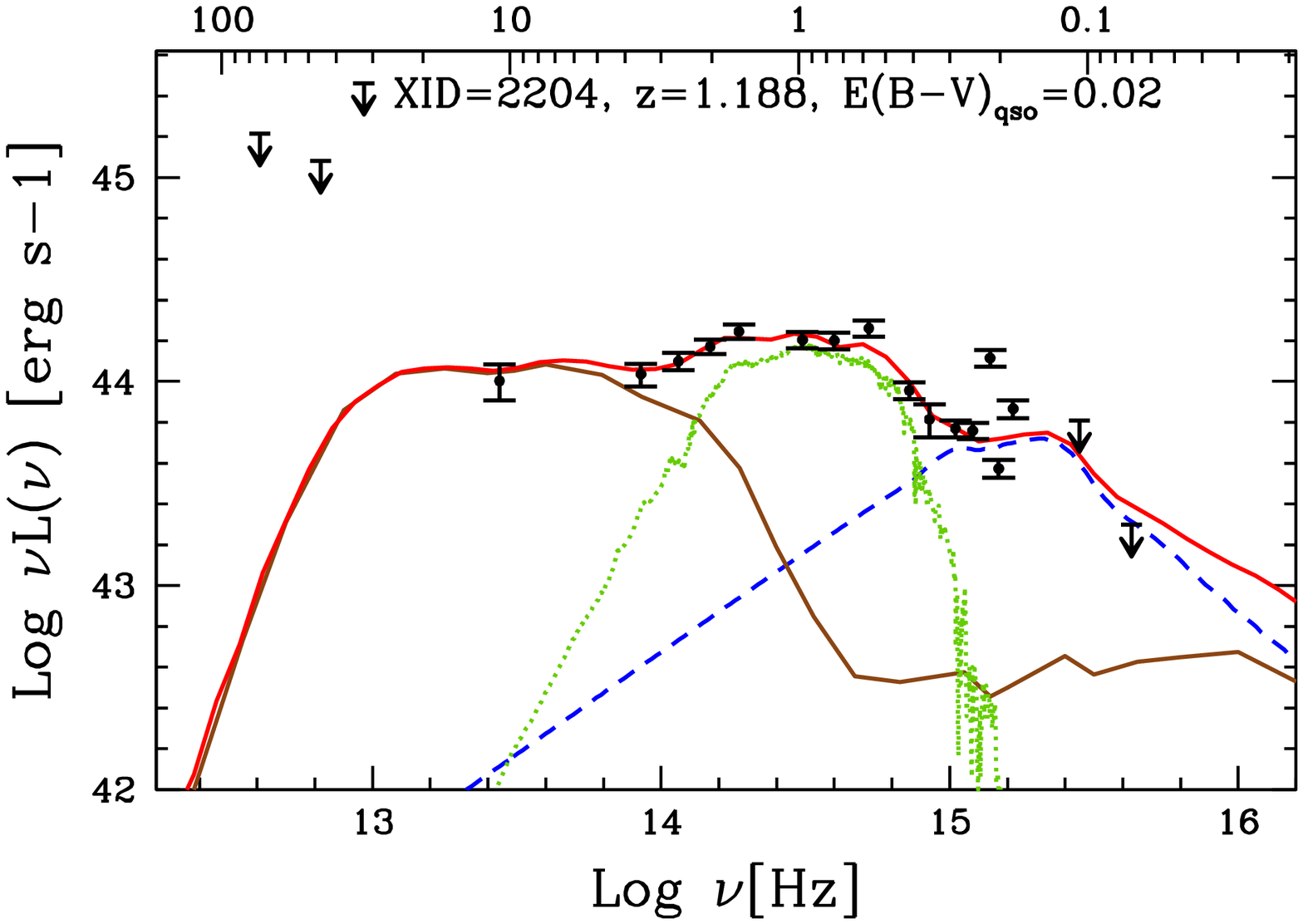}
        }\\ 
        \vspace{-2.2cm} 
        \subfigure{\label{fig:third}
            \includegraphics[width=0.4\textwidth]{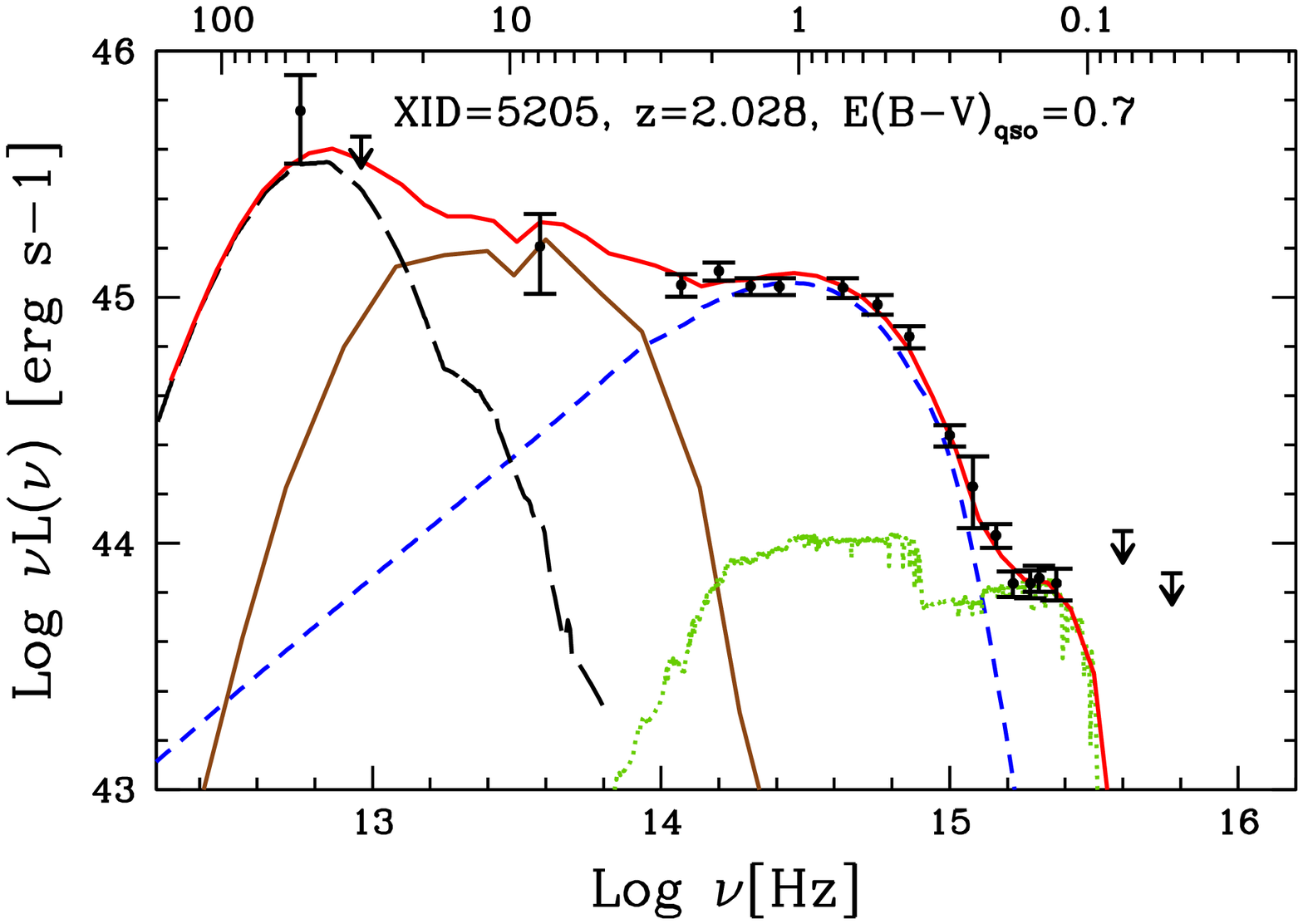}
        }
        \subfigure{\label{fig:fourth}
            \includegraphics[width=0.4\textwidth]{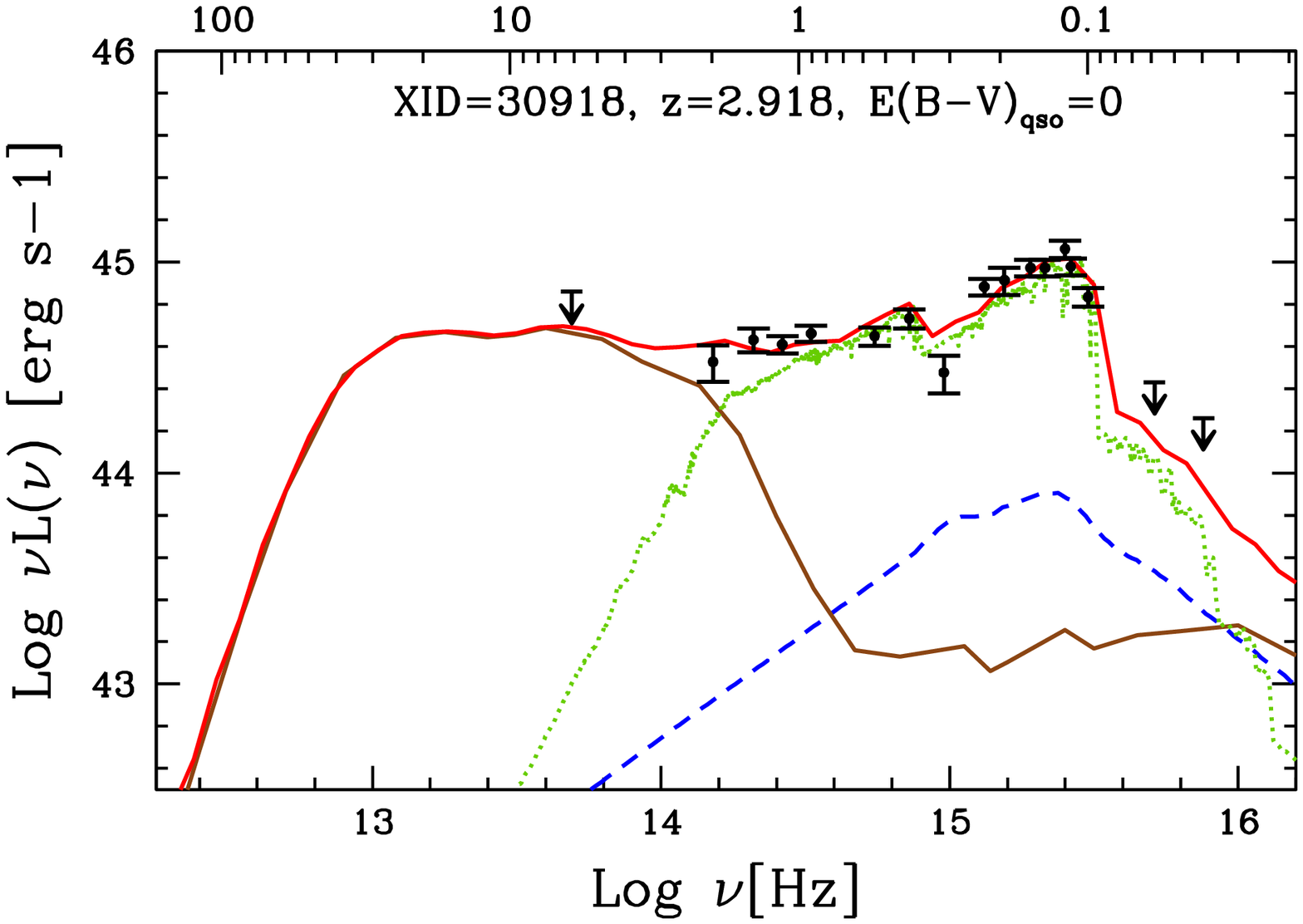}
        }
    \end{center}
    \caption{Examples of outliers in SED fitting shown in Fig.~\ref{ldderedcomp}. Keys as in Fig.~\ref{panel}. \textit{Top-middle row:} outliers at more than 3~$\sigma$ above the median. \textit{Bottom row:} outliers at more than 3~$\sigma$ below the median.}
   \label{sed_outlierslddered}
\end{figure}

The upper outliers in Fig.~\ref{ldderedcomp} are shown in the top-middle row of Fig.~\ref{sed_outlierslddered}. On average, these objects present very high reddening ($\ebvq\geq0.7$) and $\NH$ values of the order of $\sim 10^{22}$ cm$^{-2}$. 
The optical spectrum of XID=135 ($\NH\simeq1.1\times10^{22}$ cm$^{-2}$) shows a strong [\ion{O}{iii}] emission line, a faint/absent \ion{H}{$\beta$}, \revs{but the \ion{H}{$\alpha$} is broad}. The high \ion{H}{$\alpha$}/\ion{H}{$\beta$} ratio ($\gg3$) indicates high reddening, consistent with the best-fit BBB SED \revs{(see Fig.~\ref{spectra}, middle left)}.  
The spectrum of XID=447 ($\NH\simeq7\times10^{22}$ cm$^{-2}$) has a poor signal-to-noise, but it shows a broad \ion{Mg}{ii}.
\revs{
XID=2152 ($\NH\simeq1.2\times10^{21}$ cm$^{-2}$) has a strong narrow \ion{H}{$\beta$} emission line on top of a broad component with a FWHM$\sim2000$ km s$^{-1}$ (see Fig.~\ref{spectra}, bottom left). The continuum of this source shows high level of obscuration consistently with the results from our SED-fit.}
\revs{The spectrum of XID=5205 ($\NH\simeq1.2\times 10^{23}$ cm$^{-2}$) has good signal-to-noise and it shows broad emission lines, however with some level of reddening in agreement with our SED fit.}  
These four AGN might be considered intermediate Type AGN (1.5 to 1.9).
{For the other five outliers, three have a photometric redshift, while for the two with spectra similar considerations can be made: either the spectrum has a poor-signal-to-noise but with broad-lines present, and/or lines show intermediate FWHM ($\sim2000$ km s$^{-1}$).}
\par
The SED of the lower outliers in Fig.~\ref{ldderedcomp} are presented in the bottom row of Fig.~\ref{sed_outlierslddered}. 
The spectrum of XID=30918 has a \rev{good signal-to-noise with evident broad-line features (e.g., broad \ion{C}{iv})}.
XID=2204 is clearly a broad-line AGN (broad \ion{Mg}{ii}), in agreement with the best-fit, although a significant host-galaxy contamination is present.

\section{Examples of outliers in SED fitting shown in Fig.~7}
\label{appendixC}
\begin{figure}[ht!]
     \begin{center}
        \subfigure{\label{fig:first}
            \includegraphics[width=0.4\textwidth]{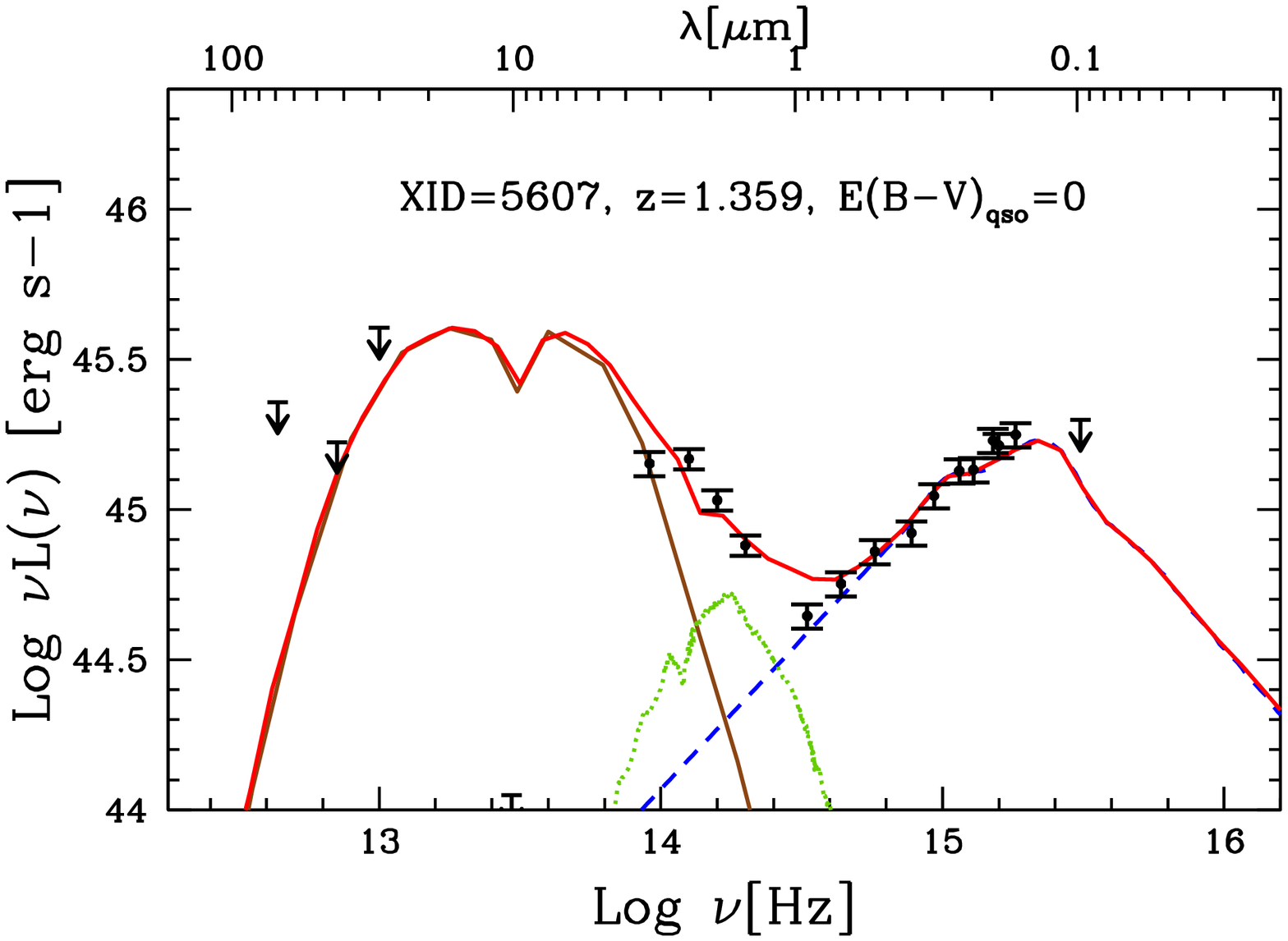}
        }
        \subfigure{\label{fig:second}
           \includegraphics[width=0.4\textwidth]{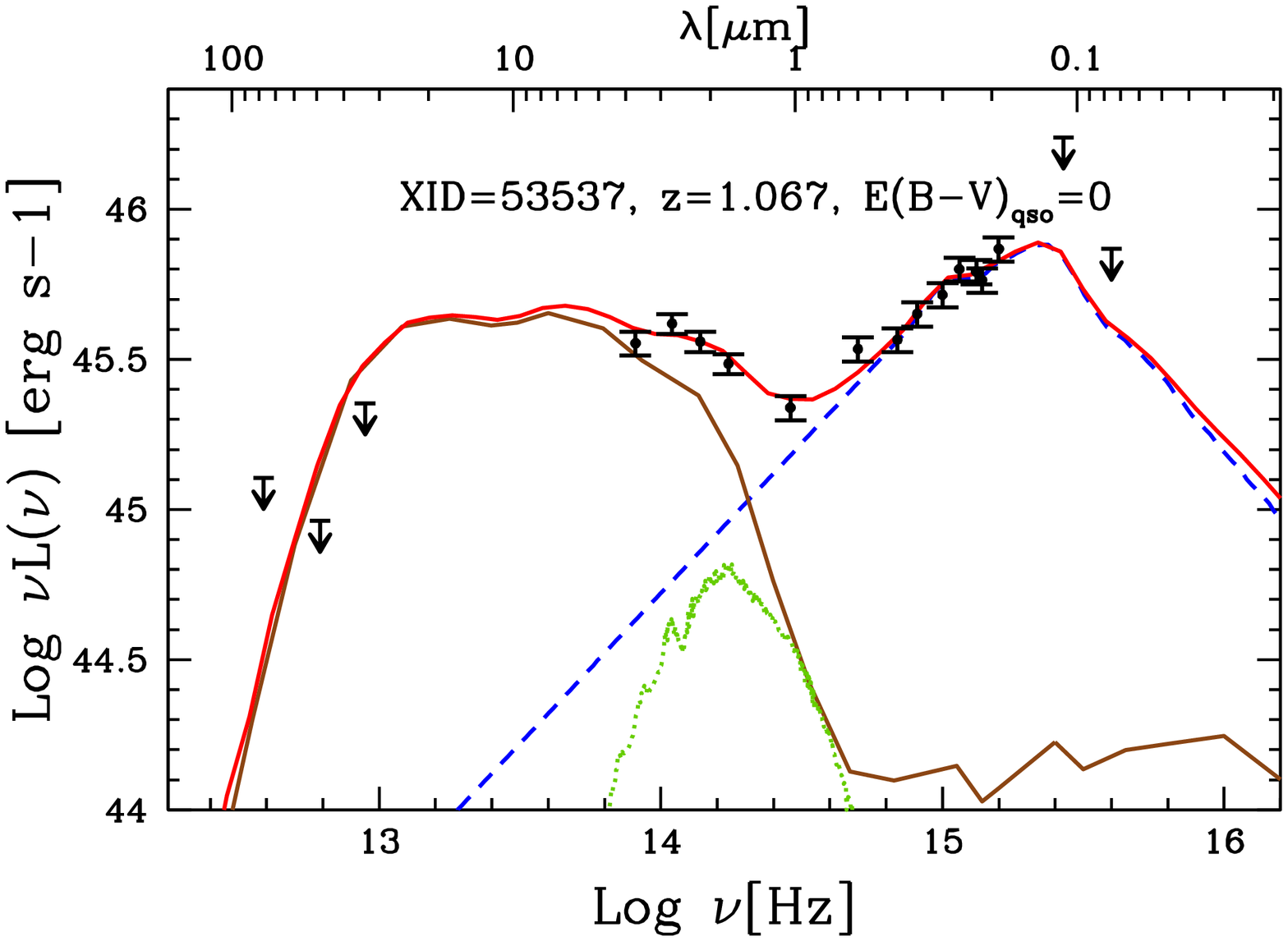}
        }\\ 
        \vspace{-2.3cm} 
        \subfigure{\label{fig:third}
            \includegraphics[width=0.4\textwidth]{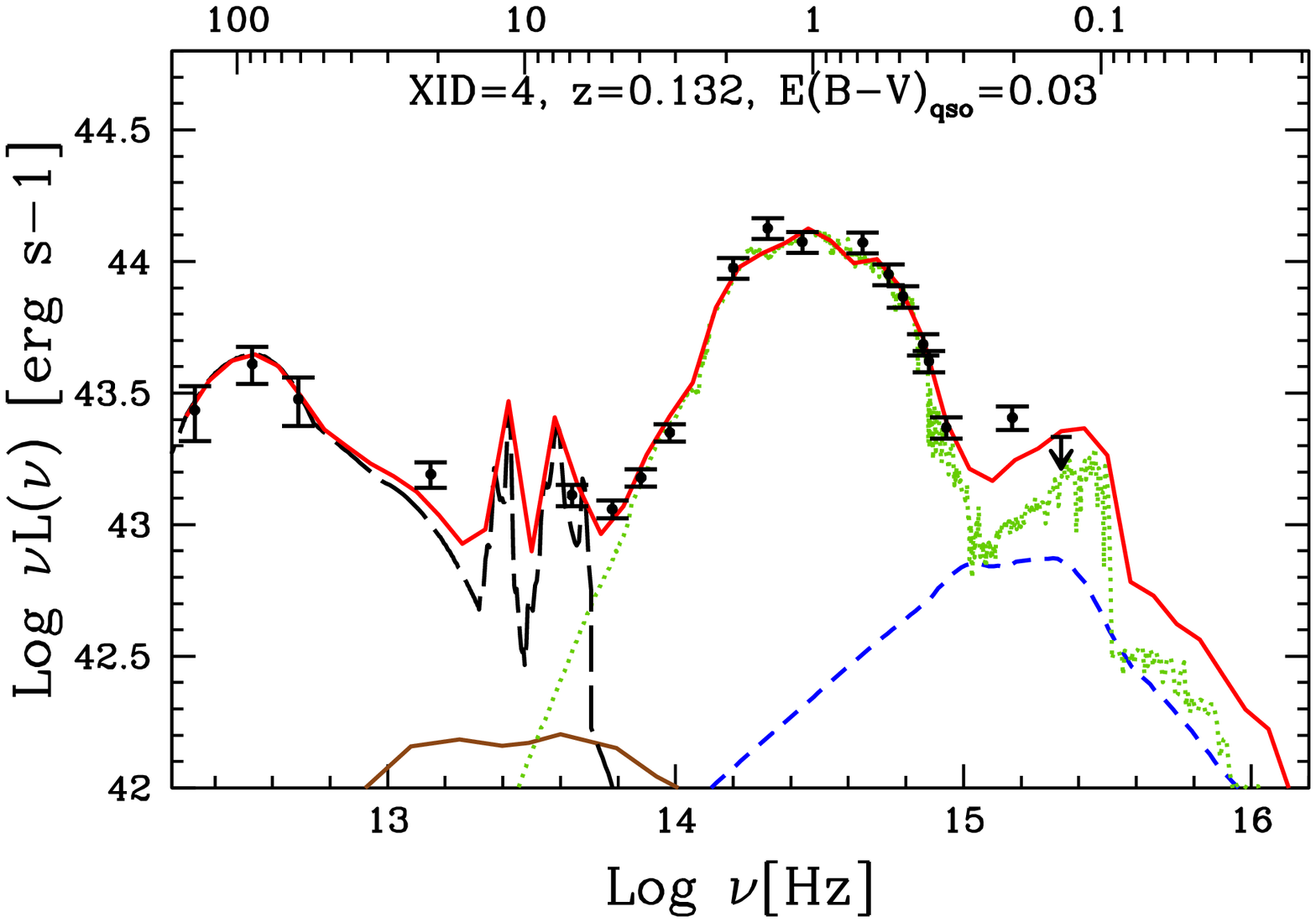}
        }
        \subfigure{\label{fig:fourth}
            \includegraphics[width=0.4\textwidth]{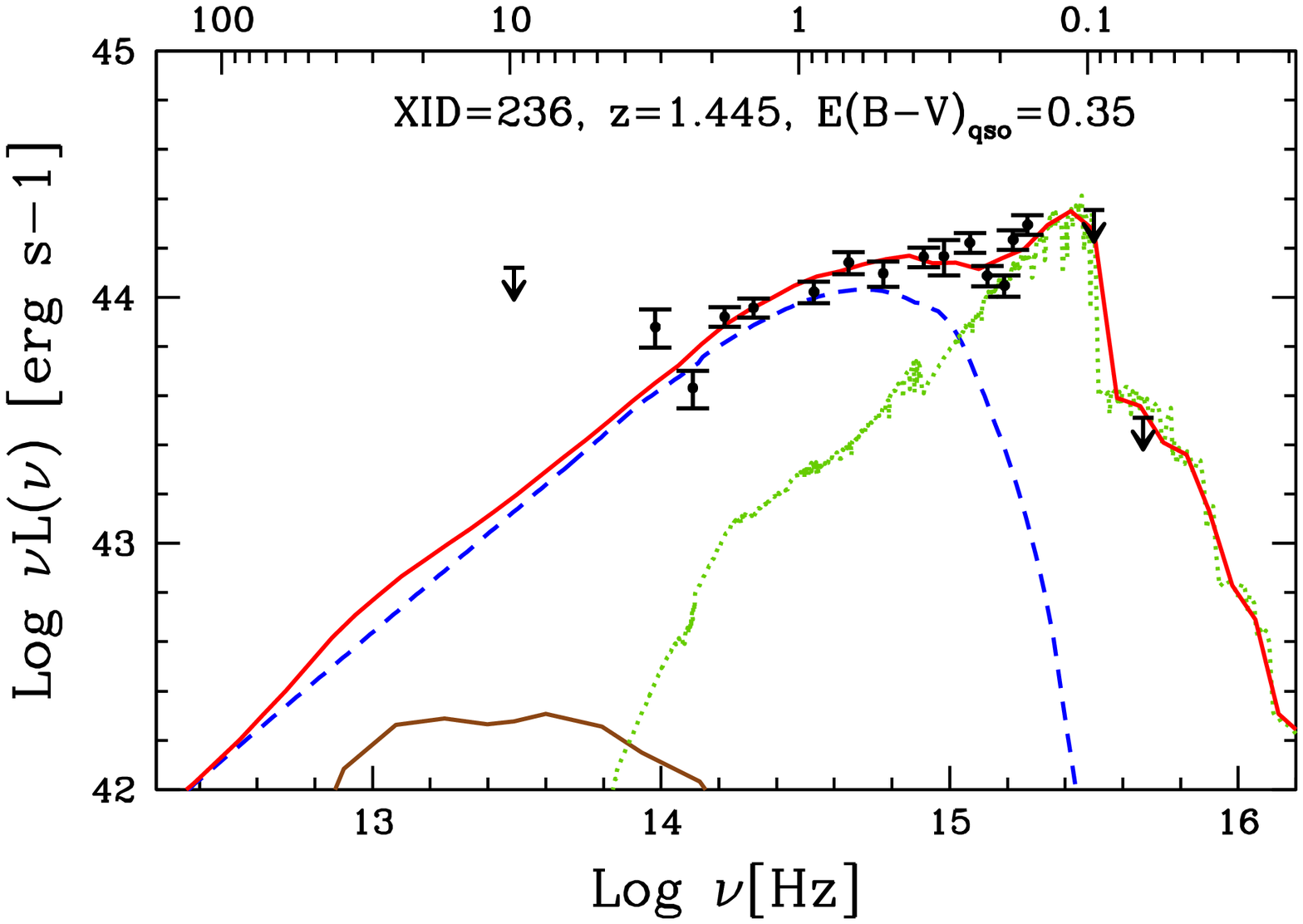}
        }\\ 
        \vspace{-2.3cm} 
        \subfigure{\label{fig:third}
            \includegraphics[width=0.4\textwidth]{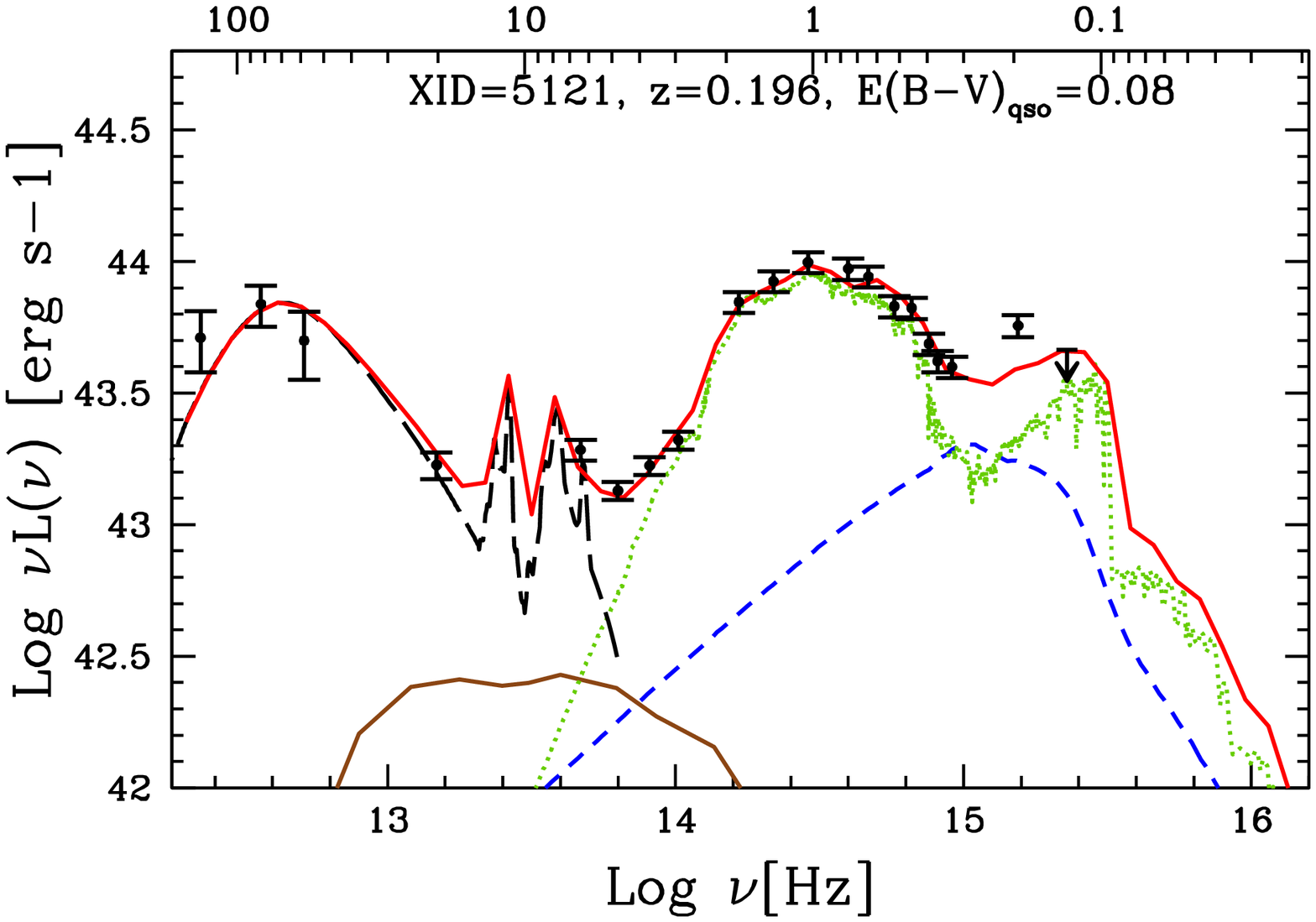}
        }
        \subfigure{\label{fig:fourth}
            \includegraphics[width=0.4\textwidth]{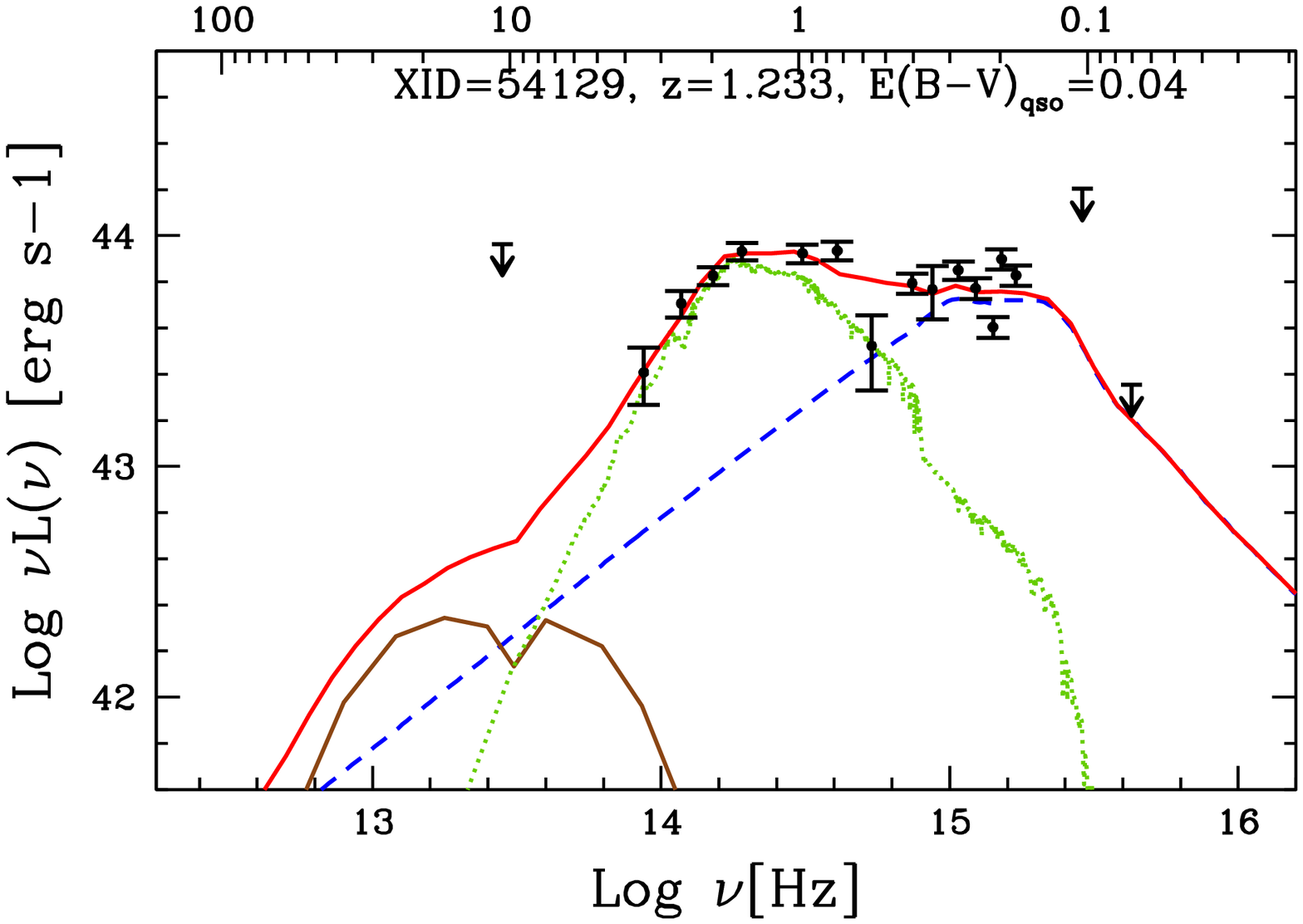}
        }\\ 
        \vspace{-2.3cm} 
        \subfigure{\label{fig:third}
            \includegraphics[width=0.4\textwidth]{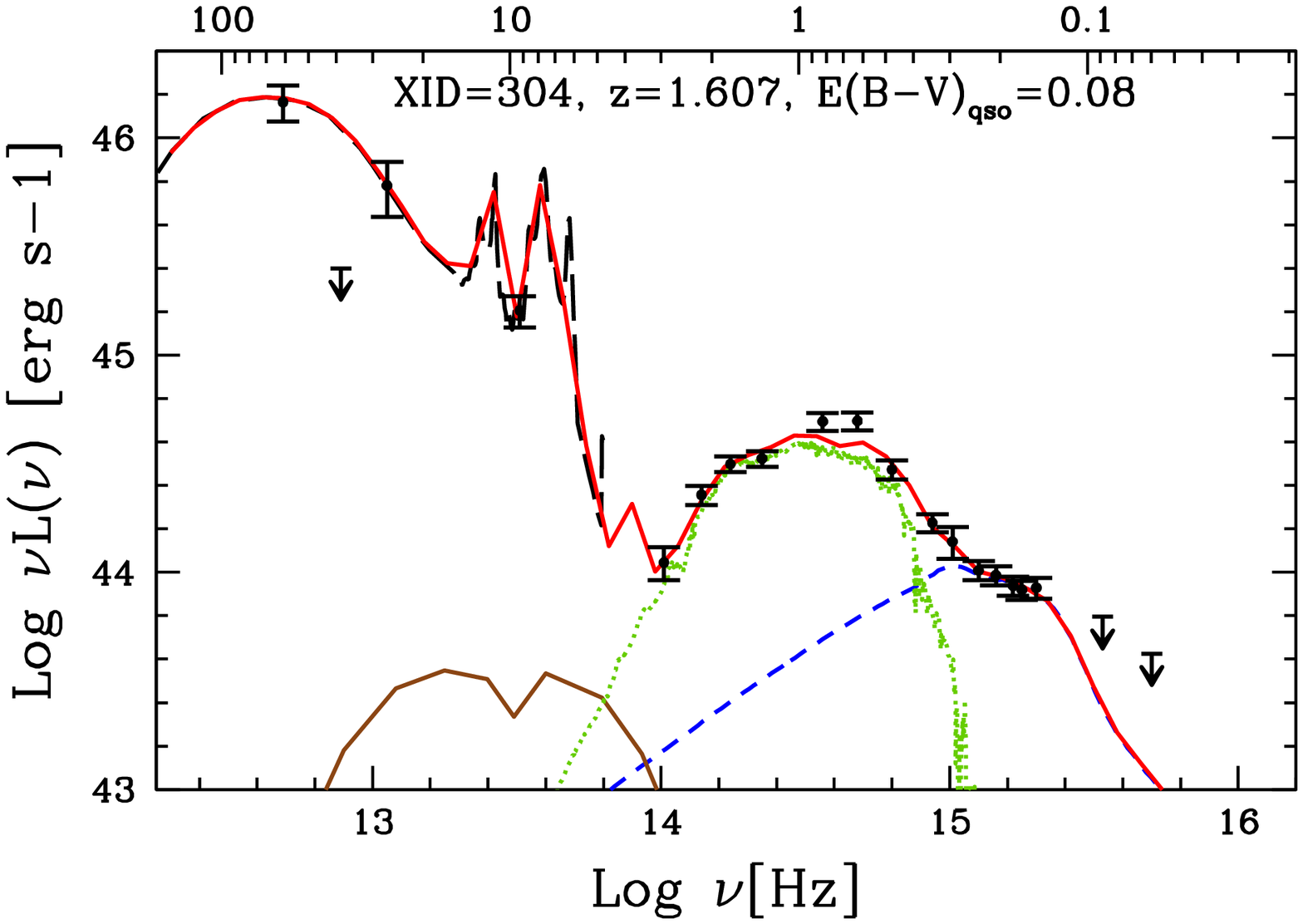}
        }
        \subfigure{\label{fig:fourth}
            \includegraphics[width=0.4\textwidth]{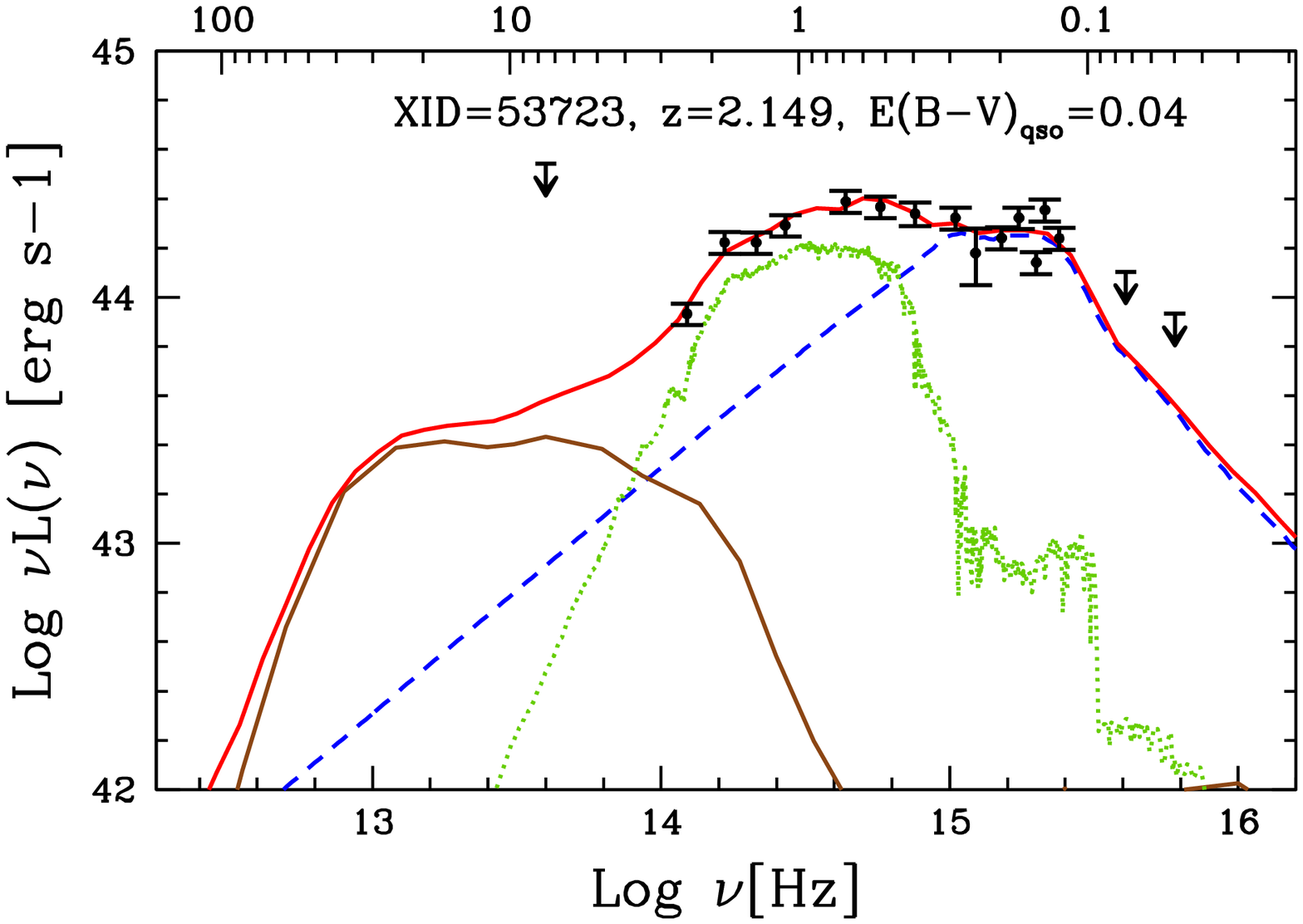}
        }
    \end{center}
    \caption{Examples of outliers in SED fitting shown in Fig.~\ref{lircomp}. Keys as in Fig.~\ref{panel}. \textit{Top row:} outlier at 3~$\sigma$ above the median of the $\langle \Log \Lir/\Ltorus\rangle$ distribution (right panel of Fig.~\ref{lircomp}). \textit{Second-third row:} outliers at 3~$\sigma$ below the median at $\Lir < 10^{44.6}$ erg s$^{-1}$. \textit{Bottom row:} outliers at 3~$\sigma$ below the median at $\Lir > 10^{44.6}$ erg s$^{-1}$.}
   \label{sed_outliers}
\end{figure}

The upper outliers in Fig.~\ref{lircomp} are presented in the top row of Fig.~\ref{sed_outliers}. These objects present an SED with a strong near-infrared bump. The SED of XID=5607 was already discussed in \citet{2012arXiv1210.3044H} and was presented as a good candidate for an AGN at the beginning of the "blow-out phase", where the nucleus emerges from its dusty cocoon and starts dominating in the optical-UV. 
However, we caution the reader that the $\Ltorus$ values for these two outliers are uncertain given that these sources do not have a 24~$\mu$m detection.
\par
Examples of outlier' SEDs at $\Lir < 10^{44.6}$ erg s$^{-1}$ are presented in the second and third row of Fig.~\ref{sed_outliers}. These sources are strongly galaxy-dominated with weak near-infrared emission coming from the torus, and all of them have MIPS detection at 24 $\mu$m. At $\Lir > 10^{44.6}$ erg s$^{-1}$ outliers still show a strong galaxy emission, but the disk is clearly present (see bottom row in Fig.~\ref{sed_outliers}). Eight sources in this region do not have $24~\mu$m detection, making these estimates uncertain (as shown by the large error bars), while 6 sources also have {Herschel} data and a SED similar to XID=304 ($z=1.607$) plotted in Fig.~\ref{sed_outliers}. The latter object has a best-fit torus luminosity of $\Ltorus = 10^{44}$ erg s$^{-1}$, much lower than what is measured from the observed SED ($\Lir = 10^{45.6}$ erg s$^{-1}$). In our fit this source appears to be a composite AGN/starburst SED, where the AGN component is extremely weak in comparison with the starburst one. This is in good agreement with the results presented by \citet{2012arXiv1210.3044H}. In this paper XID=304 has been fitted by a ULIRG Arp 220 SED \citep{polletta07}, but weak broad emission lines are present in the spectrum. 
This result has been interpreted by Hao et collaborators as an evidence for a "new born quasar", 
where the AGN starts becoming visible during a merger triggered starburst \citep{2006ApJS..163....1H}.
We cannot exclude this possibility, but a detailed separate analysis is required (Hao et al., in preparation).

\bibliographystyle{apj}
\bibliography{bibl}

\end{document}